\documentclass[twocolumn,trackchanges]{aastex701}

\usepackage{amsmath}
\usepackage{amssymb}
\usepackage{amsfonts}

\begin{document}

\title{Eclipses of Nearby Millimeter-bright Galactic Nuclei by Stars in Nuclear Star Clusters}

\author[orcid=0000-0001-6450-1187,sname='MZ']{Michal Zaja\v{c}ek}
\affiliation{Department of Theoretical Physics and Astrophysics, Faculty of Science, Masaryk University, Kotl\'{a}\v{r}sk\'{a} 2, 611 37 Brno, Czech Republic}
\email[show]{zajacek@physics.muni.cz}

\begin{abstract}

It is of a general interest to look for signatures of stellar bodies orbiting supermassive black holes (SMBHs) in galactic nuclei other than the Galactic center. Previously stellar transits were analyzed in UV, optical, and X-ray domains as well as potential microlensing signatures due to more compact bodies orbiting SMBH accretion disks. Here we complement previous studies by considering nearby ($z=0.001$) millimeter-bright active galactic nuclei targeted by different facilities in the millimeter domain. At these wavelengths the jet base and the nuclear ring are sufficiently compact so that they can be occulted by large evolved stars in dense nuclear star clusters. We find that in the millimeter domain evolved stars with stellar radii of $\gtrsim 500\,R_{\odot}$ can cause eclipses with the relative depth of $\sim 10-40\%$. Typical recurrence timescales are at least 10 years and the eclipse durations are $\sim 10$ days. Towards lower frequencies the eclipse temporal profiles become shallower and broader while towards higher frequencies they are deeper and narrower. Although expected to be rare due to selection effects and evolved stars being prone to tidal disruption, eclipses of millimeter radio cores can be applied to estimate the distance of the extragalactic star from the SMBH and its stellar radius. In case the eclipse is recurring, we provide a relation for estimating the SMBH mass.

\end{abstract}



\section{Introduction} 
\label{sec:introduction}

Supermassive black holes (SMBHs) whose presence can be detected dynamically or using electromagnetic signatures are associated with galactic nuclei \citep{2005SSRv..116..523F,2013ARA&A..51..511K,2014ARA&A..52..589H}. There is a large amount of evidence that SMBHs are typically not isolated but surrounded by dense nuclear star clusters (NSCs). These dense stellar systems are present in $\sim 70\%$ of the intermediate-mass galaxies with the total stellar mass in the range $\sim 10^8-10^{10}\,M_{\odot}$ \citep{2020AARv..28....4N}, though the occupation fraction is a function of the galaxy mass. 

Regardless of their origin \citep[in-situ star formation or infall of globular clusters;][]{2022A&A...658A.172F} NSCs are old structures with the age of $\sim 10\,{\rm Gyr}$ predominantly composed of late-type stars and compact remnants that are expected to be dynamically relaxed via two-body relaxation \citep{2020A&A...641A.102S}. The number density profile is therefore generally characterized as a simple power law with $n_{\star}\propto r^{-\gamma}$ with $\gamma \simeq 3/2$ corresponding to a relaxed stellar system with the dominant potential of the SMBH \citep{1977ApJ...216..883B}. However, the inner density profile may be flattened by different dynamical processes depleting red giants \citep[merger with another SMBH, stellar collisions, star-black hole interactions, star-disk collisions, and star-jet interactions; see][and references therein]{2020ApJ...903..140Z}. Therefore we can distinguish cuspy and core-like NSCs. Since the efficiency of red-giant depletion mechanisms often depends on the stellar radius, the cusp or the core presence can generally differ among lower-mass (fainter) and higher-mass (bright) giants in the same NSC, such as also in the case of the NSC in the Milky Way \citep{2020A&A...641A.102S}. 

NSCs are the densest stellar systems in a galaxy. Since they typically surround the SMBH, it is expected that NSC stars can occasionally obscure the background SMBH surrounded by an accretion flow \citep{2013ApJ...762...35B}. In special cases, for bound binary SMBH systems, the accretion disk around the foreground SMBH can eclipse the accretion disk around the background SMBH \citep{2021MNRAS.503.1703I}.

In addition, for radio AGN, the accreting SMBH launches the jet whose brightest part referred to as radio core, which denotes the surface where the optical depth due to synchrotron self-absorption reaches unity, can be occulted by extended gas clouds as well as stars. The radio core in radio AGN has the angular radius that is a function of the observing frequency due to variable opacity, $\theta_{\rm c}\sim (0.1-1)(\nu/1\,{\rm GHz})^{-1}\,{\rm mas}$ \citep[core-shift effect;][]{1998A&A...330...79L,2005AJ....130.2473K}. Using this relation scaled to 1 GHz, the core has a linear radius of $R_{\rm c}\approx \theta_{\rm c}D_{\rm A}(z)\sim 0.2-2\,{\rm pc}$ for the angular-diameter distance $D_{\rm A}(z)$ at the redshift of $z=0.1$. This is too large to be obscured by typical main-sequence or evolved stars in the NSC, though it could be obscured by a more extended molecular cloudlet. By going to the millimeter domain at $\nu=230\,{\rm GHz}$, the linear size decreases to $R_{\rm c}\lesssim 1700\,{\rm AU}$ for $z=0.1$. As one goes towards nearby radio AGN ($z\sim 0.001$) and increases the observing frequency to $\sim 100\,{\rm GHz}$ (3 mm), the angular millimeter core radius shrinks to $\theta_{\rm c} \sim 0.001-0.01\,{\rm mas}$ because of the core-shift effect, which becomes $R_{\rm c}\sim 920 - 9200\,R_{\odot}$. At 230 GHz the linear core radius is $R_{\rm c}\sim 400 - 4000\,R_{\odot}$, which is comparable to the radii of evolved red giant/asymptotic giant-branch stars. The stellar radius of the occulting star that can completely obscure the background core is given by $R_{\star}=R_{\rm c}(D_{\rm A}-r_{\star})/D_{\rm A}$, where $D_{\rm A}$ is the angular-diameter distance of the host galaxy and $r_{\star}$ is the distance of the star from the SMBH. For $r_{\star}\ll D_{\rm A}$ we have $R_{\star}\sim R_{\rm c}\sim 2000\,R_{\odot}$ considering the intermediate value of the radio-core radius. Large evolved stars bound to the SMBH within the NSC can thus repeatedly occult the radio core in the mm/submm domain if their radius is comparable to the length-scale of the mm radio core. The estimate for the fractional decrease in flux density is $100\% \times \Delta F_{\nu}/F_{\nu}\sim 100 \times \pi\theta_{\star}^2/(\pi \theta_{\rm c}^2) = 100 (R_{\star}/R_{\rm c})^2\sim 6.25\% (R_{\star}/500\,R_{\odot})^2(R_{\rm c}/2000\,R_{\odot})^{-2}$, where we considered $\theta_{\rm c}\sim 1(\nu/1\,{\rm GHz})^{-1}\,{\rm mas}$ and $z=0.001$. Hence, a $10\%$ variability in the millimeter radio light curves is expected with a potential periodicity when the star is bound to the SMBH.   

There is also an indication that the star formation in the Milky Way NSC is episodic rather than continuous \citep{2020A&A...641A.102S}. When generalized to other NSCs, this implies that depending on the supply of the dense molecular gas or transport mechanisms there are several stellar populations present in the sphere of the SMBH influence \citep{2020AARv..28....4N}. These stars are characterized by certain mean cross-sections given by effective stellar radii. In the Milky Way NSC, most of the mass ($\sim 80\%$) is contributed by old, evolved stars that formed 8-10 Gyr ago. Then there is a minor contribution of stars that formed 2-3 Gyr ago ($\sim 15\%$) and a few percent is contributed by stars that formed less than 100 Myr ago. The mean radius for the NSC stars is comparable to the Solar radius ($1.5-3\,R_{\odot}$) because small light stars (dwarfs) dominate the population due to the longest lifetime. Using the Milky Way NSC star formation history \citep{2020A&A...641A.102S} together with stellar-population synthesis models including the asymptotic giant branch phase \citep[e.g.][]{2017ApJ...835...77M}, only a very small fraction of stars are expected to reach radii of several hundred solar radii at any given time. Therefore, we introduce the fraction $f_{>R}$ of stars exceeding $R_{\star}\sim 100\,R_{\odot}$, i.e. those capable of the eclipse, as a free parameter in the range $f_{>R} \sim 10^{-6}-10^{-3}$. Although this appears to be a small representation, their presence makes the eclipse of the mm-emission core region statistically plausible especially when a larger number of nuclei is taken into account.

In this contribution, we elaborate on the curious jet base-star size correspondence. As there are several radio AGN within or close to $z=0.001$, a detection of a single dimming event or a SMBH eclipse (hereafter called also ``dip'') in the mm/submm light curves could turn such a source into a new dynamical laboratory where the dip duration, depth as well as its potential recurrence could be used to estimate the stellar distance, stellar radius, the SMBH mass, and to learn about the NSC composition in a nearby nucleus with a jetted AGN. Since these measurements concern nearby radio AGN observed in the mm domain, this analysis is also relevant for assessing any obscuration and potential related quasiperiodic variability detected during mm Very Long Baseline Interferometry (VLBI) observations of galactic nuclei, in particular Event Horizon Telescope (EHT) observations \citep{2019ApJ...875L...1E}. Hence, not only can light curves be affected but also visibilities and reconstructed images of galactic nuclei in the millimeter domain \citep{2008ASPC..386..186K}.

The paper is structured as follows. In Section~\ref{sec_results} we analyze the setup of a passing star in front of the radio core in mm domain and we derive the required radius of such a star to cause occultation in contrast to microlensing (Subsec.~\ref{sec:occultation vs. microlensing}). Subsequently, we derive the range of dip depths (Subsec.~\ref{sec:depth}), durations, and recurrence timescales (Subsec.~\ref{sec:duration_recurrence}). We also specifically derive a relation for the estimate of the SMBH mass (Subsec.~\ref{sec:mass}). Furthermore, a likelihood of star-related dips in the radio light curve is estimated considering general NSC properties (Subsec.~\ref{sec:likelihood}). Synthetic eclipse temporal profiles for different eccentricities, inclinations, and frequencies are presented in Section~\ref{sec:modelling} including intrinsic variability, measurement uncertainty, and the ring-like geometry of the millimeter nuclear emission. In Section~\ref{sec:discussion} we discuss further aspects of potential radio-core eclipses, namely, the obscuration due to TDE streams and a specific observed case with repeating dips; furthermore, we analyze the parameters of observing facilities in terms of the flux sensitivity, measurement uncertainty, and the cadence. We summarize the main results in Section~\ref{sec:conclusions}.

\section{Results}
\label{sec_results}

\subsection{Model set-up}

For the following calculations, we consider the SMBH mass of $M_{\bullet}=5\times 10^7\,M_{\odot}$ that is orbited by the star with $m_{\star}=1\,M_{\odot}$ unless stated otherwise. For nearest radio-loud AGN, there is a large spread in SMBH masses. We adopt the value close to the SMBH mass in Centaurus A ($z\simeq 0.00183.$) \citep{2009MNRAS.394..660C}. When we follow the general core-shift effect for the core angular radius $\theta_{\rm c} \sim 0.5 (\nu/1\,{\rm GHz})^{-1}\,{\rm mas}$, we arrive at the linear core radius at $z\sim 0.001$, $R_{\rm c}\sim \theta_{\rm c}D_{\rm A}(z)\sim 2000\,(\nu/230\,{\rm GHz})^{-1}R_{\odot}\simeq 19\,r_{\rm g}$, where $r_{\rm g}=GM_{\bullet}/c^2\simeq 106\,(M_{\bullet}/5\times 10^7\,M_{\odot})\,R_{\odot}$ is the gravitational radius of the SMBH. This implies that for the mm radio core with the length-scale of $R_{\rm c}\sim 10\,r_{\rm g}$, an evolved red giant/AGB star with $R_{\star}\sim 1000\,R_{\odot}$ can significantly eclipse it (for the uniform circular emission, we have $100\Delta F_{\nu}/F_{\nu}\sim 100(R_{\star}/R_{\rm c})^2 \sim 89\%$). Stars with smaller radii can only lead to fractional dimming characterized by shallow dips in the mm light curve.   

More precisely, in addition to the jet base, the mm emission can generally consist of the gravitationally lensed accretion inflow and/or counter-jet, which forms a characteristic ring-like feature with the linear radius corresponding to the lensed last photon orbit, $R_{\rm r}\simeq 3\sqrt{3}\,r_{\rm g}\sim 5.2\,r_{\rm g}$ in the reconstructed mm images of Sgr~A* and M87*. Furthermore, the extended jet can contribute on larger scales of $\sim 10-100\,r_{\rm g}$. In contrast to the jet base -- the radio/mm core $R_{\rm c}$, whose size is inversely proportional to the observing frequency due to variable opacity and depends linearly on the angular-diameter distance, the mm ring linear radius of $R_{\rm r}\sim 3\sqrt{3}\,GM_{\bullet}/c^2$ is only proportional to the SMBH mass. Depending on the dominant nuclear emission at millimeter wavelengths -- jet base/radio core or mm ring -- the linear size of the SMBH-associated millimeter emission is or is not angular distance-dependent, respectively, which affects the distance-limitation of the galaxy sample as well as stellar sizes that can eclipse the nuclear emission.

\begin{figure}
    \centering
    \includegraphics[width=\columnwidth]{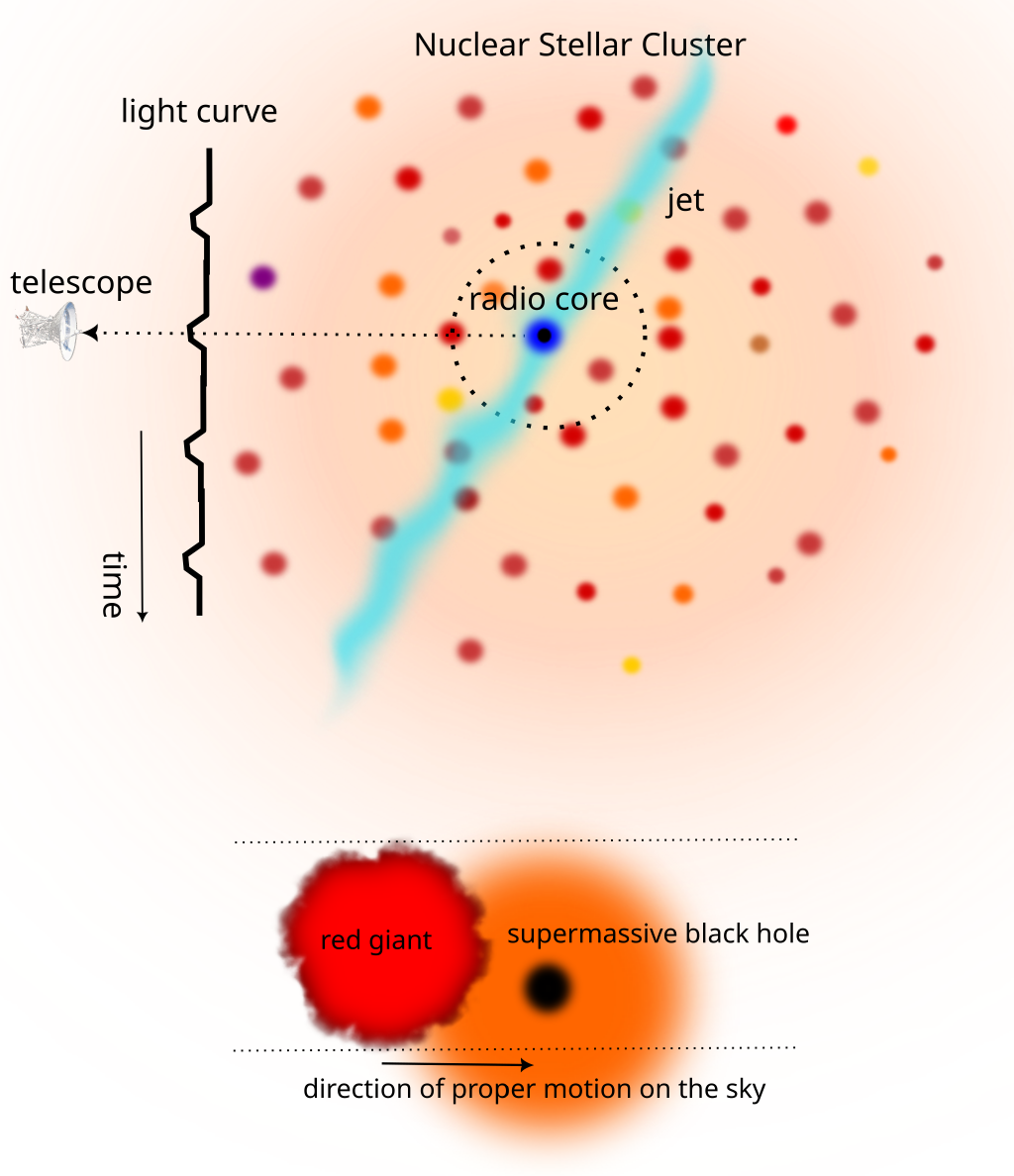}
    \caption{Illustration of a setup where a radio core (in the mm domain) is occulted by a bound evolved, red giant star on an approximately circular orbit around the SMBH. This is manifested by potentially recurring dips in the mm radio light curve that are depicted on the left. The figure inset at the bottom illustrates the basic geometry of the galactic nucleus occultation by an evolved (red-giant) star.}
    \label{fig_illustration}
\end{figure}

The star orbits the SMBH with the semi-major axis of $a_{\star}$, which for the assumed circular orbit is also its mean distance, $r_{\star}\sim a_{\star}$. We assume that the stellar orbit is well within the sphere of gravitational influence of the SMBH, which can be estimated as \citep{2013degn.book.....M},
\begin{align}
    r_{\rm SI} &=\frac{GM_{\bullet}}{\sigma_{\star}^2}\,\notag\\
    &\simeq 8.5\,\left(\frac{M_{\bullet}}{5 \times 10^7\,M_{\odot}} \right) \left(\frac{\sigma_{\star}}{159\,{\rm km\,s^{-1}}} \right)^{-2}{\rm pc}\,,
    \label{eq_sphere_SMBHinf}
\end{align}
where the stellar velocity dispersion $\sigma_{\star}\simeq 200 \times 10^{(\log{M_{\bullet}}-8.12)/4.24}\sim 159\,{\rm km\,s^{-1}}$ is estimated for $M_{\bullet}=5\times 10^7\,M_{\odot}$ from the $M_{\bullet}-\sigma_{\star}$ relation \citep{2009ApJ...698..198G}. We see that $r_{\star}\ll D_{\rm A}$ for all the sources at or within $z=0.001$. In the sphere of the SMBH influence, the stellar dynamics is dominated by the SMBH, hence a classical two-body (SMBH-star) approximation is generally applicable for a few orbits. On longer timescales, secular dynamical processes, such as vector and scalar resonant relaxation due to surrounding stars, can influence stellar orbits in terms of inclination and eccentricity (see Section~\ref{sec:discussion} for the discussion of these effects). In Fig.~\ref{fig_illustration} we illustrate the general setup for a radio core to be eclipsed by a star within the NSC along the line of sight. The figure inset at the bottom depicts the basic geometry of the eclipse of the central SMBH engine by an evolved (red-giant) star. 

For the inner radius of the potential stellar obscurer, specifically the one from the late-type stellar component, we take the distance where tidal forces from the SMBH exceed the gravitational acceleration of the star on its surface (tidal radius),
\begin{align}
    r_{\rm t}&\approx R_{\star} \left(\frac{M_{\bullet}}{m_{\star}} \right)^{1/3}\,\notag\\
    & \approx 4.15 \times 10^{-3} \left(\frac{R_{\star}}{500\,R_{\odot}}\right)\left(\frac{M_{\bullet}}{5\times 10^7\,M_{\odot}} \right)^{1/3} \times \,\notag\\
    &\times \left(\frac{m_{\star}}{1\,M_{\odot}} \right)^{-1/3}\,{\rm pc}\,,
    \label{eq_tidal_radius}
\end{align}
where we scaled the stellar radius to $R_{\star}=500\,R_{\odot}$ and the stellar mass to $m_{\star}=1\,M_{\odot}$. In terms of gravitational radii, the tidal radius is $r_{\rm t}/r_{\rm g}\approx 1739 (R_{\star}/500R_{\odot}) (M_{\bullet}/5\times 10^7\,M_{\odot})^{-2/3} (m_{\star}/1\,M_{\odot})^{-1/3}$, hence still in the weak-field limit around the SMBH. Dynamical processes in NSCs such as stellar collisions can affect the inner radius of the NSC \citep{2013ApJ...762...35B}. In addition, even the tidal disruption of a red giant at the radius given by Eq.~\eqref{eq_tidal_radius} does not have to prohibit one-time eclipses since these can also correspond to tidal stellar debris that eventually get dispersed on the orbital timescale (see more details in Subsec.~\ref{subsec_tde_streams}). We will discuss these effects on the transit duration, recurrence timescale, and the likelihood of transits in NSCs in Subsection~\ref{sec:likelihood}.

\subsection{Occultation vs. microlensing}
\label{sec:occultation vs. microlensing}

Concerning the outer radius for the stellar eclipses of galactic nuclei, we infer the distance range from the SMBH where the Einstein radius of the star is less than its physical radius. The Einstein radius of a star can be expressed as,
\begin{equation}
    R_{\rm E}=\left[4\frac{Gm_{\star}}{c^2}\frac{D_{\star}(D_{\rm c}-D_{\star})}{D_{\rm c}}\right]^{1/2}\,,
    \label{eq_Einstein_radius}
\end{equation}
where $D_{\star}$ and $D_{\rm c}$ are angular-diameter distances to the star and the radio core, respectively. Since $D_{\star}\approx D_{\rm c}$ and $D_{\rm c}-D_{\star}=r_{\star}$ we can express the distance range for occultations from the condition $R_{\rm E} < R_{\star}$,
\begin{equation}
  r_{\star} < \frac{c^2 R_{\star}^2}{4Gm_{\star}}\sim 665\,\left(\frac{R_{\star}}{500\,R_{\odot}} \right)^2 \left(\frac{m_{\star}}{1\,M_{\odot}} \right)^{-1}{\rm pc}\,, 
  \label{eq_distance_microlens}
\end{equation}
which exceeds $r_{\rm SI}$ by at least two orders of magnitude for evolved stars that can cause the dimming in the mm/submm domain. Hence, red giants and supergiants will always cause the occultation of a background radio core within the NSC. The limiting stellar radius for $M_{\bullet}=5 \times 10^7\,M_{\odot}$ is $R_{\star}\gtrsim (4Gm_{\star}r_{\rm SI}/c^2)^{1/2} \sim  57\,R_{\odot}$ for $m_{\star}=1\,M_{\odot}$ when the distance for microlensing becomes comparable to $r_{\rm SI}$.
For the distances greater than those given by Eq.~\eqref{eq_distance_microlens} the star will act as a microlens, which can generally be detected as a flux enhancement of the radio core. However, such enhancements may be more difficult to disentangle than transit dips with respect to the stochastic AGN background and hence we do not consider them here.

\begin{figure*}
    \centering
    \includegraphics[width=0.48\textwidth]{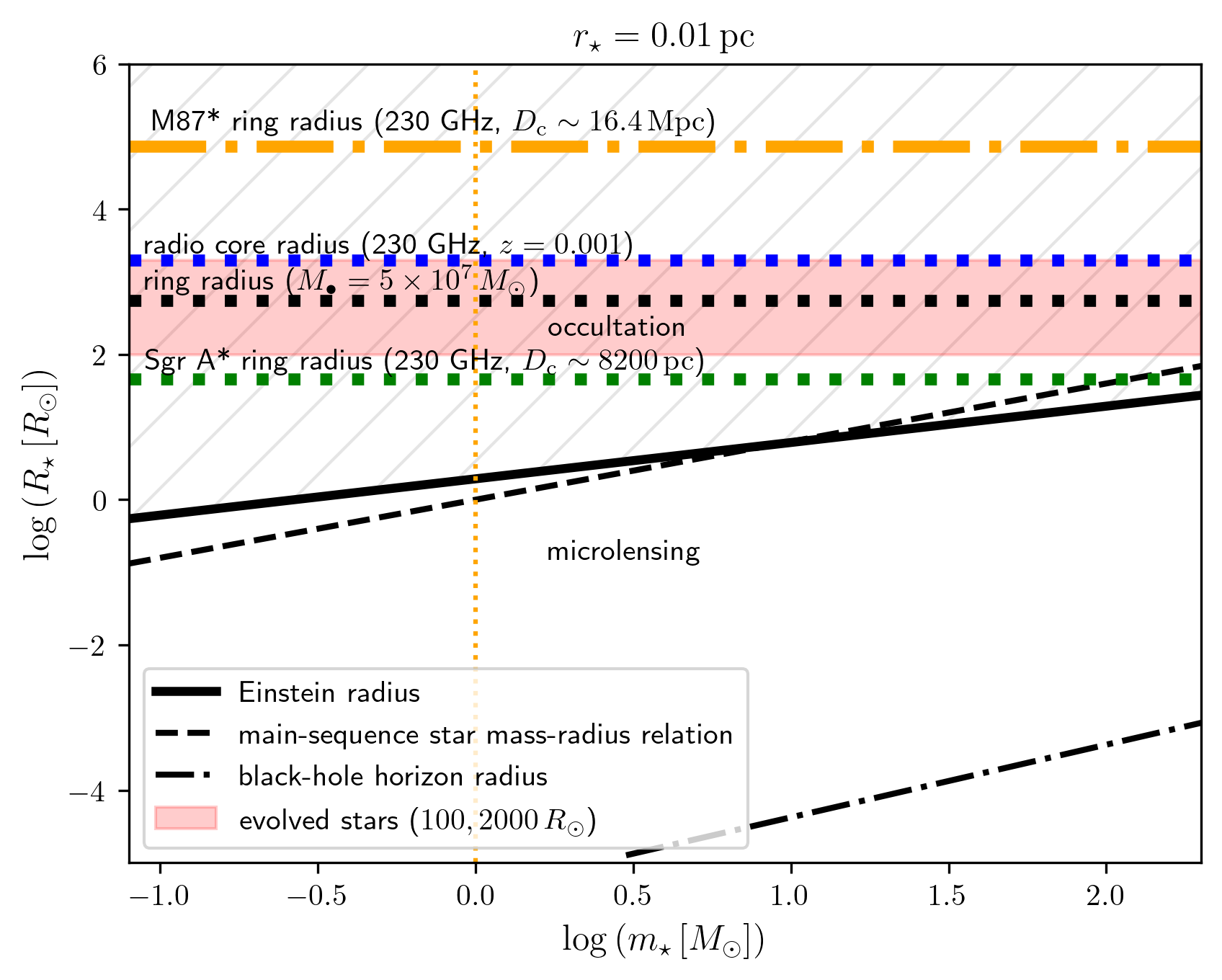}
    \includegraphics[width=0.48\textwidth]{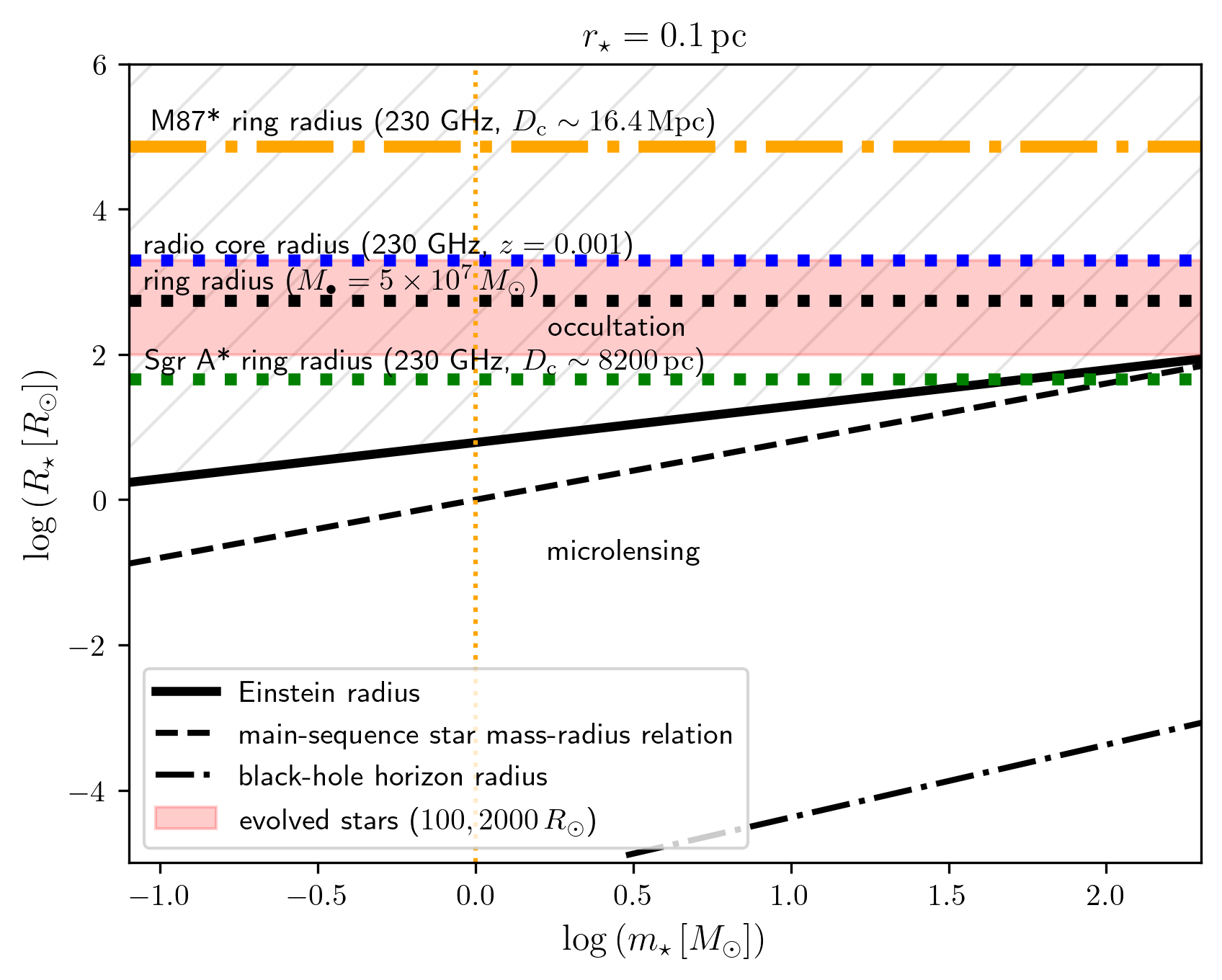}
    \caption{Stellar radius versus the stellar mass for a star orbiting the SMBH in a galactic nucleus. \textit{Left panel:} The plot of the stellar radius vs. the stellar mass where we show the Einstein radius for a star orbiting at $r_{\star}=0.01\,{\rm pc}$ from the SMBH. The region above the Einstein radius stands for the occultation parameter space while below it there is a microlensing parameter space. Evolved stars across the whole mass range can cause eclipses of the radio core in the mm domain if there are large enough ($R_{\star}\sim 100-2000\,R_{\odot}$). Main-sequence stars (dashed line) are smaller than their corresponding Einstein radii, with the exception of massive stars with $m_{\star}\gtrsim 10\,M_{\odot}$ that are short-lived, therefore they can only serve as microlenses. This is also valid for stellar black holes (dash-dotted line). \textit{Right panel:} The same as the left panel but for the star-SMBH distance of $r_{\star}=0.1\,{\rm pc}$. The main effect of the increasing stellar distance from the SMBH is the decreasing parameter space for the occultation in terms of the stellar radius -- however, red giants and supergiants ($R_{\star}\sim 100-2000\,R_{\odot}$) remain as occulting objects of the mm radio core throughout the whole NSC and even beyond -- they can cause transits all the way to the distance of at least $\sim 2700\,{\rm pc}$, see Eq.~\eqref{eq_distance_microlens}. For an easier comparison, we plot the predicted mm radio core radius at $z=0.001$ (dotted blue line) as well as the measured ring radii for Sgr~A* (dotted green line) and M87* (dash-dotted orange line). The horizontal black dotted line represents the ring radius for $M_{\bullet}=5\times 10^7\,M_{\odot}$.}    \label{fig_occultation_microlensing}
\end{figure*}


In Fig.~\ref{fig_occultation_microlensing}, we show the plots of the stellar radius versus its mass, including the limiting case dividing the occultation and the microlensing parameter space when the stellar radius is equal to the Einstein radius. We consider two cases for the distance of a star from the SMBH: $r_{\star}=0.01\,{\rm pc}$ (left panel) and $r_{\star}=0.1\,{\rm pc}$ (right panel), which are both between $r_{\rm t}$ and $r_{\rm SI}$. From Fig.~\ref{fig_occultation_microlensing} it is clear that the only stars that can cause the radio core occultation are large evolved stars -- red giants and supergiants with $R_{\star}\sim 100-2000\,R_{\odot}$. The main effect of the increasing star-SMBH distance is the shrinking parameter space for occultations, see the left and the right panels of Fig.~\ref{fig_occultation_microlensing}. However, red supergiants can cause eclipses of the radio core throughout the whole NSC and even for larger distances of a few $\sim 1000$ pc, see Eq.~\eqref{eq_distance_microlens} for the limiting distance. On the other hand, stellar radii of main-sequence stars are below their corresponding Einstein radii for the whole mass range, except for the closer massive stars at $r_{\star}\sim 0.01\,{\rm pc}$ and $m_{\star}\gtrsim 10\,M_{\odot}$. Therefore they act as microlenses, causing flux enhancements of the background emission rather than decrease. This is also valid for stellar black holes and remnants in general because of their physical compactness.

We perform analogous calculations of the occultation-microlensing distinction for stars passing in front of the Galactic center Sgr A*. For Sgr~A* we adopt $D_{\rm c}\simeq 8200\,{\rm pc}$ \citep{2019A&A...625L..10G,2023MNRAS.519..948L} and we consider $r_{\star} \in \{0.01, 0.1\}\,{\rm pc}$, while $D_{\star}=D_{\rm c}-r_{\star}\approx D_{\rm c}$, hence for stars within the NSC the relation expressed by Eq.~\eqref{eq_Einstein_radius} is approximately insensitive to the distance of the source. The angular radius of the radio ring at $\nu=230\,{\rm GHz}$ is $\theta_{\rm c}=25.9 \pm 1.2\,{\rm \mu as}$ \citep{2022ApJ...930L..12E}, which corresponds to $R_{\rm c}\simeq \theta_{\rm c}D_{\rm c}\sim 46\,R_{\odot}$. Hence, the source linear size decreases linearly with the distance, which enhances the prospects for occultation by smaller, however, still mostly evolved stars. In Fig.~\ref{fig_occultation_microlensing} we show the approximate linear radius of Sgr~A*. The evolved stars with $R_{\star}=100-2000\,R_{\star}$ would cause occultations for $m_{\star}=1\,M_{\odot}$ at both representative distances. Even smaller Solar-mass stars with $R_{\star}\sim 10\,R_{\odot}$ would cause eclipses. For completeness, in Fig.~\ref{fig_occultation_microlensing} we also indicate the expected linear scale (radius) for M87*. However, from $\theta_{\rm c}\simeq \,21{\rm \mu as}$ \citep{2019ApJ...875L...1E} and the (angular-diameter) distance of M87 of $D_{\rm c}\sim 16.4\,{\rm Mpc}$, we get the radio core radius of $R_{\rm c}\simeq \theta_{\rm c}D_{\rm c}\sim 74000\,R_{\odot}\sim 0.002\,{\rm pc}$. This is far bigger than the radii of evolved stars and therefore only more extended circumnuclear gas clouds could in principle cause the eclipses of M87*.  

Here we also stress that while the jet base/radio core linear size is proportional to the angular-diameter distance at a fixed frequency, which in principle limits the eclipses to the sources at $z\lesssim 0.001$, the millimeter ring feature is proportional to the SMBH mass, $R_{\rm r}=3\sqrt{3}\,GM_{\bullet}/c^2$, see Fig.~\ref{fig_occultation_microlensing} for the ring radius corresponding to $M_{\bullet}=5\times 10^7\,M_{\odot}$ (dotted black line). In this sense, in case the dominant millimeter nuclear emission comes from the ring, the angular-diameter distance (redshift) is not a limiting factor anymore. Instead, the occultation criterion is set by the minimum and the maximum SMBH mass (for the fixed stellar mass) and by the presence of the surrounding (foreground) emission that limits the detection of especially shallow eclipses. As a specific example, let us consider one Solar-mass star orbiting at $r_{\star}=0.01\,{\rm pc}$ from the SMBH. The eclipse criterion $R_{\star}\gtrsim R_{\rm E}=[4(Gm_{\star}/c^2)r_{\star}]^{1/2}\sim 1.94\,R_{\odot}$ sets the smallest radius of the star. Considering the 10\% eclipse depth, the limiting SMBH mass can be estimated from, 
\begin{align}
    M_{\rm \bullet,lim}&\simeq \frac{c^2 R_{\rm \star, lim}}{3\sqrt{3}G \left(\frac{\Delta F_{\nu}}{F_{\nu}}\right)^{1/2}}\,\notag\\
    &\sim 5.6 \times 10^5\,\left(\frac{R_{\rm \star,lim}}{1.94\,R_{\odot}}\right) \left(\frac{\Delta F_{\nu}/F_{\nu}}{0.1}\right)^{-1/2}\,M_{\odot}\,,
    \label{eq_SMBH_limit}
\end{align}
which expresses the lower SMBH mass limit of the galaxy sample. The upper limit is set by the size of red supergiants, $R_{\rm \star,max}\sim 2000\,R_{\odot}$, which for the 10\% dip yields the upper SMBH mass of $M_{\rm \bullet, max}\sim 5.7\times 10^8\,M_{\odot}$. This is consistent with the previous estimate for the M87* ring size, which is too large to be effectively eclipsed by any star, see Fig.~\ref{fig_occultation_microlensing}. We stress that as long as the jet base contributes significantly to the millimeter emission, then the redshift limit $z\lesssim 0.001$ is still relevant for this core emission to be small enough. Otherwise it would tend to outshine any eclipse signal.

\subsection{Relative Decrease in Flux: Dip depth}
\label{sec:depth}

For the uniform-brightness radio cores and stellar obscurers, the relative decrease in the flux density can be expressed as,
\begin{equation}
    \frac{\Delta F_{\nu}}{F_{\nu}}\simeq \frac{\pi \theta_{\star}^2}{\pi \theta_{\rm c}^2}=\left(\frac{R_{\star}}{R_{\rm 0c}(\nu/\nu_0)^{-1}} \right)^2 = \left(\frac{R_{\star}}{R_{\rm 0c}}\right)^2 \left(\frac{\lambda}{\lambda_0} \right)^{-2}\,,
    \label{eq_eclipse_depth}
\end{equation}
which implies that for the fixed obscurer radius of $R_{\star}$ the eclipse depth at millimeter wavelengths is $\sim 100$-times deeper than at centimeter wavelengths.

\begin{figure*}
    \centering
    \includegraphics[width=0.49\textwidth]{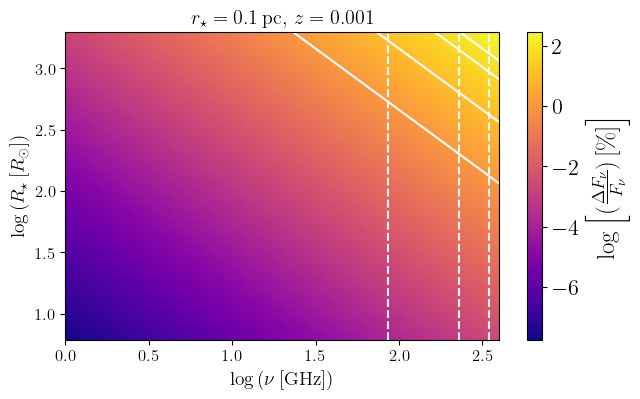}
     \includegraphics[width=0.49\textwidth]{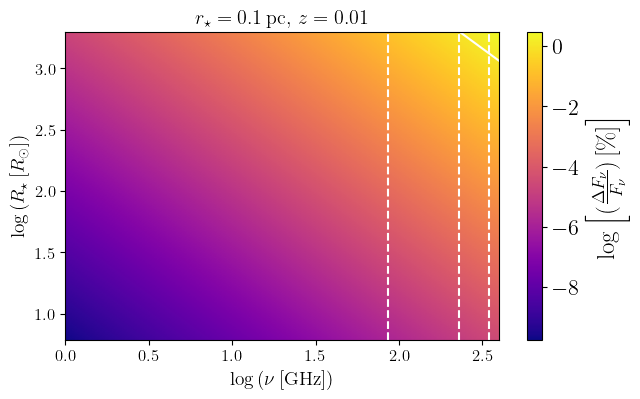}
    \caption{Relative occultation depths as a function of the stellar radius and the observing frequency. \textit{Left panel:} The radio core is located at $z=0.001$ and the star orbits the SMBH at $r_{\star}=0.1\,{\rm pc}$. The white solid diagonal lines denote the relative eclipse depths of $1\%$, $10\%$, $50\%$, and $100\%$, while the vertical dashed lines represent the typical VLBI frequencies of $86\,{\rm GHz}$ (3.5 mm), $230\,{\rm GHz}$ (1.3 mm), and $345\,{\rm GHz}$ (0.87 mm). \textit{Right panel:} The same as in the left panel but for the radio core redshift of $z=0.01$. The white diagonal line depicts the occultation depth of $1\%$ in this case. }
    \label{fig_occultation_depth}
\end{figure*}

We estimate example relative occultation depths for a range of frequencies $\nu \in (1, 400)\,{\rm GHz}$ for the cores located at $z=0.001$ and $z=0.01$. We assume that the star orbits the SMBH at the distance of $r_{\star}=0.1\,{\rm pc}$, i.e. well within the NSC as well as the sphere of influence of the SMBH of $M_{\bullet}=5\times 10^7\,M_{\odot}$. In Fig.~\ref{fig_occultation_depth}, we show the colour-coded relative eclipse depths, $100\Delta F_{\nu}/F_{\nu}$ (in \%), as a function of the stellar (obscurer) radius (in Solar radii) and the observing frequency (in GHz). The values along the both axes are expressed in decadic logarithms. We see that to reach measurable relative depths of at least a few percent, the stellar radius must exceed $100\,R_{\odot}$ for the observing wavelength in the millimeter/submillimeter domain (frequencies of 86 GHz, 230 GHz, and 345 GHz are indicated by vertical dashed lines). For the radio core angular radius, we adopt the approximate core-shift scaling of $\theta_{\rm c}\sim 0.5 (\nu/1\,{\rm GHz})^{-1}\,{\rm mas}$. The minimum stellar radius is set to $R_{\star}=6.13\,R_{\odot}$, which corresponds to the Einstein radius, $R_{\rm E}\simeq (4Gm_{\star}r_{\star}/c^2)^{1/2}$, of a $1\,M_{\odot}$ star at $r_{\star}=0.1\,{\rm pc}$ from the SMBH. The maximum stellar radius of $R_{\star}=2000\,R_{\odot}$ is the value corresponding to red supergiants and asymptotic giant branch stars.   

\begin{table}[]
    \centering
     \caption{Relative occultation depths of radio cores with the angular radius of $\theta_{\rm c}=0.5\,(\nu/1\,{\rm GHz})^{-1}{\rm mas}$ (in percent) for different stellar radii and observing frequencies (in GHz) in the millimeter domain. The source redshift is set to z=0.001.}
    \begin{tabular}{c|c|c|c}
    \hline
    \hline
      $R_{\star}/R_{\odot}$ & 86 GHz & 230 GHz & 345 GHz \\
    \hline  
     100    &  0.035   & 0.25   &  0.56  \\
     500    &  0.87   &  6.26  &   14.07  \\
     1000   &  3.50   &  25.02  &  56.30   \\
     2000   &  13.99   & $100.09^{*}$   &  $225.19^{*}$   \\
    \hline  
    \end{tabular}   
    $^{*}$Values more than 100\% indicate the area ratio here.
    \label{tab_occultation_depths}
\end{table}

In Table~\ref{tab_occultation_depths} we provide the overview of occultation depths (in percent) for the radio core redshift of $z=0.001$ since the detection of a significant decrease in flux density due to the passage of a star is more likely in this case than for more distant targets. For $\nu=230\,{\rm GHz}$ the passages of evolved stars with $R_{\star}=500\,R_{\odot}$ and $R_{\star}=1000\,R_{\odot}$ can lead to the eclipses with the depth of $\sim 6\%$ and $\sim 25\%$, respectively. For these radii and $\nu=345\,{\rm GHz}$, the depth increases by slightly more than a factor of two. For $R_{\star}=2000\,R_{\odot}$ (red supergiants and asymptotic giant branch stars) the eclipse can be complete for an exact overlap of the orbit (i.e. the edge-on orbital configuration; values more than 100\% indicate the area ratio here).

\subsection{Recurrence timescale and eclipse duration}
\label{sec:duration_recurrence}

\begin{figure}
    \centering
    \includegraphics[width=\columnwidth]{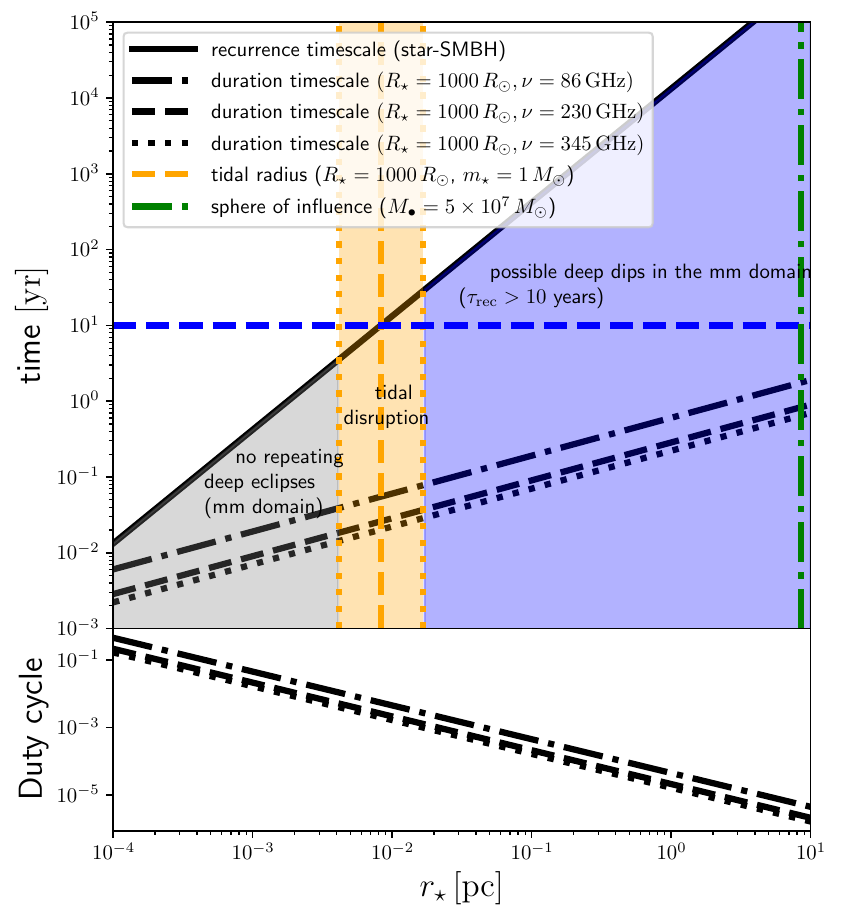}
    \caption{Recurrence and duration timescales of galactic nuclei eclipses (in years) as a function of the distance of a star from the SMBH (in parsecs). In the upper panel, the black solid line represents the eclipse recurrence timescale (orbital period), while the dash-dotted, dashed, and dotted lines stand for the occultation duration timescale at 86, 230, and 345 GHz, respectively (for the stellar radius of $1000\,R_{\odot}$). The vertical orange dotted and dashed lines mark the tidal disruption region of red giants/supergiants (vertical lines correspond to $500$, $1000$, and $2000\,R_{\odot}$). The vertical dash-dotted green line marks the outer radius of the SMBH sphere of influence. The dotted blue horizontal line stands for the timescale of 10 years. For all the calculations, the SMBH mass is set to $M_{\bullet}=5\times 10^7\,M_{\odot}$. In the lower panel, we show the duty cycle of the eclipses, $D=\tau_{\rm dur}/\tau_{\rm rec}$, as a function of the distance of the star from the SMBH for the three corresponding frequencies (see the legend in the upper panel).}
    \label{fig_rec_timescale}
\end{figure}

From the observational perspective, recurrent occultations of the central SMBH engine are the most relevant since they enable to infer the SMBH mass (see Section~\ref{sec:mass}) as well as the content of the NSC depending on the (non)-detection of eclipses (see Section~\ref{sec:likelihood}). Regular repeating eclipses imply that the body is bound to the SMBH inside its sphere of influence, see Eq.~\eqref{eq_sphere_SMBHinf}. Then the recurrence timescale is given by the orbital timescale, $\tau_{\rm rec}\simeq P_{\rm orb}$, to the first approximation (neglecting orbital precession and relaxation mechanisms on longer timescales). From the two-body dynamics we obtain,
\begin{align}
    \tau_{\rm rec}&\simeq 2\pi \frac{r_{\star}^{3/2}}{(GM_{\bullet})^{1/2}}\,\notag\\
    &\sim 419\,\left(\frac{r_{\star}}{0.1\,{\rm pc}} \right)^{3/2}\left(\frac{M_{\bullet}}{5\times 10^7\,M_{\odot}} \right)^{-1/2}\text{years}\,.
    \label{eq_recurrence_timescale}
\end{align}
This is a longer timescale than the current monitoring duration of nearby AGN. In order to have the recurrence timescale of the order of $\tau_{\rm rec}\sim 10$ years, the star needs to be located at $r_{\star}\sim 0.01\,{\rm pc}$ for $M_{\bullet}=5\times 10^7\,M_{\odot}$, i.e. within the region comparable to the S cluster of the Milky Way \citep{2022RvMP...94b0501G}. At the same time, the star of the given radius $R_{\star}$ cannot orbit closer than its tidal radius, see Eq.~\eqref{eq_tidal_radius}, where its envelope would get disrupted. In Fig.~\ref{fig_rec_timescale}, we show the basic relation between the recurrence timescale (in years; black solid line), the distance of a star from the SMBH (in parsecs), tidal disruption zone of red giants/supergiants (orange-shaded rectangle), and the radius of the SMBH sphere of influence (vertical green dot-dashed line).

We see that to measure significant eclipse dips at the level of $\sim 25\%$ at $\nu=230\,{\rm GHz}$ at least every 10 years (see Table~\ref{tab_occultation_depths}), the star with $R_{\star}\sim 1000\,R_{\odot}$ would need to orbit at $r_{\star}\sim 0.01\,{\rm pc}$ where it is still tidally stable. Hence, the radial zone for repeating deeper eclipses ($100\Delta F_{\nu}/F_{\nu}\gtrsim10\%$) is quite narrow. On the other hand, if we relax the recurrence timescale to $>10$ years, then the eclipse zone encompasses essentially the whole NSC (blue-shaded region) except for the inner region where the evolved stars get tidally disrupted. From Fig.~\ref{fig_rec_timescale} we can also infer that for $M_{\bullet}\sim 5\times 10^7\,M_{\odot}$ it is unlikely to have a deeper eclipse in the millimeter domain involving a larger, evolved star that would repeat on the timescale of a few years and less (gray-shaded region).  

The duration of the eclipse can be estimated from the general expression for the temporal separation of the obscurer and radio core centers as seen on the sky: $d(t)=\sqrt{v_{\rm sky}^2(t-t_0)^2+b^2}$, where $v_{\rm sky}$ is the projection of the velocity of the star on the sky, $t_0$ is the time of the closest approach, and $b$ is the impact parameter. For the projection of the velocity on the sky we have $v_{\rm sky}=(GM_{\bullet}/r_{\star})^{1/2}\sin{\iota}$, where $\iota=90^{\circ}$ for an edge-on orbit and $\iota=0^{\circ}$ for a face-on orbit. For the start and the end of the occultation we have $d_{\rm start}=R_{\star}+R_{\rm c}=d_{\rm end}$. From $d(t)=d_{\rm start}$, we obtain the general expression for the total duration of the occultation, which is,
\begin{equation}
    \tau_{\rm dur}=\frac{2\sqrt{(R_{\star}+R_{\rm c})^2-b^2}}{v_{\rm sky}}\,.
    \label{eq_eclipse_duration}
\end{equation}
For $b=0$ (overlap of centers) we have
\begin{equation}
\tau_{\rm dur}=\frac{2(R_{\star}+R_{\rm c})}{v_{\rm sky}}\propto \lambda\,,
\label{eq_eclipse_duration}
\end{equation}
hence towards shorter wavelengths occultations in the radio domain tend to be deeper, see Eq.~\eqref{eq_eclipse_depth}, and shorter and they are shallower and longer at longer wavelengths. At $\nu=230\,{\rm GHz}$, for $R_{\star}=1000\,R_{\odot}$ and $z=0.001$ we can express the duration timescale for the edge-on, center-aligned orbits as follows,
\begin{equation}
    \tau_{\rm dur}\simeq 10.4\left(\frac{r_{\star}}{0.01\,{\rm pc}} \right)^{1/2} \left(\frac{M_{\bullet}}{5\times 10^7\,M_{\odot}} \right)^{-1/2}\,\text{days}\,
\end{equation}
which becomes $\tau_{\rm dur}\simeq 22.0$ days at $\nu=86\,{\rm GHz}$ and $\tau_{\rm dur}\simeq 8.1$ days at $\nu=345\,{\rm GHz}$ for the same parameters (stellar distance and radius). When we fix the frequency at $\nu=230\,{\rm GHz}$ and we keep the stellar radius at $R_{\star}=1000\,R_{\odot}$, the eclipse duration is $\tau_{\rm dur}\simeq 32.9$ days at $r_{\star}\simeq 0.1\,{\rm pc}$ and $\tau_{\rm dur}\simeq 104.2$ days at $r_{\star}\simeq 1\,{\rm pc}$. In Fig.~\ref{fig_rec_timescale} we show duration timescales for the three different frequencies (86, 230, and 345 GHz) as a function of the stellar distance from the SMBH (see dash-dotted, dashed, and dotted lines, respectively). We see that while at $r_{\star}=0.01\,{\rm pc}$ the dip duration is of the order of several days to $\sim 10$ days, the duration increases to about a year at the outer edge of the Nuclear Star Cluster. 

When it comes to the duty cycle of the eclipses defined as $D=\tau_{\rm dur}/\tau_{\rm rec}$, we obtain the following relation for edge-on orbits,
\begin{equation}
    D=\frac{\tau_{\rm dur}}{\tau_{\rm rec}}=\frac{R_{\star}+R_{\rm c}}{\pi r_{\star}}\propto \frac{1}{\nu r_{\star}}\,,
    \label{eq_duty_cycle}
\end{equation}
which increases for smaller distances and lower frequencies, which can also be inferred from Fig.~\ref{fig_rec_timescale}. For $\nu=230\,{\rm GHz}$, $z=0.001$, $r_{\star}=0.01\,{\rm pc}$, and $R_{\star}=1000\,R_{\odot}$, the duty cycle is $D\sim 0.0022$. In Fig.~\ref{fig_rec_timescale}, in the lower panel, we show the duty cycle as a function of the distance of the star with $R_{\star}=1000\,R_{\odot}$ from the SMBH for the three chosen frequencies (86, 230, and 345 GHz). The maximum duty cycle for the $\sim 6\%$-deep occultation at $\nu=230\,{\rm GHz}$ is $D\sim 0.0043$, which is given by the tidal radius of the $R_{\star}=500\,R_{\odot}$ star, see Eq.~\eqref{eq_tidal_radius}.

The information that can be obtained from a single eclipse is enough to estimate the distance of a star from the SMBH and its stellar radius. From the relative eclipse depth $\Delta F_{\nu}/F_{\nu}$, see Eq.~\eqref{eq_eclipse_depth}, one can obtain the stellar radius given the measured jet base angular radius $\theta_{\rm c}$,
\begin{equation}
    R_{\star}=\theta_{\rm c}D_{\rm c}\left(\frac{\Delta F_{\nu}}{F_{\nu}} \right)^{1/2}\,.
    \label{eq_stellar_radius_single}
\end{equation}
Then it is possible to infer the distance of a star from the SMBH using Eq.~\eqref{eq_eclipse_duration} and assuming $v_{\rm sky}\sim \sqrt{GM_{\bullet}/r_{\star}}$, from which,
\begin{equation}
    r_{\star}=\frac{GM_{\bullet}\tau_{\rm dur}^2}{4\theta_{\rm c}^2D_{\rm c}^2\left[1+(\Delta F_{\nu}/F_{\nu})^{1/2} \right]^2}\,.
    \label{eq_distance_single}
\end{equation}
Relation~\eqref{eq_distance_single} assumes that we have independent constraints on the SMBH mass. Both relations also assume that the stellar orbit is circular and exactly edge-on -- in particular, for a non-zero impact parameter $b$, relation~\eqref{eq_stellar_radius_single} between the eclipse depth and the stellar radius is not exactly valid, see also Subsection~\ref{eq_eclipse_profiles}, where it is shown that for higher inclinations of the star, the eclipse becomes shallower and narrower.

\subsection{SMBH mass estimate relation for repeating eclipses}
\label{sec:mass}

Using the occultation recurrence timescale $\tau_{\rm rec}$, see Eq.~\eqref{eq_recurrence_timescale}, and its duration timescale $\tau_{\rm dur}$, see Eq.~\eqref{eq_eclipse_duration}, we can estimate the SMBH mass $M_{\bullet}$. This is simply based on the fact that from the recurrence timescale relation we have $M_{\bullet}\propto r_{\star}^3$ while from the duration timescale equation we obtain $M_{\bullet}\propto r_{\star}$. Hence it is possible to factor out the stellar distance and obtain the SMBH mass relation as a function of basic observables.

From the recurrence timescale based on the bound star scenario, see Eq.~\eqref{eq_recurrence_timescale}, we can obtain the mass as follows,
\begin{equation}
    M_{\bullet}=\frac{4 \pi^2 }{G \tau_{\rm rec}^2} r_{\star}^3\,,
    \label{eq_mass_Kepler}
\end{equation}
which can be transformed to the following form by expressing $r_{\star}$ using the duration timescale formula, see Eq.~\eqref{eq_eclipse_duration} (aligned edge-on transit),
\begin{equation}
    M_{\bullet}=\frac{4[R_{\star}+R_{\rm c}(\nu)]^3\tau_{\rm rec}}{\pi G \tau_{\rm dur}^3}\,.
    \label{eq_mass_no_dist}
\end{equation}
Eq.~\eqref{eq_mass_no_dist} can further be rewritten using the relative eclipse depth $\Delta F_{\nu}/F_{\nu}$ as,
\begin{equation}
    M_{\bullet}=\frac{4 \theta_{\rm c}^3(\nu)D_{\rm A}^3(z)[1+(\Delta F_{\nu}/F_{\nu})^{1/2}]^3\tau_{\rm rec}}{\pi G \tau_{\rm dur}^3}\,.
     \label{eq_mass_obs}
\end{equation}
For a specific numerical estimate we use the core-shift relation $\theta_{\rm c}\sim 0.5\,(\nu/1\,{\rm GHz})^{-1}\,{\rm mas}$, the source redshift of $z=0.001$, and we consider the occultation of the nucleus by the star with $R_{\star}=1000R_{\odot}$. The observing frequency is set to $\nu=230\,{\rm GHz}$ and the measured recurrence timescale is $\tau_{\rm rec}=10.1$ years and the eclipse duration is $\tau_{\rm dur}=8$ days. Then the SMBH mass is,
\begin{align}
    M_{\bullet}&=8.4 \times 10^7\left(\frac{[1+(\Delta F_{\nu}/F_{\nu})^{1/2}]}{1.5}\right)^3\times \,\notag\\
    &\times \left(\frac{\tau_{\rm rec}}{10.1\,{\rm yr}}\right)\left(\frac{\tau_{\rm dur}}{8\,{\rm days}}\right)^{-3}\,M_{\odot}\,,
    \label{eq_mass_num}
\end{align}
where we scaled the relative eclipse depth to $\Delta F_{\nu}/F_{\nu}\sim 0.25$, which corresponds to the occultation by the star with $R_{\star}\sim 1000\,R_{\odot}$ at $\nu=230\,{\rm GHz}$ (see also Table~\ref{tab_occultation_depths}).

\begin{figure}
    \centering
    \includegraphics[width=\columnwidth]{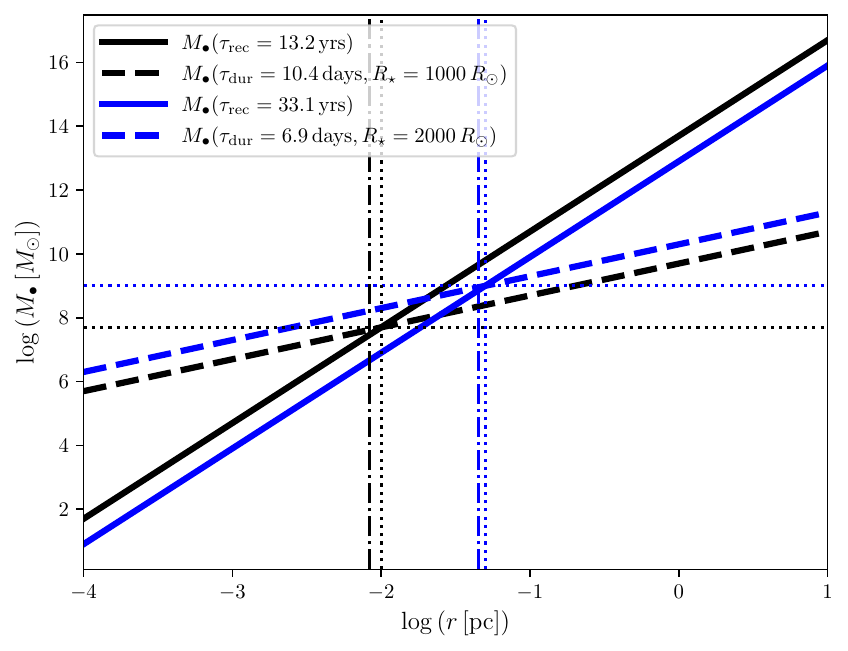}
    \caption{Dependency of the SMBH mass on the distance of a star from the SMBH. The steeper solid lines represent the SMBH mass derived from the recurrence timescale when $M_{\bullet}\propto r_{\star}^3$ while shallower dashed lines depict the SMBH mass based on the duration timescale, $M_{\bullet}\propto r_{\star}$. Black lines represent the case of a bound star with $R_{\star}=1000\,R_{\odot}$ while blue lines stand for $R_{\star}=2000\,R_{\star}$. The intersections of the same-color line pairs mark unique SMBH mass solutions for a given combination of the eclipse recurrence and the duration timescales (dotted horizontal lines; see also Eq.~\eqref{eq_mass_num}). The vertical black and blue dash-dotted lines stand for the tidal radii for the $1000\,R_{\odot}$ and $2000\,R_{\odot}$ stars, respectively, orbiting around the SMBH with the inferred masses. These tidal radii are only slightly smaller than the actual stellar orbits in these setups shown by dotted vertical lines.}
    \label{fig_Mbh_dist}
\end{figure}

In Fig.~\ref{fig_Mbh_dist} we plot the dependency of the SMBH mass on the stellar distance for the recurrence timescale relation, $M_{\bullet}\propto r_{\star}^3$ (solid lines), and also using the duration timescale relation, $M_{\bullet}\propto r_{\star}$ (dashed lines). As a specific example, we show the case of the star with $R_{\star}=1000\,R_{\odot}$ causing significant dips in the mm light curves ($\nu=230\,{\rm GHz}$) with $\tau_{\rm rec}\sim 13.2$ years and $\tau_{\rm dur}\sim 10.4$ days (black lines). For comparison, we also show the case of a star with $R_{\star}=2000\,R_{\odot}$ with $\tau_{\rm rec}\sim 33.1$ years and $\tau_{\rm dur}\sim 6.9$ days (blue lines). For these combinations of the recurrence and duration timescales, it is possible to infer the unique SMBH mass marked by dotted horizontal lines, which is in agreement with the SMBH mass relation given by Eq.~\eqref{eq_mass_num}. Vertical dash-dotted lines show the corresponding tidal radii for these stars around the corresponding SMBHs, which implies that they are close to being tidally disrupted. 

Relation~\eqref{eq_mass_obs} is practically difficult to apply since there are different sources of uncertainties. The main practical difficulty also lies in the determination of the recurrence timescale $\tau_{\rm rec}$, which requires the detection of at least two occultations but their typical separation is at least $\sim 10$ years, see Eq.~\eqref{eq_recurrence_timescale}. In terms of the propagation of relative uncertainties, the relative uncertainty for the SMBH mass is, 
\begin{equation}
    \eta_{M_{\bullet}}=\sqrt{9 \eta_{\theta_{\rm c}}^2+\frac{9\delta}{4(1+\delta^{1/2})^2}\eta_{\delta}^2+\eta_{\tau_{\rm rec}}^2+9\eta_{\tau_{\rm dur}}^2}\,.
    \label{eq_rel_uncertainty_SMBH}
\end{equation}
We see that both $\theta_{\rm c}$ and $\tau_{\rm dur}$ enter the SMBH mass relative uncertainty with the weights of three while $\tau_{\rm rec}$ has only the factor of unity. The smallest factor is associated with the eclipse relative depth -- for $\delta=0.5$, the factor is only $\sim 0.62$. Overall, if the relative uncertainties for all the observables are $\sim 10\%$, then $\eta_{M_{\bullet}}\sim 0.44$ according to Eq.~\eqref{eq_rel_uncertainty_SMBH} for $\delta=0.5$.

\subsection{Likelihood of radio core transits}
\label{sec:likelihood}

Here we assess the chance how likely it is that an evolved star (red giant/supergiant) with a sufficiently large radius within the NSC causes a potentially detectable eclipse of the mm core in a galactic nucleus. At $\nu=230-345\,{\rm GHz}$, stars with $R_{\star}\gtrsim 100\,R_{\odot}$ will cause eclipses with the relative depths of at least a few percent, see Fig.~\ref{fig_occultation_depth}. We assume that stars follow the power-law number density distribution within the NSC with $n_{\star}=n_0 (r/r_{\rm SI})^{-\gamma}$, where $\gamma>0$ and $n_0$ is the number density at the sphere of influence $r_{\rm SI}$, which is defined according to Eq.~\eqref{eq_sphere_SMBHinf}. We find the normalization from the definition of the sphere of influence, $M_{\star}(r<r_{\rm SI})=2M_{\bullet}$ \citep{2013degn.book.....M}, i.e. the total mass of stars within the sphere of influence is equal to $2M_{\bullet}$. When we consider the mean stellar mass $\overline{m}_{\star}$, we find that $\overline{m}_{\star}N(r<r_{\rm SI})=\overline{m}_{\star}\int_0^{r_{\rm SI}} n_{\star}(r) 4\pi r^2 \mathrm{d} r=2M_{\bullet}$. From this relation we obtain the normalization coefficient,
\begin{equation}
    n_0=\frac{M_{\bullet}(3-\gamma)}{2 \pi \overline{m}_{\star} r_{\rm SI}^3}\,.
    \label{eq_normalization}
\end{equation}

\begin{figure}
    \centering
    \includegraphics[width=\columnwidth]{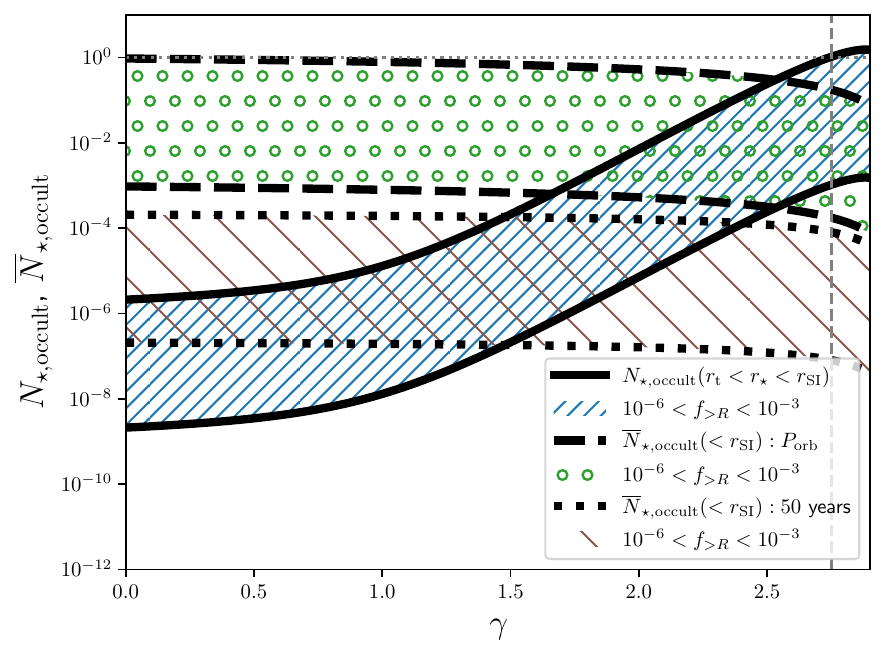}
    \caption{Number of occulting stars within the SMBH sphere of influence as a function of the power-law index $\gamma$ of the radial stellar number density (black solid lines). For this particular NSC setup, we adopted $M_{\bullet}=5\times 10^7\,M_{\odot}$, $\nu=230\,{\rm GHz}$, $z=0.001$, $\overline{m}_{\star}=1\,M_{\odot}$, $R_{\star}=100\,R_{\odot}$, and the fraction of large, occulting stars in the range of $f_{>R}=10^{-6}-10^{-3}$. The vertical dashed line denotes the power-law index $\gamma=2.75$ of the stellar distribution, for which $N_{\rm \star, occult}\sim 1$. The black dashed lines mark the mean number of occultations for the same parameters within the sphere of influence per stellar orbital period. The dotted black lines represent the mean number of occultations within the SMBH sphere of influence considering the monitoring period of 50 years while the other parameters are kept the same. The horizontal dotted line marks the unity.}
    \label{fig_number_stars_occulting}
\end{figure}
The number of evolved occulting stars within the cylinder along the line of sight (adopting the radial range $r_{\rm t}<r_{\star}<r_{\rm SI}$) with the volume of $\sim \pi R_{\rm c}^2 \int_{r_{\rm t}}^{r_{\rm SI}} \mathrm{d}r$ can be expressed as follows,
\begin{equation}
    N_{\star,{\rm occult}}\simeq \frac{f_{>R} \pi R_{\rm c}^2 M_{\bullet} (3-\gamma)(r_{\rm SI}^{1-\gamma}-r_{\rm t}^{1-\gamma})}{2\pi \overline{m}_{\star} r_{\rm SI}^{3-\gamma}(1-\gamma)}\,,
    \label{eq_Number_stars_occulting}
\end{equation}
where $f_{>R}$ is the fraction of sufficiently large evolved stars ($R_{\star}\gtrsim 100\,R_{\odot}$) within the NSC. Based on the representative Milky Way NSC star-formation history \citep{2020A&A...641A.102S} and stellar-population synthesis models including evolved stars \citep{2017ApJ...835...77M}, we adopt the values in the range $f_{>R}\sim 10^{-6}-10^{-3}$, i.e. most stars within the NSC are smaller since the composition is dominated by lighter and smaller stars. When we consider the following host properties -- $M_{\bullet}=5\times 10^7\,M_{\odot}$, $\gamma=3/2$, $r_{\rm t}=8.31\times 10^{-4}\,{\rm pc}$, $r_{\rm SI}=8.5\,{\rm pc}$, $\overline{m}_{\star}=1\,M_{\odot}$ -- and the observing frequency of $\nu=230\,{\rm GHz}$, we get $N_{\star,{\rm occult}}\simeq 2\times 10^{-4}$. The number of potentially occulting stars is, of course, quite uncertain as it depends on the several parameters of the stellar radial distribution within the NSC as well as its age distribution. There is, in particular, a steep dependence on the stellar number-density power-law index $\gamma$. In Fig.~\ref{fig_number_stars_occulting} we plot $N_{\star,{\rm occult}}$ as a function of the power-law index in the interval $\gamma\in (0, 2.9)$. In particular, for the particular set-up with $M_{\bullet}=5\times 10^7\,M_{\odot}$, $z=0.001$, and $\nu=230\,{\rm GHz}$ the number of occulting stars is proportional to the fraction of stars within the NSC, $f_{>R}$, that can cause significant eclipses, see Eq.~\eqref{eq_Number_stars_occulting}. For $f_{>R}=10^{-3}$ and $\gamma=2.75$, $N_{\rm \star, occult}$ reaches unity. Hence whether we detect at a given observing epoch a sign of the occultation or not can be used to constrain the radial distribution within the NSC, especially for the NSCs consisting of mostly late-type stars, which can be constrained from their integrated spectral energy distributions.  

\begin{table*}[h!]
    \centering
    \caption{Overview of nearby mm-bright nuclei and the evidence for the NSC presence.}
    \begin{tabular}{c|c|c|c|c|c}
    \hline
    \hline 
    Galaxy/nucleus     & distance [Mpc]  & $z$ & $M_{\bullet}\,[M_{\odot}]$ & NSC & Ref.   \\
    \hline
    Milky Way (Sgr~A*)     & 0.008    &  0   & $\sim 4\times 10^6$    & Yes  &  \citet{2020AA...634A..71G} \\
    Centaurus A (NGC 5128) & 3.8 & 0.0018 & $\sim 5 \times 10^7$ & Probably  &  \citet{2010PASA...27..449N} \\
    M81 (NGC 3031) & 3.6 & 0.0001 & $\sim 7 \times 10^7$ & Yes & \citet{2020AARv..28....4N}\\
    M87 (NGC 4486) & 16.8 & 0.0043 & $\sim 6.5 \times 10^9$ & No  & \citet{2020AARv..28....4N}\\
    NGC 4258 (M106) & 7.6 & 0.0015 & $3.9\times 10^7$  & weak & \citet{2020AARv..28....4N}\\
    Circinus & 4.2 & 0.0014 & $\sim 1.7 \times 10^6$ & Yes &\citet{2020AARv..28....4N}\\    
    \hline     
    \end{tabular}  
    \label{tab_NSCs_local}
\end{table*}

Another way of estimating the number of (bound) occulting stars within a certain volume around the SMBH is to use the rate of occultations at a certain distance from the SMBH, $\dot{N}_{\star,{\rm occult}}\approx f_{>R}n_{\star}(r)\sigma_{\star}(r) S_{\rm LOS}$, where the local velocity dispersion is estimated by the local Keplerian velocity, $\sigma_{\star}(r)\sim (GM_{\bullet}/r_{\star})^{1/2}$, and the surface area of the line-of-sight cylinder enveloping the radio core is $S_{\star}\approx 2R_{\rm c}r$. Putting all the terms together, we get,
\begin{equation}
    \dot{N}_{\star,{\rm occult}}\approx f_{>R} \frac{M_{\bullet}(3-\gamma)r_{\rm SI}^{\gamma-3}R_{\rm c}(GM_{\bullet})^{1/2}r_{\star}^{1/2-\gamma}}{\pi \overline{m}_{\star}}\,.
\end{equation}
The number of occultations during the orbital period at a given radius is, $N_{\star,{\rm occult}}(r_{\star})\sim \dot{N}_{\star,{\rm occult}}P_{\rm orb}(r_{\star})$,
\begin{equation}
    N_{\star,{\rm occult}}(r_{\star})\simeq f_{>R} \frac{2M_{\bullet}(3-\gamma)r_{\rm SI}^{\gamma-3}R_{\rm c}r_{\star}^{2-\gamma}}{\overline{m}_{\star}}\,.
\end{equation}
Finally, the mean number of occultations within the sphere of influence with the radius $r_{\rm SI}$ given by Eq.~\eqref{eq_sphere_SMBHinf} is given by the relation $\overline{N}_{\star,{\rm occult}}(<r_{\rm SI})=(1/V_{\rm SI})\int_{0}^{r_{\rm SI}} N_{\star,{\rm occult}}(r_{\star})4\pi r^2_{\star}\mathrm{d}r_{\star}$, where $V_{\rm SI}=4\pi r_{\rm SI}^3/3$. By combining with the previous expression the volume integral yields,
\begin{equation}
    \overline{N}_{\star,{\rm occult}}(<r_{\rm SI})=f_{>R}R_{\rm c}\frac{6\sigma_{\star}^2}{G\overline{m}_{\star}} \left(\frac{3-\gamma}{5-\gamma} \right)\,.
    \label{eq_mean_number_occultations}
\end{equation}
In Fig.~\ref{fig_number_stars_occulting} we plot $\overline{N}_{\star,{\rm occult}}(<r_{\rm SI})$ as a function of the power-law index $\gamma$, on which it depends only mildly. For $\gamma=3/2$ and $f_{>R}=10^{-3}$ we get  $\overline{N}_{\star,{\rm occult}}(<r_{\rm SI})\sim 0.68$, which expresses the mean number of radio core occultations by late-type stars per orbital period within the sphere of influence. Therefore, $N_{\star,{\rm occult}}$ as given by Eq.~\eqref{eq_Number_stars_occulting} and the mean number of occultations $\overline{N}_{\star,{\rm occult}}(<r_{\rm SI})$ given by Eq.~\eqref{eq_mean_number_occultations} are the occultation-number proxies that correspond to different timescales and are therefore not in contradiction (the first expression is a rather instantaneous measure while the latter corresponds to the orbital -- dynamical timescale of the NSC). One can also estimate the number of radio core occultations at a given radius for a fixed timescale of monitoring, $\tau_{\rm mon}$, i.e. $N_{\star,{\rm occult}}(r_{\star})\approx \dot{N}_{\star,{\rm occult}}\tau_{\rm mon}$. Then the corresponding mean number of occultations within the sphere of the SMBH influence is,
\begin{equation}
       \overline{N}_{\star,{\rm occult}}^{\tau}(<r_{\rm SI})=\frac{3f_{>R}\tau_{\rm obs} R_{\rm c}}{\pi \overline{m}_{\star}G^2 M_{\bullet}} \left(\frac{3-\gamma}{7/2-\gamma} \right)\sigma_{\star}^5\,.
    \label{eq_mean_number_occultations_tau}
\end{equation}
We plot $\overline{N}_{\star,{\rm occult}}^{\tau}(<r_{\rm SI})$ in Fig.~\ref{fig_number_stars_occulting} using a dotted black line considering the source monitoring timescale of 50 years. As for the case when we considered the orbital period, the dependency on the power-law index of the stellar radial distribution is only mild and for $\gamma=3/2$ and $f_{>R}=10^{-3}$ we get $\overline{N}_{\star,{\rm occult}}^{\tau}(<r_{\rm SI})=1.82 \times 10^{-4}$. For the monitoring timescale of $100$ years, we get $\overline{N}_{\star,{\rm occult}}^{\tau}(<r_{\rm SI})=3.64 \times 10^{-4}$.

The estimates above assume that the power-law stellar density cusp formed within the NSC via dynamical relaxation processes. For the nearby millimeter-bright nuclei, there is at least an indication that there are NSCs for most of the hosts, see Table~\ref{tab_NSCs_local}. For four nearby sources, the presence of a compact NSC is established, while NGC 4258 has only a weak indication and M87 seems to lack a compact NSC as it is the case for other massive ellipticals \citep{2020AARv..28....4N}. The presence of the NSC is a necessary condition for the occultations to take place. The rate of eclipses is then further affected by the NSC geometry -- either isotropic as we assumed or a flattened/disk-like stellar system (and their combination).

For estimating the enhancement factor for the nearly edge-on disk-like configuration, in comparison with an isotropic one, we first derive the corresponding eclipse probabilities. When the radius of the occultation cross-section is $R_{\rm occ}=R_{\rm c}+R_{\star}$, then the probability of an eclipse at the distance $r_{\star}$ from the SMBH for an isotropic distribution is,
\begin{equation}
    p_{\rm iso}(r_{\star})\simeq \frac{\pi R_{\rm occ}^2}{4\pi r_{\star}^2}=\frac{R_{\rm occ}^2}{4r_{\star}^2}\,.
\end{equation}
For the disk-like configuration, the eclipse takes place when the projected separation $d$ is smaller than $R_{\rm occ}$. The eclipse probability for the uniform phase distribution along the disk then is,
\begin{equation}
    p_{\rm disk}(r_{\star})\simeq \frac{1}{\pi} \arcsin{\left(\frac{R_{\rm occ}}{r_{\star}} \right)}\sim \frac{R_{\rm occ}}{\pi r_{\star}}\,.
\end{equation}
The local enhancement factor between the edge-on and isotropic configurations is,
\begin{equation}
    \mathcal{E}(r_{\star})=\frac{p_{\rm disk}(r_{\star})}{p_{\rm iso}(r_{\star})}=\frac{4r_{\star}}{\pi R_{\rm occ}}\,.
\end{equation}
For $z=0.001$, $R_{\star}=1000\,R_{\odot}$, and $r_{\star}=0.01\,{\rm pc}$, we get $\mathcal{E}\sim 190$ while for $r_{\star}=0.1\,{\rm pc}$, the enhancement is as much as $\mathcal{E}\sim 1900$. Hence for a razor-thin edge-on disk, the enhancement is by 2-3 orders of magnitude. On the other hand, for a face-on thin disk, the probability of an eclipse drops to zero. Furthermore, for the enhancement to take place for a thin disk, its inclination needs to be close to $90^{\circ}$ (edge-on), which stems from the geometrical condition of a star orbiting the SMBH on a circular orbit: for the minimal projected separation, one has the condition for the eclipse, $r_{\star}|\cos{\iota}|<R_{\rm occ}$, from which $|\cos{\iota}|<R_{\rm occ}/r_{\star}$ and $\iota\sim 89.61^{\circ}$ for $r_{\star}=0.01\,{\rm pc}$. Hence, for the eclipses to take place, the disk can only be inclined by $\sim 0.4^{\circ}$ with respect to the edge-on configuration. For larger inclinations, the eclipse probability goes to zero unless the disk is geometrically thicker (in the limit it can approach an isotropic distribution). The impact of the orbit (disk) inclination on the depth and the shape of the eclipse temporal profiles is analyzed in Subsection~\ref{eq_eclipse_profiles}.

For a small number of nearby mm-bright nuclei, the NSC configuration -- isotropic or disk-like -- is crucial for the rate of eclipses. On the other hand, for a potentially larger sample of sources, the geometrical effects will cancel out and the mean eclipse rate for a sample of galaxies will approach the one as for isotropic NSCs derived earlier.

\section{Modelling eclipse temporal profiles} \label{sec:modelling}

The star-SMBH setup studied in the previous section is analogous to exoplanet transits, i.e. the star--SMBH system is an scaled-up version of the exoplanet--star system with typically longer orbital and eclipse/transit timescales involved. The exoplanet transits typically repeat with the orbital period of $\sim 1-10$ days (here we have tens of years), while the transit durations are $1-10$ hours (here we have the eclipse durations of $\sim 10$ days). There are, however, several other qualitative differences with respect to planetary transits. One of them is the characteristic AGN variability with the fractional rms of $\sim 10-15\%$ and timescales ranging from hours to weeks. Another difference is that unlike the planet-star eclipse case, the SMBH does not necessarily have to be at the exact geometric center of the NSC or the galaxy bulge, which can complicate the search for SMBH eclipses. The latter point can usually be anticipated for merging and post-merger systems with disturbed morphologies. Such offsets may arise from gravitational-wave recoil, SMBH binary orbital motion, asymmetric jet thrust or perturbations by massive objects, for instance globular clusters \citep{2008ApJ...678..780G,2010ApJ...717L...6B,2014ApJ...795..146L}. An infalling or a recoiling SMBH is, however, expected to retain a compact population of bound stars within its sphere of influence, producing an offset secondary photometric nucleus or a hypercompact stellar system \citep{2009ApJ...699.1690M}, where eclipses by stars still take place.

In this section, we show the expected temporal profiles of jet base/ring eclipses by orbiting stars. In Subsec.~\ref{eq_eclipse_profiles} we demonstrate different effects (eccentricity, inclination, observing frequency) on the eclipse profile depth and width. Subsequently, in Subsec.~\ref{subsec_var_meas} we include intrinsic AGN variability and measurement uncertainties to yield realistic eclipse profiles. The eclipse of the millimeter ring geometry is analyzed in Subsec.~\ref{subsec_radio_core_shapes} where we show how to use the eclipse substructure to infer the stellar distance from the SMBH and the stellar radius.

\subsection{Expected eclipse profiles: different effects}
\label{eq_eclipse_profiles}

Here we consider the purely geometrical overlap between the uniformly bright radio core and the red giant/supergiant orbiting the SMBH. We calculate the radio flux density with respect to the base flux as follows,
\begin{equation}
    \frac{F(t)}{F_0}=1-\frac{S_{\rm block}}{\pi R_{\rm c}^2}\,,
    \label{eq_flux_density}
\end{equation}
where $S_{\rm block}$ is the area of the radio core blocked by the star. This simplification breaks down when the radio core is resolved as we discuss in Subsection~\ref{subsec_radio_core_shapes}. As the star orbits the SMBH nearly edge-on with respect to the observer, the blocked area can be expressed depending on the projected distance (separation) between the radio-core and stellar centroids as follows \citep[see e.g.][]{2002ApJ...580L.171M},
\begin{itemize}
    \item when $d\geq R_{\rm c}+R_{\star}$, then $S_{\rm block}=0$,
    \item when $d\leq |R_{\rm c}-R_{\star}|$, then $S_{\rm block}=\pi \text{min}(R_{\star},R_{\rm c})^2$,
    \item when $|R_{\rm c}-R_{\star}| < d < R_{\rm c}+R_{\star}$, i.e. during the partial overlap, the blocked area can be calculated as
    \begin{equation}
        S_{\rm block}=R_{\rm c}^2\alpha + R_{\star}^2\beta-\mathcal{S}\,,
    \end{equation}
    where $\alpha$, $\beta$, and $\mathcal{S}$ are
    \begin{align}
        \alpha &= \arccos{\left(\frac{d^2+R_{\rm c}^2-R_{\star}^2}{2dR_{\rm c}} \right)}\,,\\
        \beta &= \arccos{\left(\frac{d^2+R_{\star}^2-R_{\rm c}^2}{2dR_{\star}} \right)}\,,\\
        \mathcal{S}&=\frac{1}{2}(-d+R_{\rm c}+R_{\star})^{1/2}(d+R_{\rm c}-R_{\star})^{1/2}\times\,\notag\\
        &\times (d-R_{\rm c}+R_{\star})^{1/2}(d+R_{\rm c}+R_{\star})^{1/2}\,.
    \end{align}
\end{itemize}

\begin{figure*}
    \centering
    \includegraphics[width=0.6\columnwidth]{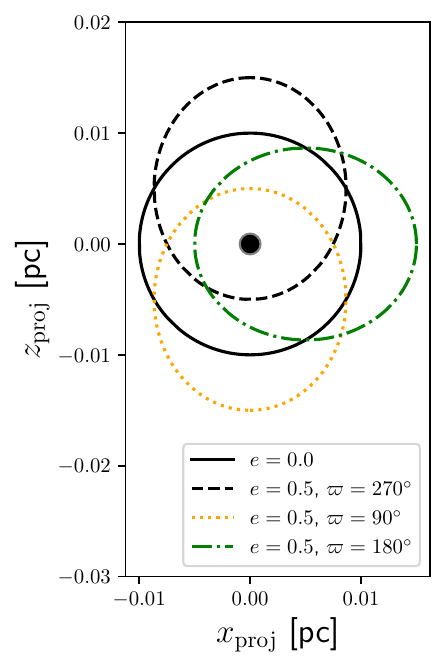}
     \includegraphics[width=1.2\columnwidth]{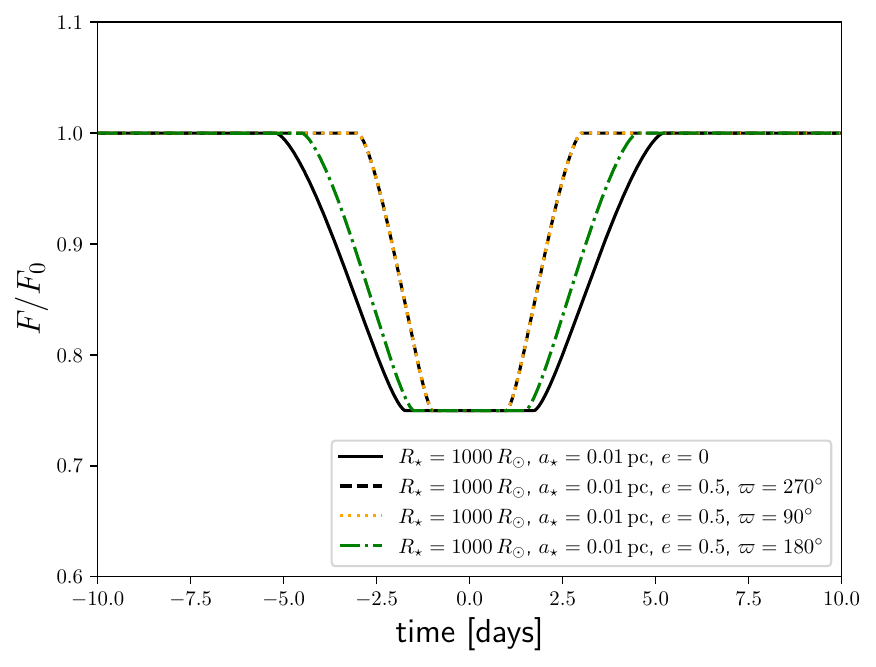}
    \caption{Effect of orbital eccentricity on the eclipse duration. We consider $e=0.0$ (circular orbit) and $e=0.5$ (eccentric orbit) with three orientations with respect to the observer (see the left panel; the observer is to the bottom). The eccentricity generally causes the eclipse shape to be narrower -- the eclipse becomes shorter with respect to the circular orbit and the orientation of the ellipse does not play a role when the major axis is aligned with respect to the observer (see the right panel); however, the eclipse shape becomes wider for the perpendicular orientation of the ellipse major axis with respect to the line of sight.}
    \label{fig_orbit_light_curve_ecc}
\end{figure*}

\begin{figure*}
    \centering
    \includegraphics[width=0.8\columnwidth]{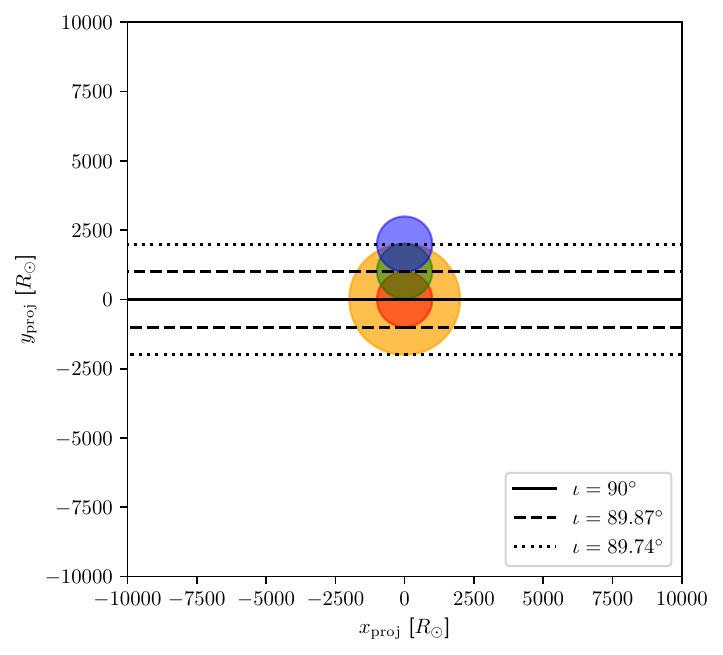}
     \includegraphics[width=1.1\columnwidth]{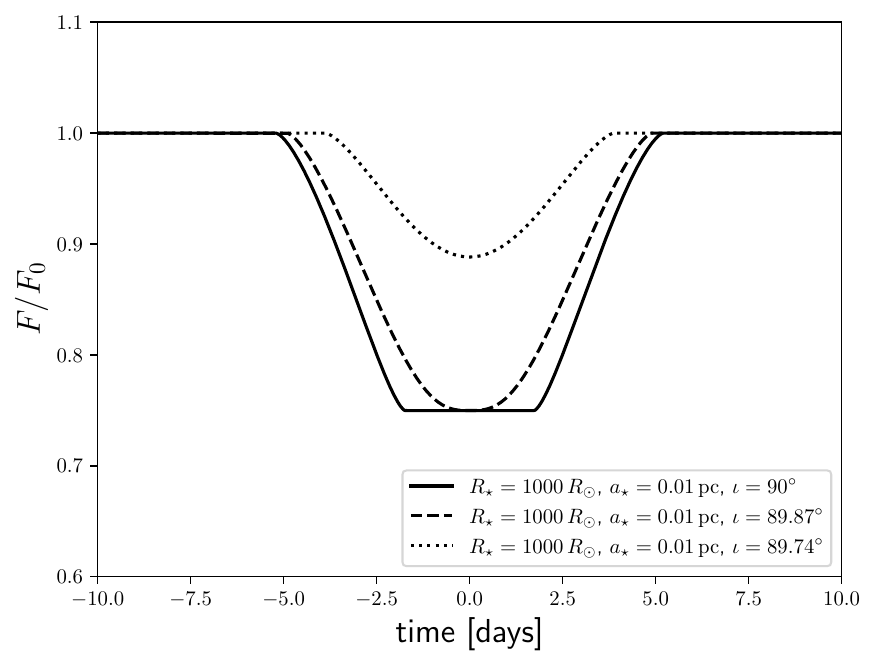}
    \caption{Effect of orbital inclination on the eclipse shape and depth. We consider three inclinations (left panel) with the fixed stellar radius of $R_{\star}=1000\,R_{\odot}$, including the edge-on orbit ($\iota=90^{\circ}$; solid line), the minimum inclination for the occultation to take place ($\iota_{\rm min}=89.74^{\circ}$; dotted line), and the intermediate value of $\iota=89.87^{\circ}$ (dashed line). The decreasing inclination causes the eclipse of the radio core to be shallower (right panel). While for the edge-on orbit ($\iota=90^{\circ}$) the eclipse depth reaches $100 \Delta F/F\sim 25.03\%$ and the light curve is flat-bottomed, for the smaller inclination ($\iota=89.87^{\circ}$) we get a sine-like obscuration profile with the same relative depth. Finally, for the minimum inclination of $\iota=89.74^{\circ}$ we obtain a shallow eclipse with the relative depth of $\sim  11.18^{\circ}$.}
    \label{fig_orbit_light_curve_inc}
\end{figure*}

First, we consider the effect of mild eccentricity. For the red giant star with $R_{\star}=1000\,R_{\odot}$ and orbiting at $r_{\star}=a_{\star}=0.01\,{\rm pc}$ (here considered as the semi-major axis) from the SMBH, we have the orbital period of $P_{\rm orb}=13.25$ years. For a circular orbit, the eclipse duration is $\tau_{\rm dur}\simeq 10.4$ days, see Fig.~\ref{fig_orbit_light_curve_ecc} for the orbit (left) and the approximate eclipse shape (right). For the eccentric orbit ($e=0.5$) the eclipse shape is narrower and the duration is $\tau_{\rm dur}\simeq 5.98$ days, hence it is shortened by $42\%$. The eclipse shape does not depend on the orientation of the ellipse with respect to the observer when the major axis is oriented towards the observer -- in projection, the orbit is narrower than the circular one. The shape becomes a bit wider for the perpendicular orientation ($\varpi=180^{\circ}$), however it is still narrower than the circular case ($\tau_{\rm dur}=8.99$ days). 

Another relevant effect is associated with the orbital inclination with respect to the line of sight. Because of the geometrical configuration the obscuration of the radio core by an orbiting star is maximized for the edge-on configuration ($\iota=90^{\circ}$). The eclipse is detectable until a certain minimum inclination is reached ($\iota_{\rm min}=90^{\circ}-\gamma$), where $\gamma\simeq \arctan{(R_{\rm c}/r_{\star})}$. For $r=0.01\,{\rm pc}$ and $\nu=230\,{\rm GHz}$, we get $\gamma\sim 0.26^{\circ}$ and hence $\iota_{\rm min}\sim 89.74^{\circ}$. We decrease the orbital inclination of a transiting star with $R_{\star}=1000\,R_{\odot}$ first by $\gamma/2$ and then by $\gamma$. This is illustrated in Fig.~\ref{fig_orbit_light_curve_inc} (left panel) by red ($\iota=90^{\circ}$), green ($\iota=89.87^{\circ}$), and blue circles ($\iota=88.74^{\circ}$) during the phase with the minimum separation $d$. In the right panel of Fig.~\ref{fig_orbit_light_curve_inc} we show the effect of the decreasing inclination on the eclipse shapes. With the decreasing inclination, the eclipse becomes shallower -- for the edge-on orbit the transit curve is clearly flat-bottomed with the relative depth of $25.03\%$, then for $\iota=89.87^{\circ}$ it becomes sine-like with the same maximum relative depth, and for $\iota_{\rm min}=89.74^{\circ}$ it gets significantly shallower with the relative depth of $\sim 11.18^{\circ}$.  

\begin{figure}
    \centering
    \includegraphics[width=\columnwidth]{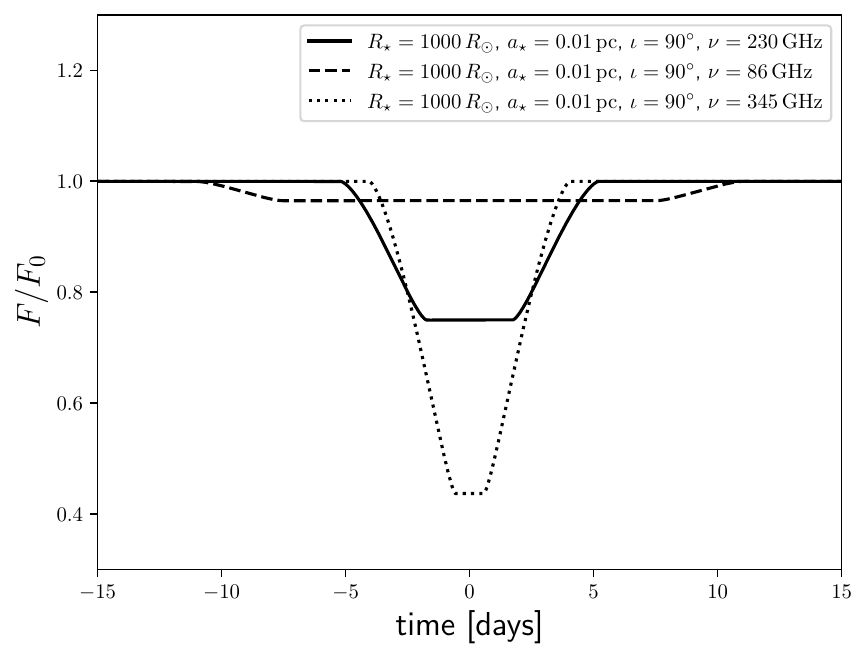}
    \caption{Effect of the observing frequency on the radio-core eclipse temporal profile caused by an orbiting red supergiant with the radius of $R_{\star}=1000\,R_{\odot}$. We show eclipse profiles for the three observing frequencies, 86, 230, and 345 GHz, which are depicted with black dashed, solid, and dotted lines, respectively.}
    \label{fig_orbit_light_curve_nu}
\end{figure}

A third relevant effect for a stellar obstacle is the change of the eclipse profile depth and width with the observing frequency, which can be used to probe whether the obstacle is of a stellar origin with the fixed radius at different frequencies. In Fig.~\ref{fig_orbit_light_curve_nu} we compare temporal eclipse profiles for three different frequencies, $\nu \in (86, 230, 345)\,{\rm GHz}$, adopting the previous parameters for the star and its orbit ($r_{\star}=0.01\,{\rm pc}$, $R_{\star}=1000\,R_{\odot}$, $\iota=90^{\circ}$, $z=0.001$). As previously derived, at lower frequencies, the eclipse is longer and shallower with the depth of $3.50\%$ at 86 GHz, while towards higher frequencies, it gets deeper (with the depth of 25.03\% at 230 GHz and 56.31\% at 345 GHz) and shorter. This prominent chromatic effect can in principle be utilized to observationally probe the nature of the obscuring stellar body.

\subsection{Effect of intrinsic variability and measurement uncertainties}
\label{subsec_var_meas}

In addition to variations due to eclipses by stars, AGN are intrinsically variable on comparable timescales of days and the variability amplitudes can reach $\sim 10\%$ \citep{2023ApJ...951...93C}. The power density spectrum of this variability can be described as a broken power law with the break timescale depending on the SMBH mass and the Eddington ratio \citep{2024A&A...684A.133A}.   

\begin{figure}
    \centering    \includegraphics[width=\columnwidth]{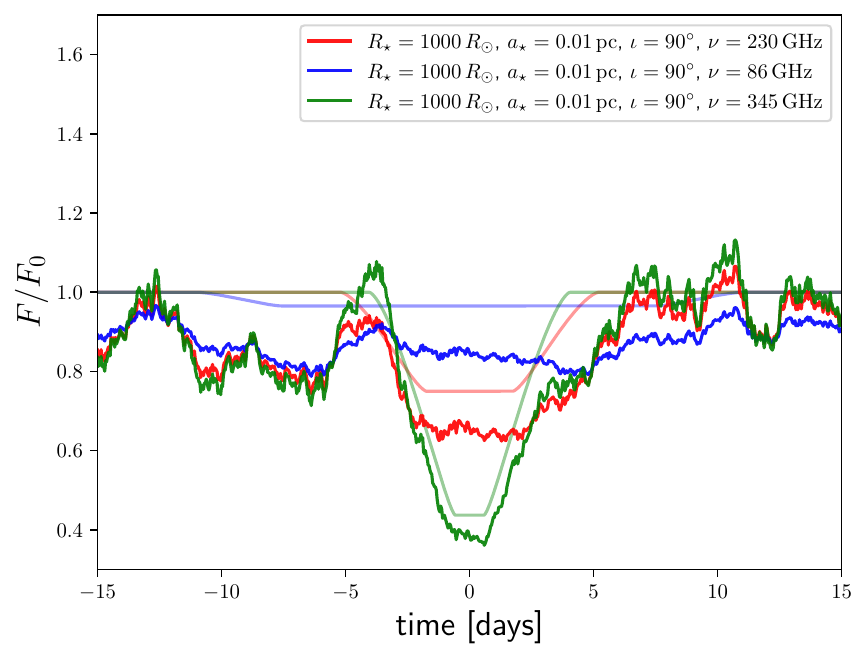}    \includegraphics[width=\columnwidth]{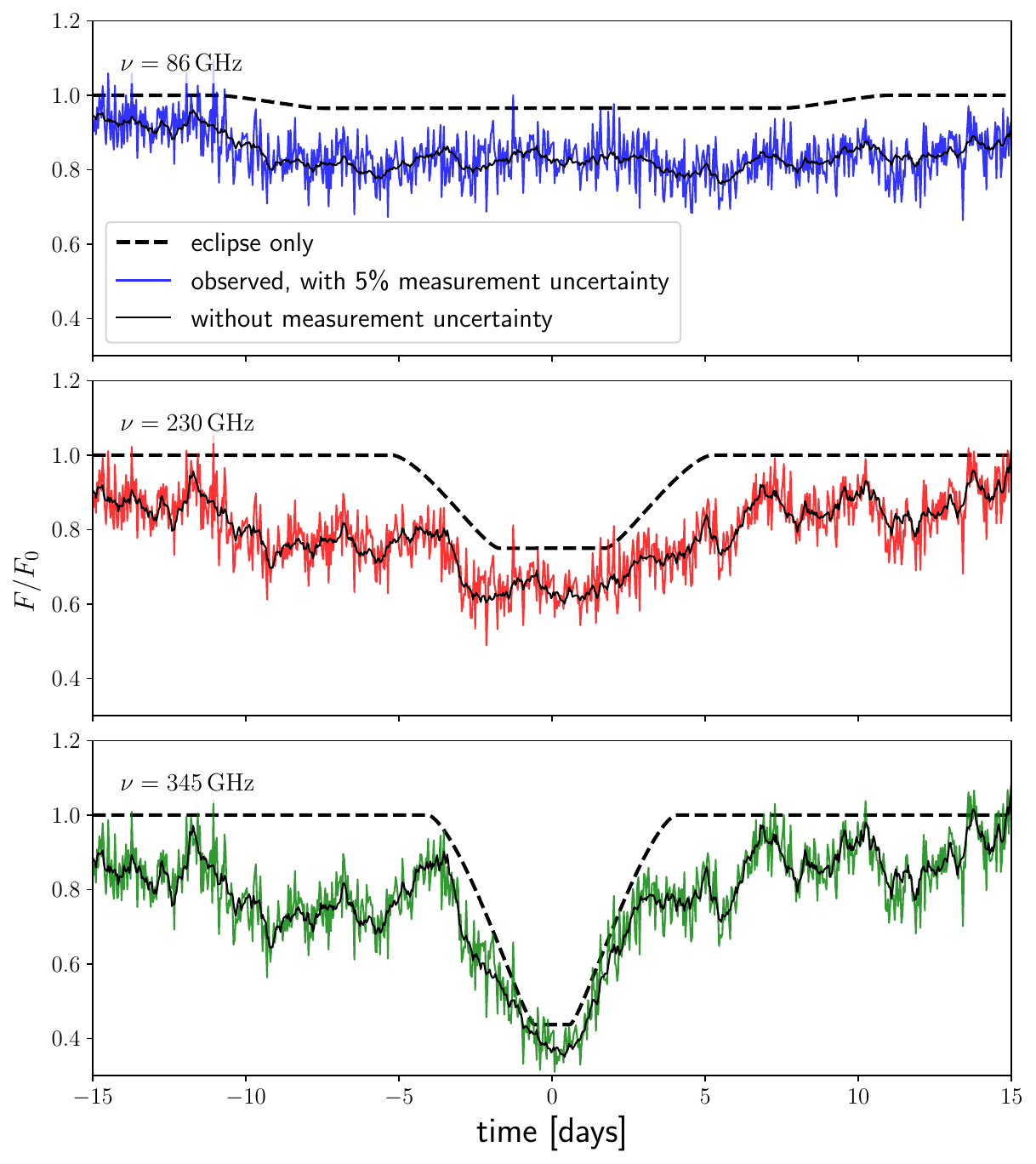}
    \caption{Demonstration of the source intrinsic variability and measurement uncertainties. \textit{Top panel:} Frequency-dependent eclipses by the star of $R_{\star}=1000\,R_{\odot}$, the same as in Fig.~\ref{fig_orbit_light_curve_nu}, with the inclusion of the source intrinsic variability. The variability is modeled as the damped random walk with the fractional rms variability of 15\%, characteristic timescale of 10 days, and the fractional rms and timescale slopes of 0.25 and -1.0 with respect to the reference frequency of 230 GHz. \textit{Lower panel:} The setup with the same intrinsic variability properties as in the upper panel but with the inclusion of the 5\% Gaussian measurement uncertainty per epoch, which is depicted by colored lines. The black solid lines represent intrinsic variability without measurement uncertainties while dashed lines show the pure eclipse profile.}
    \label{fig_eclipse_var_meas}
\end{figure}

Although there can be differences in terms of the stochastic variability in the radio/mm/submm domain among different galactic nuclei, we analyze the eclipse impact on the variable background using the damped random walk \citep[DRW;][]{2009ApJ...698..895K,2017A&A...597A.128K} that can capture basic characteristics of the AGN variability; see also Appendix~\ref{sec_appendix_drw} for a more detailed, mathematical description of the approach. 

In Fig.~\ref{fig_eclipse_var_meas} (top panel) we show the eclipse profiles at different frequencies (86, 230, and 345 GHz) including intrinsic variability amplitude of 15\%, the characteristic timescale of 10 days, and the amplitude and the timescale frequency slopes of 0.25 and -1.0, respectively. We see that unless the eclipse relative depth reaches at least a few 10\%, and thus significantly exceeding the intrinsic variability, it cannot be recovered on a significant level. The best prospects for a clear detection are therefore at the submillimeter wavelengths (345 GHz) and for the largest giants ($R_{\star}=1000\,R_{\odot}$). 

Adding the Gaussian-like measurement noise of 5\% corresponding to the well-calibrated millimeter facilities does not impact the eclipse recovery significantly, see Fig.~\ref{fig_eclipse_var_meas} (lower panel). The eclipses with relative depths of $\gtrsim 50\%$ can clearly be recovered, while those with the comparable amplitude as the intrinsic variability are masked by the intrinsic variations and the measurement noise. 

\subsection{Radio-core and ring geometry}
\label{subsec_radio_core_shapes}

In real cases, the radio core deviates from uniformly bright circular disks and hence it is necessary to take into account the substructure of the emission region for the calculation of the eclipse temporal profiles. For the millimeter jet base, the substructures could be due to the intermittent component ejection and jet instabilities.

For the millimeter ring emission, as suggested by the reconstructed images of Sgr A* and M87* \citep{2019ApJ...875L...1E,2022ApJ...930L..12E}, the central brightness depression would result in the intermittent flux enhancement as the star would be located in the shadow region in projection -- this would cause the difference with respect to the flat-bottomed temporal profile, see Subsec.~\ref{eq_eclipse_profiles}. In addition, any asymmetries in the ring azimuthal emissivity profile, e.g. due to the Doppler boosting of the accretion flow or hot spots would result in the eclipse temporal profile asymmetries.

\begin{figure}
    \centering
    \includegraphics[width=\columnwidth]{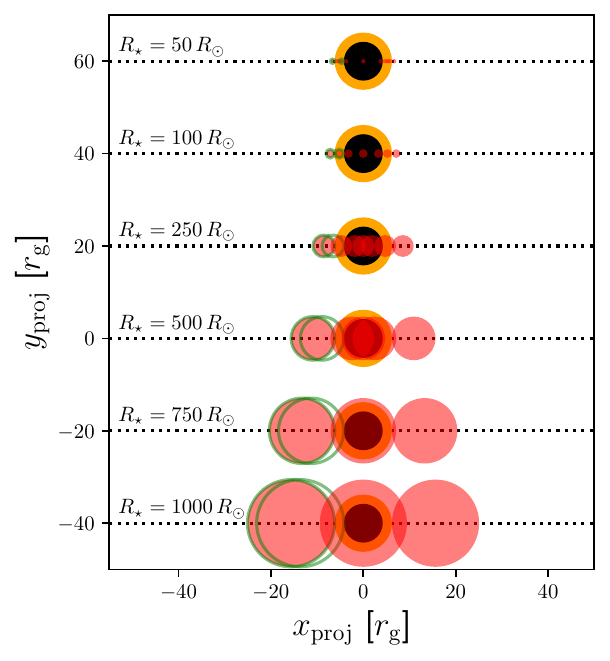}
    \caption{Edge-on view of the transit of a red-giant star in front of the nuclear ring-like emission associated with the SMBH at millimeter wavelengths. The temporal profiles of the eclipse are shown in Fig.~\ref{fig_light_curve_ring}. From the top to the bottom, the radius of the star is increased from $R_{\star}=50\,R_{\odot}$ to $R_{\star}=1000\,R_{\odot}$. The star orbits the SMBH on a circular orbit with $r_{\star}=0.01\,{\rm pc}$. The ring has an inner radius of $4.2\,r_{\rm g}$ and an outer radius of $6.2\,r_{\rm g}$. The green circles mark the contact points of the star with the outer and the inner ring radii -- the contact times, $t_2$ and $t_1$, respectively, -- are also depicted in Fig.~\ref{fig_light_curve_ring}.}
    \label{fig_orbit_xy_ring}
\end{figure}

\begin{figure}
    \centering
    \includegraphics[width=\columnwidth]{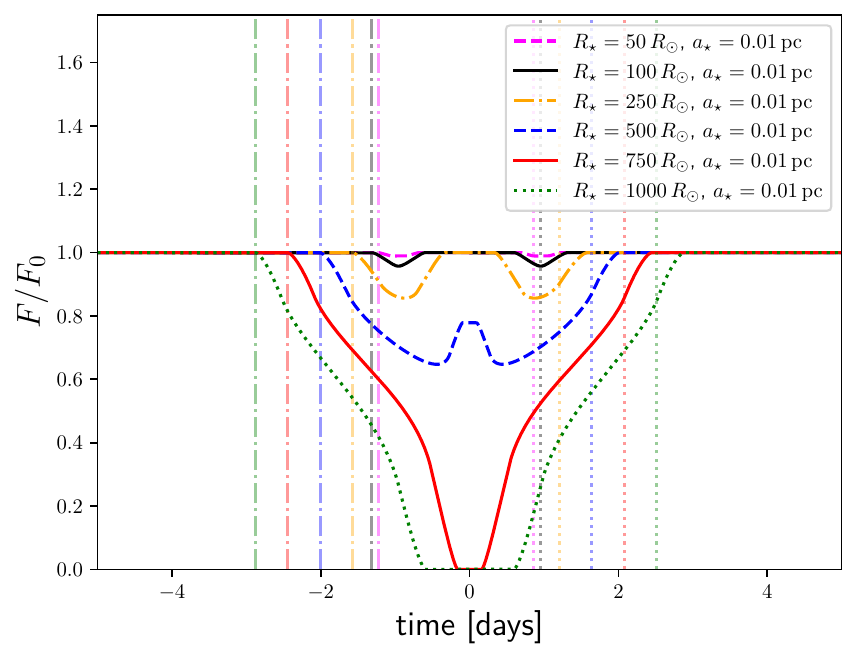}
    \caption{Temporal profiles of single stellar eclipses for the case when the millimeter nuclear emission is associated with a uniformly bright ring with the inner radius of $4.2\,r_{\rm g}$, the outer radius of $6.2\,r_{\rm g}$, and hence the width of $2r_{\rm g}$ and the mean radius of $5.2\,r_{\rm g}$. We vary the stellar radius from $50\,R_{\odot}$ to $1000\,R_{\odot}$. Vertical dash-dotted lines mark the time of the first outer circle crossing (during ingress) while the dotted lines represent the last inner ``shadow" contact (during egress).}
    \label{fig_light_curve_ring}
\end{figure}

We demonstrate the impact of the millimeter jet base/ring substructure for the case of a concentric, uniformly bright ring with the inner radius of $R_{\rm r1}=4.2\,r_{\rm g}$, the outer radius of $R_{\rm r2}=6.2\,r_{\rm g}$, and hence the width of $2\,r_{\rm g}$, which is expected for the lensed radiating inflow with some variance \citep{2020SciA....6.1310J}. This millimeter ring thus represents the case of a gravitationally lensed accretion inflow with the mean radius of $R_{\rm r}=0.5(R_{\rm r1}+R_{\rm r2})\sim 5.2r_{\rm g}$. The physical scale is set by the SMBH mass, which we set to $M_{\bullet}=5\times 10^7\,M_{\odot}$ as in the previous estimates and therefore $r_{\rm g}=GM_{\bullet}/c^2\sim 106\,(M_{\bullet}/5\times 10^7\,M_{\odot})\,R_{\odot}$. The occulting star is assumed to orbit the SMBH on a circular orbit that is exactly edge-on with respect to the observer, with $r_{\star}=0.01\,{\rm pc}$, which implies the eclipse recurrence timescale of $\tau_{\rm rec}=13.25\,(r_{\star}/0.01\,{\rm pc})^{3/2}(M_{\bullet}/5\times 10^7\,M_{\odot})^{-1/2}\,{\rm years}$. Although it is quite long, here we are interested in the single eclipse profile, its substructure, and what properties about the stellar orbit we can infer from it. The total duration of the eclipse is proportional to the sum of the stellar and the outer ring radii and inversely proportional to the velocity component along the sky, see Eq.~\eqref{eq_eclipse_duration}:
\begin{align}
   \tau_{\rm dur}&=\frac{2(R_{\star}+R_{\rm r2})r_{\star}^{1/2}}{(GM_{\bullet})^{1/2}}\,\notag\\
   &\simeq 4.02 \left(\frac{R_{\star}+R_{\rm r2}}{500\,R_{\odot}+6.2\times 106\,R_{\odot}} \right) \left(\frac{r_{\star}}{0.01\,{\rm pc}} \right)^{1/2} \times \,\notag\\ &\times \left(\frac{M_{\bullet}}{5\times 10^7\,M_{\odot}} \right)^{-1/2}\,\text{days}.    
\end{align}
Hence, to detect and sample the substructure information, it is necessary to perform a high-cadence observation with the sub-day sampling. In particular, the inner shadow region is crossed in $\sim 3.3\,[(R_{\star}+R_{\rm r1})/(500\,R_{\odot}+4.2\times 106\,R_{\odot})](r_{\star}/0.01\,{\rm pc})^{1/2}(M_{\bullet}/5\times 10^7\,M_{\odot})^{-1/2}$ days. The short duration of the stellar eclipse justifies the negligence of the intrinsic variability of the source in the following estimates since there are not expected to be significant changes during one eclipse.

To calculate normalized temporal profiles of stellar eclipses (neglecting intrinsic variability), we modify the relation in Eq.~\eqref{eq_flux_density} for the ring geometry,
\begin{equation}
    \frac{F(t)}{F_0}=1-\frac{S_{\rm block}(R_{\rm r2})-S_{\rm block}(R_{\rm r1})}{\pi (R_{\rm r2}^2-R_{\rm r1}^2)}\,,
    \label{eq_flux_density_ring}
\end{equation}
where the areas $S_{\rm block}(R_{\rm r2})$ and  $S_{\rm block}(R_{\rm r1})$ blocked by the star with the radius $R_{\star}$ are each calculated as for the corresponding uniformly bright circles.

\begin{figure}
    \centering
    \includegraphics[width=\columnwidth]{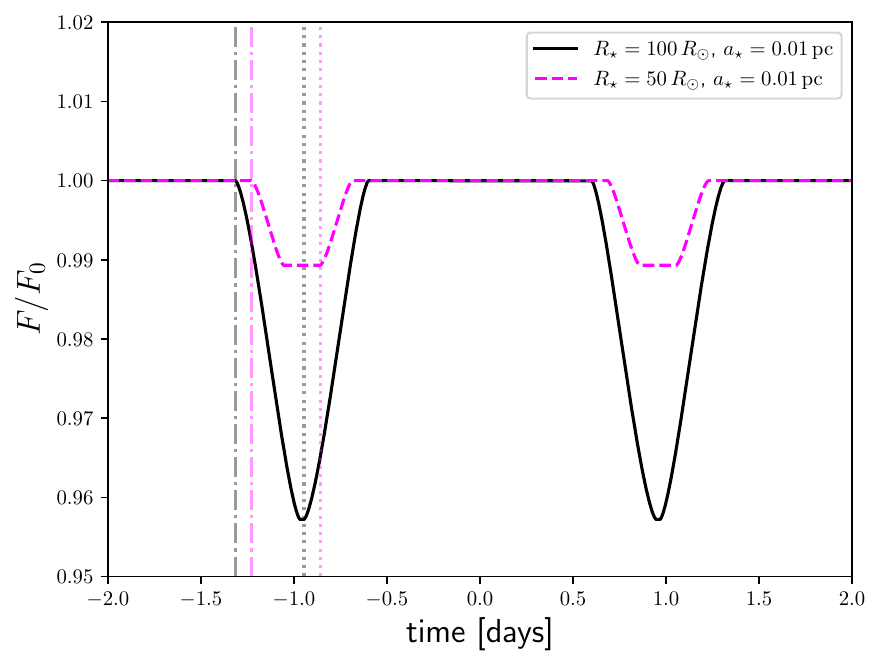}
    \caption{Temporal profiles of single stellar eclipses for the cases of smaller evolved stars with $R_{\star}=50-100\,R_{\odot}$ that cause eclipses with the relative flux density drop of only a few percent for the ring radii of $4.2-6.2\,r_{\rm g}$. Vertical dash-dotted lines mark the time of the first outer circle crossing while the dotted lines represent the first inner ``shadow" contact (i.e. both times are during the ingress).}
    \label{fig_light_curve_ring50-100}
\end{figure}

We vary the stellar radius using the following values $R_{\star} \in \{50, 100, 250, 500, 750, 1000\}\,R_{\odot}$, see Fig.~\ref{fig_orbit_xy_ring} for a graphical representation of the eclipses (edge-on view). There is a qualitative difference in the temporal eclipse profiles, see Fig.~\ref{fig_light_curve_ring}, depending on the ratios between the ring width, central depression radius, and the stellar radius. For stellar radii smaller than the shadow radius, $R_{\star}\lesssim 4.2\times 106\sim 445\,R_{\odot}$, the flux enhancement at the eclipse center reaches the nominal flux. For larger stellar radii, i.e. $445\,R_{\odot}\lesssim R_{\star} < 657\,R_{\odot}$, the central flux enhancement is progressively smaller. For $R_{\star} \gtrsim 657\,R_{\odot}$ the millimeter ring emission is intermittently fully blocked. This can be explicitly seen by comparing eclipse profiles in Fig.~\ref{fig_light_curve_ring}. As estimated before in Subsec.~\ref{sec:depth}, the larger the stellar radius, the larger the relative drop in the flux density for the fixed frequency, though the relation is more complex due to the ring shape of the millimeter emission region in comparison with a uniformly bright circle. In Table~\ref{tab_flux_drop_ring} we summarize the maximum relative flux density drops (in percent) for individual stellar radii.
\begin{table}[h!]
    \centering
     \caption{Relative maximum flux density drops (in percent) for stars of different radii in the range between $50\,R_{\odot}$ and $1000\,R_{\odot}$ passing in front of the nuclear ring with the mean radius of $\sim 5.2\,r_{\rm g}$.}
    \begin{tabular}{c|c}
    \hline
    \hline
     $R_{\star}$ [$R_{\odot}$]    &   max $\Delta F_{\nu}/F_{\nu}$ [$\%$]\\
     \hline
     50    &  1.07  \\
     100   &  4.28   \\
     250   &  14.39   \\
     500   &  35.26   \\
     750   &  100.00   \\
     1000  &  100.00    \\
     \hline 
    \end{tabular}   
    \label{tab_flux_drop_ring}
\end{table}
Since for $R_{\star}=50-100\,R_{\odot}$ the relative flux density drop reaches only a few percent, we show the corresponding eclipse profiles separately in Fig.~\ref{fig_light_curve_ring50-100} for clarity.

When we assume that the SMBH mass $M_{\bullet}$ is inferred by an independent method and hence we can estimate the ring radius $R_{\rm r}$ and its width $w$, it is possible to estimate the stellar radius $R_{\star}$ and its distance $r_{\star}$ from mock light curves as those calculated earlier when compared with the single observed eclipse profile. The forward model $F(t; R_{\star},v_{\rm sky},t_0)$ contains the three free parameters -- $R_{\star}$, the projection of the stellar velocity on the sky $v_{\rm sky}$ (and hence the distance $r_{\star}$ in the central potential of the SMBH, neglecting the inclination), and the time of the closest transit $t_0$.

For practical estimates, we can also calculate the stellar radius and its distance from the SMBH from the eclipse profile alone by identifying the two key times: $(t_2-t_0)$ and $(t_1-t_0)$, which stand for the times corresponding to the first contact points with the outer radius $R_{\rm r2}$ and the inner radius $R_{\rm r1}$, respectively, with respect to the time of the closest transit $t_0$. For the first contact we have $t_2-t_0=(R_{\rm r2}+R_{\star})/v_{\rm sky}$ while for the second contact $t_1-t_0=(R_{\rm r1}+R_{\star})/v_{\rm sky}$ where $v_{\rm sky}=(GM_{\bullet}/r_{\star})^{1/2}$ is the orbital velocity. For their difference we have $\Delta t=t_2-t_1=W/v_{\rm sky}$, where $W$ is the ring width. From these relations we obtain the distance of the eclipsing star from the SMBH,
\begin{equation}
    r_{\star}=GM_{\bullet}\left(\frac{\Delta t}{W}\right)^2\,.
    \label{eq_distance}
\end{equation}
The stellar radius can be inferred from the relation for any contact time in combination with Eq.~\eqref{eq_distance},
\begin{equation}
    R_{\star}=W\frac{\overline{t}}{\Delta t}-R_{\rm r}\,,
    \label{eq_stellar_radius}
\end{equation}
where $\overline{t}=0.5(t_2+t_1)$ is the mean contact time. It is more convenient to express the ring radius as well as its width in gravitational radii, $R_{\rm r}=\xi_{\rm r}r_{\rm g}$ and $W=wr_{\rm g}$, respectively. Then the stellar distance and its radius in gravitational radii are, 
\begin{equation}
    \frac{r_{\star}}{r_{\rm g}}=w^{-2}\left(\frac{\Delta t}{t_{\rm g}} \right)^2\,,
    \label{eq_distance_rg}
\end{equation}
\begin{equation}
    \frac{R_{\star}}{r_{\rm g}}=w \frac{\overline{t}}{\Delta t} -\xi_{\rm r}\,,
    \label{eq_stellar_radius_rg}
\end{equation}
where $t_{\rm g}=r_{\rm g}/c\simeq 4.1 (M_{\bullet}/5\times 10^7\,M_{\odot})$ min is the gravitational time of the SMBH.

As an example, we will attempt to recover stellar distances and radii for $\sim 4\%$ and $35\%$ eclipses, see Table~\ref{tab_flux_drop_ring}. The time of the first contact $t_2$ is relatively straightforward to infer -- this is when the drop in flux density begins. The time of the contact with the central depression is more difficult to estimate but is the time when the slope during the flux decrease changes from a steeper one to less steep. In the estimates, we fix the SMBH mass to $M_{\bullet}=5\times 10^7\,M_{\odot}$, the ring radius to $R_{\rm r}\simeq 5.2\,r_{\rm g}$, and the ring width to $W\simeq 2\,r_{\rm g}$. In Table~\ref{tab_eclipse_param_star}, we summarize the results based on the visual inspection of Fig.~\ref{fig_light_curve_ring} and Fig.~\ref{fig_light_curve_ring50-100}. We see that even though we use crude visual estimates for the timing, the distance is determined with $\sim 10\%$ precision and the stellar radius is inferred within $\sim 20\%$ precision. A direct model fit using e.g. the least-squares method would be even more precise. On the other hand, there are several systematic uncertainties, such as the SMBH mass, the orbital eccentricity, the orbital inclination (impact parameter), orbital phase, and the ring width (unless resolved) that effectively enlarge the uncertainty range or induce certain degeneracies (in particular the ring width directly affects the stellar radius and the distance from the SMBH). However, despite these uncertainties, the ring geometry provides more constraints on the stellar parameters due to more contact points in comparison with a simple (uniformly bright) disk. 

\begin{table}[h!]
    \centering
    \caption{Stellar distance from the SMBH and radius determination based on the timing of the single eclipse of the nuclear ring in the millimeter domain.}
    \begin{tabular}{c|c|c|c|c}
    \hline
    \hline
    Eclipse [\%]  & $t_2-t_0$ [day]  & $t_1-t_0$ [day]  & $r_{\star}$ [pc]  & $R_{\star}\,[R_{\odot}]$  \\
    \hline
    4.28   & -1.32  & -0.96  & 0.0096  &  120 \\
    35.26  & -2.01  & -1.63   & 0.011   & 464  \\
    \hline     
    \end{tabular}   
    \label{tab_eclipse_param_star}
\end{table}

\begin{figure*}
    \centering    \includegraphics[width=\columnwidth]{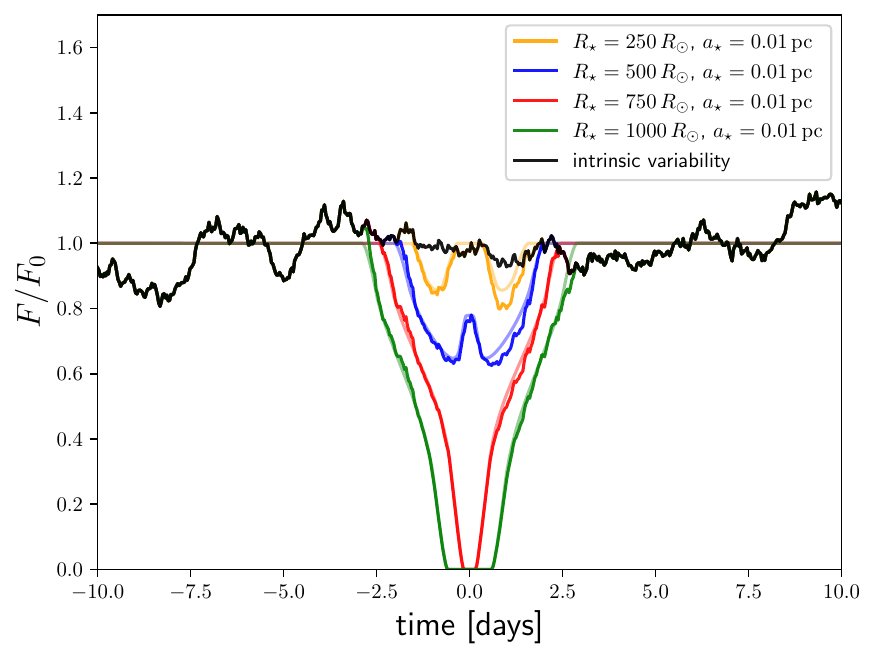}    \includegraphics[width=\columnwidth]{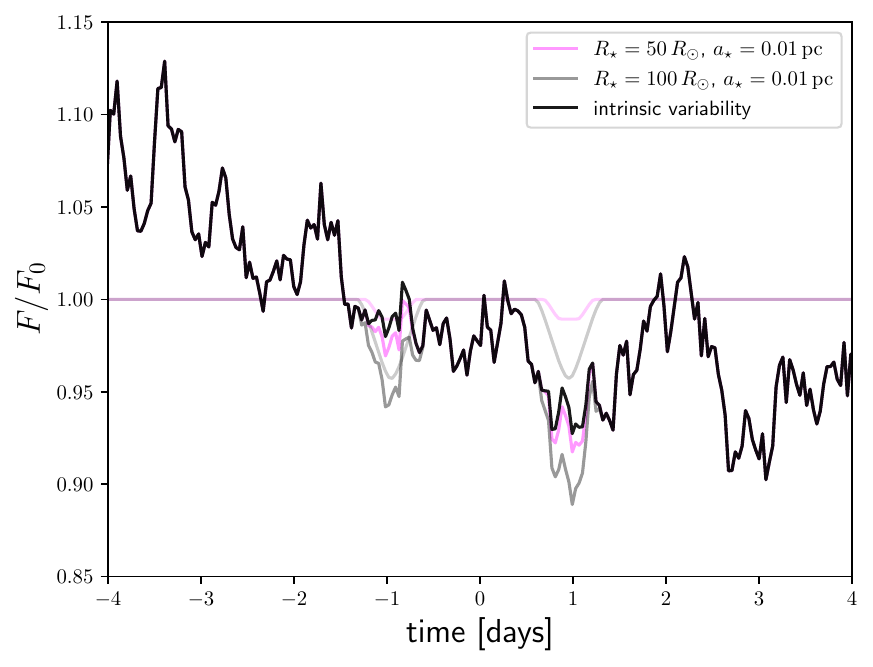}
    \caption{Eclipse profiles with the inclusion of intrinsic source variability. \textit{Left panel:} The same as in Fig.~\ref{fig_light_curve_ring} for $R_{\star}\in\{250, 500, 750, 1000\,R_{\odot}\}$ with the DRW parameters of $\sigma_{X0}=0.15$, $\tau_0=10$ days, $\alpha_{\sigma}=0.25$, and $\alpha_{\tau}=-1.0$. The ring eclipse can be traced well for the stellar radii in this range, with the maximum eclipse depth of $\gtrsim 14\%$. \textit{Right panel:} The same as in the left panel but for the stellar radii of $R_{\star}=50\,R_{\odot}$ and $100\,R_{\odot}$. Since the eclipse depth reaches 1.07\% and 4.28\%, respectively, the eclipse profile is hidden within the intrinsic variability.}
    \label{fig_eclipse_var_ring}
\end{figure*}

The precision of determining contact timescales with the outer and inner ring radius depends on the eclipse recovery within the variable AGN emission. When we calculate synthetic light curves according to fiducial DRW parameters as in Subsec.~\ref{subsec_var_meas}, for the adopted ring radii of $4.2-6.2\,r_{\rm g}$, we see in Fig.~\ref{fig_eclipse_var_ring} that eclipses for stars $R_{\star}\gtrsim 250\,M_{\odot}$ (maximum relative depths of $\gtrsim 14\%$) can be traced well within the intrinsic variable emission (left panel). This is due to the fact that the eclipse amplitude is comparable to or greater than the intrinsic variability amplitude. For smaller stars (right panel), the eclipse impact on the light curve is hidden since it is typically smaller than the intrinsic variations.   

In the future model calculations of stellar eclipses, especially for the case of the Sgr~A*-like millimeter ring, we will consider the asymmetric ring-like geometries for the eclipse temporal profiles that can naturally arise due to Doppler-boosting of the approaching hot plasma or due to the intrinsic variability and instabilities in the lensed accretion inflow. It also remains to be investigated whether the occultation event by evolved stars, especially those whose size is comparable to the ring radius, $R_{\star}\gtrsim R_{\rm r}\simeq 551\,R_{\odot}$, could be accompanied by a lensing event half an orbit later. This would correspond to the phase when the star is behind the SMBH. Although the thermal millimeter emission of the star could be strongly magnified at this phase, the lensed continuum is expected to remain several orders of magnitude fainter than the millimeter-bright nuclear source for galaxies at distances of a few megaparsecs. This is due to the fact that the mm flux of star behaves as $F_{\star,{\rm mm}}\propto T_{\star} R_{\rm mm}^2D_{\star}^{-2}$ in the Rayleigh-Jeans regime where $R_{\rm mm}$ is the stellar photosphere in the millimeter domain \citep{1997ApJ...476..327R,2020A&A...638A..65O}. The effect is therefore unlikely to contaminate the eclipse signal.

\section{Discussion}
\label{sec:discussion}

We analyzed the possibility of the eclipses of radio cores by stars in NSCs. We found that the only possibility for a significant occultation ($100 \Delta F_{\nu}/F_{\nu}>10\%$) is in the mm domain ($\nu= 86-345$ GHz) by evolved stars in the red-giant and supergiant stage of evolution ($R_{\star}=500-2000\,R_{\odot}$). In addition, the mm radio cores need to be nearby ($z\sim 0.001$) since for more distance sources ($z\sim 0.01$) only percent-level eclipses by stars are possible. If the millimeter emission is dominated by the lensed accretion flow/jet (ring), the distance is in principle not a limiting criterion. Instead it is the SMBH mass, see Eq.~\eqref{eq_SMBH_limit}. We also demonstrated specific eclipse temporal profiles, including intrinsic variability and measurement uncertainties. Here we discuss further details, such as the occultations due to TDE streams, one specific source with recurring dips, and the requirements for observational facilities.

\subsection{Occultations by TDE streams}
\label{subsec_tde_streams}

It is clear that the observational requirement of repeating nuclear occultations in the millimeter domain pushes the stars towards their tidal radii. For instance, for the red supergiant with $R_{\star}=1000R_{\odot}$ causing occultations every $\tau_{\rm rec}=10$ years, we have the orbital distance simply following from Eq.~\eqref{eq_recurrence_timescale} (two-body approximation),
\begin{align}
  r_{\star}&\simeq 8.29 \times 10^{-3} \left(\frac{\tau_{\rm rec}}{10\,\text{years}} \right)^{2/3} \times\,\notag\\
  &\times\left(\frac{M_{\bullet}}{5\times 10^7\,M_{\odot}}\right)^{1/3}\,{\rm pc}\,,  
\end{align}
while its tidal radius is, see Eq.~\eqref{eq_tidal_radius},
\begin{align}
r_{\rm t}&\simeq 8.31 \times 10^{-3}\left(\frac{R_{\star}}{1000\,R_{\odot}} \right) \times \,\notag\\
&\times \left(\frac{M_{\bullet}}{5\times 10^7\,M_{\odot}} \right)^{1/3} \left(\frac{m_{\star}}{1\,M_{\odot}} \right)^{-1/3}\,{\rm pc}\,.
\end{align}
Since in this case $r_{\star}\lesssim r_{\rm t}$, the star would get disrupted at some point and form a tidal stream. Initially, when the stream is dense enough, it would also cause the obscuration of the radio core with the duration typically longer than the one given by the star of a given radius, see Eq.~\eqref{eq_eclipse_duration}. The debris is stretched by a factor of two, i.e. from $R_{\star}$ to $2R_{\star}$, on the tidal timescale \citep{2012ApJ...755..155S,2022ApJ...931...39M}
\begin{align}
    \tau_{\rm tidal} &\approx \frac{r_{\rm t}^{3/2}}{(GM_{\bullet})^{1/2}}\,\notag\\
    &=1.6\,\left(\frac{r_{\rm t}}{8.3\times 10^{-3}\,{\rm pc}} \right)^{3/2} \left(\frac{M_{\bullet}}{5\times 10^7\,M_{\odot}} \right)^{-1/2}\text{years}\,,    
\end{align}
hence during one orbital timescale the debris gets stretched to $R_{\rm stream}\sim 6.25\times 2\sim 12.5\,R_{\star}$ and the total duration of the eclipse can reach (at $\nu=230\,{\rm GHz}$ and $z=0.001$),
\begin{align}
    \tau_{\rm dur}^{\rm stream} &\simeq \frac{2(R_{\rm stream}+R_{\rm c})r_{\star}^{1/2}}{(GM_{\bullet})^{1/2}}\,\notag\\
    &=46\,\left(\frac{r_{\rm t}}{8.3\times 10^{-3}\,{\rm pc}} \right)^{1/2}\left(\frac{M_{\bullet}}{5\times 10^7\,M_{\odot}} \right)^{-1/2}\text{days}\,,
\end{align}
under the assumption that the debris remains optically thick at this frequency. The duration of the eclipse due to the TDE stream is thus longer by a factor of $\sim 5$ than the eclipse due to the star of $R_{\star}=1000\,R_{\odot}$. However, as the stream is being stretched, it is effectively diluted on the orbital timescale and hence the prolonged obscurations due to TDE streams are essentially non-periodic. 

\subsection{Case of PKS 1413+135 with recurrent dips}

The blazar PKS 1413+135 (J1420+1315) ($z\sim 0.25$, $D_{\rm A}=806.5\,{\rm Mpc}$) is a source that exhibits repeating prolonged dips in the radio light curve with the recurrence timescale of $\tau_{\rm rec}\sim 4$ years and the duration of $\tau_{\rm dur}\sim 0.8$ years in the rest frame \citep{Vedanthamb}. The depth of the dips is $\Delta F_{\nu}/F_{\nu}\sim 0.5$ and the core size at 15 GHz is of the order of $\theta_{\rm c}\sim 1\,{\rm mas}$. From $\Delta F_{\nu}/F_{\nu}\sim (R_{\rm obs}/R_{\rm c})^2$ we obtain an estimate of the obscurer size, $R_{\rm obs}\sim 2.8\,{\rm pc}$, which is far larger than any stellar object. If the dips in the radio light curve are due to eclipses, the occulting bodies are extended gaseous clouds and streamers. If we assume that a single cloud bound to the SMBH causes the eclipses, then we can use Eq.~\eqref{eq_mass_obs} to infer the SMBH mass of the blazar. We get $M_{\bullet}\sim 6.6\times 10^{17}\,M_{\odot}$, which is almost eight orders of magnitude larger than a realistic SMBH mass. Therefore we conclude that the dips in PKS 1413+135 are caused by a different mechanism than by a bound dense gaseous material. They could be caused by the string of clouds with a certain separation further away from the SMBH, such as in the central molecular zone of the host galaxy. 

\subsection{Requirements for eclipse observations}

In this subsection we focus on the possibility of detecting one-epoch eclipses in the millimeter/submillemeter light curves of nearby AGN. As long as one eclipse profile is well measured, for the independently determined SMBH mass it is possible to estimate the stellar radius and its distance from the SMBH, see Eqs.~\eqref{eq_stellar_radius_single} and \eqref{eq_distance_single} for the circular-like radio core and Eqs.~\eqref{eq_distance} and \eqref{eq_stellar_radius} for the ring-like millimeter emission. This way one gets an estimate for the recurrence timescale using Eq.~\eqref{eq_recurrence_timescale} and the next observational campaign can be planned on the timescale of $\sim 10$ years. In case the second eclipse of the same nucleus is detected with a comparable depth and duration, which makes it more likely that it comes from the same bound star, one can use Eq.~\eqref{eq_mass_num} to independently constrain the SMBH mass.

For the flux density sensitivity, the minimal requirement is that the millimeter emission of nearby AGN is detected, including those that are classified as low-luminosity or accreting significantly below their Eddington limit. For these nuclei the mm flux density is in the range $F_{\nu}\sim 0.1-5\,{\rm Jy}$ \citep{2021ApJ...910L..14G,2025ApJ...981..126F} but can be even lower, e.g. NGC 4258 \citep{2013ApJ...765...63D}, or higher \citep[Cen A with 5.6 Jy][]{2021NatAs...5.1017J}. Therefore we adopt the limit of $F_{\nu}\gtrsim 100\,{\rm mJy}$ as the minimal requirement. In addition, the eclipse profile with the maximum relative depth of $\delta=\Delta F_{\nu}/F_{\nu}$ needs to be detected with a high enough significance $q$, where we set $q=5$ $\sigma$. Hence the basic requirement for the instrumental uncertainty at a given significance level is,
\begin{equation}
    \sigma \leq \frac{\delta F_{\nu}}{q}\,.
\end{equation}
We pick two representative values for the relative eclipse depth, $\delta=0.25$ and $\delta=0.5$. Since these eclipses exceed the typical AGN variability of $\sim 10-15\%$ they can be resolved against the intrinsic variable emission, see also Figs.~\ref{fig_eclipse_var_meas} and \ref{fig_eclipse_var_ring} for the simulated light curves. In Table~\ref{tab_uncertainty} we summarize the required instrumental rms for the $5\sigma$ detection of the eclipse of galactic nuclei with a different millimeter brightness.

\begin{table}[h!]
    \centering
    \caption{Required observational rms uncertainties for the 5$\sigma$ detection of eclipses with relative depths of 25\% (second column) and 50\% (third column). Galactic nuclei with different mm flux densities are in different rows (from the top to the bottom with the increasing flux density from 10 to 1000 mJy).}
    \begin{tabular}{c|c|c}
    \hline
    \hline
    $F_{\nu}$ [mJy]  & rms for $\delta=0.25$ [mJy] & rms for $\delta=0.5$ [mJy]  \\
    \hline
    10     &  0.5   &  1.0    \\
    100    &  5.0   &  10.0     \\
    500    &  25.0   &  50.0    \\
    1000   &  50.0   &  100.0   \\
    \hline
    \end{tabular}
    \label{tab_uncertainty}
\end{table}

Current mm/submm arrays in the northern and southern hemispheres meet the basic thermal rms noise criteria even for faint sources. In case of Atacama Large Millimeter/submillimeter Array (ALMA), the rms thermal noise is $\sigma_{\rm ALMA}\approx 0.05\,(t/\text{10 min})^{-1/2}{\rm mJy}$ for a 10-minute on-source integration time\footnote{See \url{https://almascience.eso.org/proposing/sensitivity-calculator} for details}. Similarly, for the Northern Extended Millimeter Array (NOEMA), we have $\sigma_{\rm NOEMA}\approx 0.17(t/\text{10 min})^{-1/2}\,{\rm mJy}$ \footnote{See \url{https://www.iram.fr/GENERAL/calls/w24/NOEMACapabilities.pdf} for details.}. Another relevant facility is the Submillimeter array (SMA) with $\sigma_{\rm SMA}\approx 1.1\,(t/\text{10 min})^{-1/2}{\rm mJy}$\footnote{See \url{https://lweb.cfa.harvard.edu/sma/memos/165.pdf} for details.}. From these rms uncertainties the minimal source flux can be estimated so that the 10-minute on-source integration would result in the 5$\sigma$ detection of the 50\%-deep eclipse: $F_{\rm \nu, min}\approx q\sigma/\delta$, which yields $\sim 0.5\,{\rm mJy}$, $1.7\,{\rm mJy}$, and $\sim 11\,{\rm mJy}$ for ALMA, NOEMA, and SMA facilities. Hence for the sources that are relatively millimeter-bright in the range $F_{\nu}\sim 100-1000\,{\rm mJy}$ or more the eclipse detection is not thermal-noise limited. The detection of the nuclear occultation by a star will instead depend primarily on cadence, relative flux calibration, separation of compact and extended emission (contamination), and the potential smearing of the eclipse signal by intrinsic variability. The latter can be mitigated by focusing on deep enough eclipses $\delta\gtrsim 0.3$, i.e. those exceeding by at least a factor of two the intrinsic variability amplitude. However, the trade-off for $\delta\gtrsim 0.3$ is that $R_{\star}=\sqrt{\delta}R_{\rm c}\gtrsim \sqrt{0.3}R_{\rm c}\sim 1095\,R_{\odot}$ ($z=0.001$ and $\nu=230$ GHz) or $R_{\star}\gtrsim 500\,R_{\odot}$ for the jet base and ring geometries, respectively, hence these events are observationally cleaner but statistically rarer.

\begin{figure}
    \centering
    \includegraphics[width=\columnwidth]{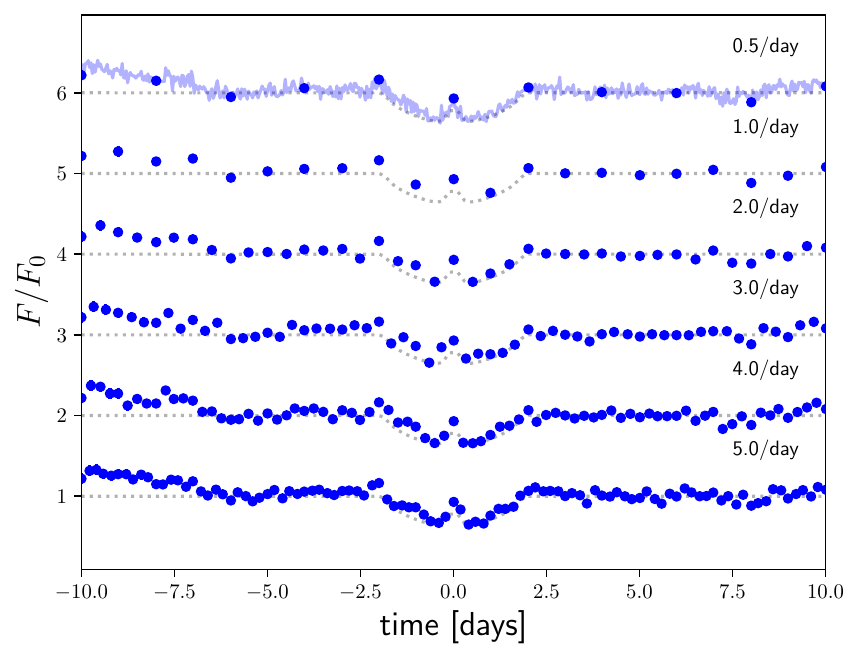}
    \caption{Sampling of the eclipse of the ring with the mean radius $R_{\rm r}=5.2\,r_{\rm g}$ and the width of $2\,r_{\rm g}$ by the star with $R_{\star}=500\,R_{\odot}$ for different cadences. From the top to the bottom the cadence increases from 0.5 visit per day to 5 visits per day. We see that for the case of 0.5/day there is no apparent eclipse in the discrete time series, hence the cadence of 1 visit per day is necessary to detect it. The eclipse substructure -- the central peak -- is recovered for the cadence of 2 visits/day and higher. The mean reference level is arbitrary. The dotted line depicts the clean transit curve of the eclipse while the variable light curve at the top shows the total eclipse light curve including intrinsic variability and measurement uncertainties.}
    \label{fig_cadence}
\end{figure}

Concerning the cadence, the observations should repeat several times during the eclipse duration, which is $\sim 4$ days for the star of $R_{\star}=500\,R_{\odot}$ crossing the ring with the mean radius of $5.2\,r_{\rm g}$ and the width of $2\,r_{\rm g}$, see Fig.~\ref{fig_light_curve_ring}. For this case, we test the eclipse detection, including its substructure, for several cadences, including the intrinsic variability and the measurement uncertainty, see Fig.~\ref{fig_cadence}. While one visit of the source per two days is not enough to clearly detect the occultation, one visit per day is sufficient to detect the basic eclipse event. With two visits per day and more the substructure -- the central peak during the eclipse -- becomes more and more apparent.

\section{Conclusions} \label{sec:conclusions}

In this work we studied conditions for the occultations of galactic nuclei in the radio domain, especially with the focus on potential stellar transits within the Nuclear Stellar Cluster (NSC). We found that deep enough eclipses are possible for nearby sources ($z\lesssim 0.001$) in the millimeter domain ($\nu\gtrsim 86\,{\rm GHz}$). For the eclipse to be deep enough (relative flux decrease by at least a few percent), the star needs to be evolved with the radius of at least a few $100\,R_{\odot}$.

A nominal studied model case was for the red supergiant/asymptotic giant-branch star with $R_{\star}=1000\,R_{\odot}$ orbiting at $r_{\star}=0.01\,{\rm pc}$ around the SMBH with $M_{\bullet}=5\times 10^7\,M_{\odot}$. For the host galaxy at $z=0.001$ and the observing frequency of $\nu=230\,{\rm GHz}$, the eclipse depth reaches $\sim 25\%$ for the jet base and it can lead to the complete eclipse of the compact millimeter ring emission. Its recurrence timescale is $\sim 13.2$ years and the total eclipse duration is $\sim 10.4$ days. For smaller frequencies, the eclipse becomes shallower and longer while for higher frequencies the occultation becomes deeper and shorter. We also studied the effect of eccentricity and inclination of the stellar orbit on the eclipse profile. For eccentric orbits, the eclipse is shorter (i.e. narrower) with respect to the circular case. For higher inclinations of the star with respect to the line of sight, the eclipse becomes shallower.

Furthermore, we included the intrinsic nuclear variability driven by the damped random walk. In addition, measurement uncertainties and a fixed monitoring cadence were considered while calculating synthetic light curves. For the $\sim 15\%$ intrinsic variability amplitude, eclipses comparable to the stochastic variations cannot be recovered; however, those that typically exceed the intrinsic variability by a factor of two ($\Delta F_{\nu}/F_{\nu}\sim 0.3$) can be detected with well-calibrated millimeter facilities with the observing cadence of at least once per day.

Moreover, we derived simple formulae for the stellar distance and its radius based on the detection of a single jet base/ring eclipse. For periodic recurring eclipses, it is possible to estimate the SMBH mass with some limitations. This relation depends on the basic observables for repeating eclipses, such as the recurrence timescale $\tau_{\rm rec}$ and the eclipse duration $\tau_{\rm dur}$, the eclipse relative depth $\Delta F_{\nu}/F_{\nu}$, the angular radio core size $\theta_{\rm c}$, and the host angular-diameter distance. On the other hand, in case the SMBH mass is tightly constrained by other methods, the equation can be used to verify whether the eclipse is caused by a stellar body bound to the SMBH or by some other, more extended object.  

We also analyzed the likelihood of the galactic nuclei eclipses by evolved stars and the instantaneous number of occulting stars is typically less than unity unless the NSC has a very steep cuspy profile with the power-law index close to $\gamma\sim 2.75$. The mean number of occultations for the monitoring period of 50 years is $\sim 10^{-4}$, which gives a rough estimate that we need to monitor $\gtrsim 10^4$ galaxies in that time frame to detect at least one eclipse, although chance alignments with nuclear disk-like stellar populations can enhance the probability of an eclipse by two to three orders of magnitude. Current and future observations in the millimeter domain, such as with the ALMA, NOEMA, SMA, the Event Horizon Telescope (EHT) and next-generation EHT \citep{2023Galax..11...61J} have a chance to detect a nuclear eclipse by a star serendipitously for the daily monitoring cadence of millimeter-bright nuclei.

\begin{acknowledgments}
The author is grateful to an anonymous referee who provided useful and constructive comments that significantly improved the manuscript.
MZ acknowledges the financial support of the GA\v{C}R Junior Star grant no. GM24-10599M (``Stars in galactic nuclei: interrelation with massive black holes"). MZ is grateful to J\'an Budaj for discussion and useful comments. MZ also appreciates the hospitality of the Astronomical Institute of the Slovak Academy of Sciences in Tatransk\'{a} Lomnica, where the idea of radio core eclipses due to orbiting evolved stars was first presented during the Tatra Astro Summit 2025. 
\end{acknowledgments}

%



\appendix

\section{Modeling stochastic AGN variability and measurement uncertainties}
\label{sec_appendix_drw}

AGN emission is stochastically variable across all wavelengths, including the radio and mm domain. Typical variability timescales reach $\sim$ hours to weeks and the fractional variability amplitude is $\sim 10-15\%$ \citep{2023ApJ...951...93C,2024A&A...684A.133A}. The correlated stochastic variability across frequencies appears to be a function of the SMBH mass, relative accretion rate as well as the observing frequency -- at higher frequencies when we penetrate deeper layers within the jet and eventually reach the accretion flow in the mm/submm domain, the variability timescale is decreasing while the fractional rms amplitude is expected to mildly increase.

Following observational results in the mm/submm domain \citep{2023ApJ...951...93C}, we model the stochastic intrinsic variability using the damped random walk \citep[DRW;][]{2009ApJ...698..895K,2017A&A...597A.128K} prescribed mathematically by the Ornstein-Uhlenbeck process \citep[OU;][]{1930PhRv...36..823U}. For the normalized flux density we have,
\begin{equation}
    F_{\rm int}(t)=\exp{\left[X(t)-\frac{\sigma_{X}^2}{2}\right]}\,,
    \label{eq_intrinsic_flux}
\end{equation}
where $X(t)$ is the OU process and equation~\eqref{eq_intrinsic_flux} ensures that the flux remains positive with the mean value of unity. At the same time $\log{F_{\rm int}}\propto X$, which is consistent with the findings that the variability of low-luminosity nuclei, including the Galactic center, follows the lognormal behaviour \citep{2018ApJ...863...15W}. The OU follows the differential equation,
\begin{equation}
    \mathrm{d}X(t)=-\frac{X(t)}{\tau}\mathrm{d}t+\sqrt{\frac{2\sigma_{X}^2}{\tau}}\mathrm{d}W(t)\,,    
\end{equation}
where $\tau$ is the DRW characteristic timescale, $\sigma_{\rm X}$ is the DRW variability amplitude, and $\mathrm{d}W$ is the increment of the Wiener process, $\mathrm{d}W=\sqrt{\mathrm{d}t}G$, where $G=\mathcal{N}(0,1)$ is the Gaussian distribution with the zero mean and unity variance. For the discrete sampling in time with the time-step of $\Delta t$ $X(t)$ follows the recursive relation,
\begin{equation}
    X_{i+1}=X_{i}\exp{(-\Delta t/\tau)} + \sigma_{X}\sqrt{1-\exp{(-2\Delta t/\tau)}} G_{i}\,.
\end{equation}
The DRW parameters, $\sigma_{X}$ and $\tau$, follow the power-law dependence on frequency,
\begin{align}    \sigma_{X}&=\sigma_{X0}\left(\frac{\nu}{\nu_0} \right)^{\alpha_{\sigma}}\,\notag\\
\tau&=\tau_{0}\left(\frac{\nu}{\nu_0} \right)^{\alpha_{\tau}}\,,
\end{align}
where $\nu_0=230$ GHz is the reference frequency. For the nominal DRW model of the AGN intrinsic millimeter variability, we adopt the fiducial values $\sigma_{X0}=0.15$, $\tau_0=10$ days, $\alpha_{\sigma}=0.25$, and $\alpha_{\tau}=-1.0$ that are based on the statistical analysis of the mm/submm variability of several low-luminosity AGN \citep{2015ApJ...802...69B,2023ApJ...951...93C}.

For the occultation -- eclipse normalized transmission factor, we can express it for the uniformly bright circles as,
\begin{equation}
    T(t)=1-\frac{S_{\rm block}}{\pi R_{\rm c}^2}\,,
\end{equation}
where $S_{\rm block}$ is the area of the radio core blocked by the star and $R_{\rm c}$ is the characteristic radius of the core. The total signal can be written as the multiplication of $F_{\rm int}$ and $T$,
\begin{equation}
    F_{\rm tot}(t)=F_{\rm int}(t) T(t)\,,
\end{equation}
which reflects both the intrinsic variability and the eclipse event.

Finally, the observations are conducted with a certain measurement uncertainty. The measurement noise is added to the total signal as,
\begin{equation}
    F_{\rm obs}(t)=F_{\rm tot}+\epsilon(t)\,,
\end{equation}
which is approximated by the Gaussian measurement noise, $\epsilon(t)=\mathcal{N}(0,\sigma_{\rm meas}^2F_{\rm tot}^2)$ with the variance $\sigma_{\rm meas}^2F_{\rm tot}^2$. We set $\sigma_{\rm meas}=0.05$, i.e. 5\% fractional measurement uncertainty, which is typical for well-calibrated mm/submm facilities.

\bibliography{references}{}

@ARTICLE{2005SSRv..116..523F,
       author = {{Ferrarese}, Laura and {Ford}, Holland},
        title = "{Supermassive Black Holes in Galactic Nuclei: Past, Present and Future Research}",
      journal = {\ssr},
         year = 2005,
        month = feb,
       volume = {116},
       number = {3-4},
        pages = {523-624},
          doi = {10.1007/s11214-005-3947-6},
archivePrefix = {arXiv},
       eprint = {astro-ph/0411247},
 primaryClass = {astro-ph},
       adsurl = {https://ui.adsabs.harvard.edu/abs/2005SSRv..116..523F}
}

@ARTICLE{2013ARA&A..51..511K,
       author = {{Kormendy}, John and {Ho}, Luis C.},
        title = "{Coevolution (Or Not) of Supermassive Black Holes and Host Galaxies}",
      journal = {\araa},
         year = 2013,
        month = aug,
       volume = {51},
       number = {1},
        pages = {511-653},
          doi = {10.1146/annurev-astro-082708-101811},
archivePrefix = {arXiv},
       eprint = {1304.7762},
 primaryClass = {astro-ph.CO},
       adsurl = {https://ui.adsabs.harvard.edu/abs/2013ARA&A..51..511K}
}

@ARTICLE{2014ARA&A..52..589H,
       author = {{Heckman}, Timothy M. and {Best}, Philip N.},
        title = "{The Coevolution of Galaxies and Supermassive Black Holes: Insights from Surveys of the Contemporary Universe}",
      journal = {\araa},
         year = 2014,
        month = aug,
       volume = {52},
        pages = {589-660},
          doi = {10.1146/annurev-astro-081913-035722},
archivePrefix = {arXiv},
       eprint = {1403.4620},
 primaryClass = {astro-ph.GA},
       adsurl = {https://ui.adsabs.harvard.edu/abs/2014ARA&A..52..589H}
}

@ARTICLE{2020AARv..28....4N,
       author = {{Neumayer}, Nadine and {Seth}, Anil and {B{\"o}ker}, Torsten},
        title = "{Nuclear star clusters}",
      journal = {\aapr},
         year = 2020,
        month = jul,
       volume = {28},
       number = {1},
          eid = {4},
        pages = {4},
          doi = {10.1007/s00159-020-00125-0},
archivePrefix = {arXiv},
       eprint = {2001.03626},
 primaryClass = {astro-ph.GA},
       adsurl = {https://ui.adsabs.harvard.edu/abs/2020A&ARv..28....4N}
}

@ARTICLE{1997ApJ...476..327R,
       author = {{Reid}, Mark J. and {Menten}, Karl M.},
        title = "{Radio Photospheres of Long-Period Variable Stars}",
      journal = {\apj},
         year = 1997,
        month = feb,
       volume = {476},
       number = {1},
        pages = {327-346},
          doi = {10.1086/303614},
       adsurl = {https://ui.adsabs.harvard.edu/abs/1997ApJ...476..327R}
}

@ARTICLE{2020A&A...638A..65O,
       author = {{O'Gorman}, E. and {Harper}, G.~M. and {Ohnaka}, K. and {Feeney-Johansson}, A. and {Wilkeneit-Braun}, K. and {Brown}, A. and {Guinan}, E.~F. and {Lim}, J. and {Richards}, A.~M.~S. and {Ryde}, N. and {Vlemmings}, W.~H.~T.},
        title = "{ALMA and VLA reveal the lukewarm chromospheres of the nearby red supergiants Antares and Betelgeuse}",
      journal = {\aap},
         year = 2020,
        month = jun,
       volume = {638},
          eid = {A65},
        pages = {A65},
          doi = {10.1051/0004-6361/202037756},
archivePrefix = {arXiv},
       eprint = {2006.08023},
 primaryClass = {astro-ph.SR},
       adsurl = {https://ui.adsabs.harvard.edu/abs/2020A&A...638A..65O}
}

@ARTICLE{2020SciA....6.1310J,
       author = {{Johnson}, Michael D. and {Lupsasca}, Alexandru and {Strominger}, Andrew and {Wong}, George N. and {Hadar}, Shahar and {Kapec}, Daniel and {Narayan}, Ramesh and {Chael}, Andrew and {Gammie}, Charles F. and {Galison}, Peter and {Palumbo}, Daniel C.~M. and {Doeleman}, Sheperd S. and {Blackburn}, Lindy and {Wielgus}, Maciek and {Pesce}, Dominic W. and {Farah}, Joseph R. and {Moran}, James M.},
        title = "{Universal interferometric signatures of a black hole's photon ring}",
      journal = {Science Advances},
         year = 2020,
        month = mar,
       volume = {6},
       number = {12},
        pages = {eaaz1310},
          doi = {10.1126/sciadv.aaz1310},
archivePrefix = {arXiv},
       eprint = {1907.04329},
 primaryClass = {astro-ph.IM},
       adsurl = {https://ui.adsabs.harvard.edu/abs/2020SciA....6.1310J}
}

@ARTICLE{2024A&A...684A.133A,
       author = {{Ar{\'e}valo}, P. and {Churazov}, E. and {Lira}, P. and {S{\'a}nchez-S{\'a}ez}, P. and {Bernal}, S. and {Hern{\'a}ndez-Garc{\'\i}a}, L. and {L{\'o}pez-Navas}, E. and {Patel}, P.},
        title = "{The universal power spectrum of quasars in optical wavelengths. Break timescale scales directly with both black hole mass and the accretion rate}",
      journal = {\aap},
         year = 2024,
        month = apr,
       volume = {684},
          eid = {A133},
        pages = {A133},
          doi = {10.1051/0004-6361/202347080},
archivePrefix = {arXiv},
       eprint = {2306.11099},
 primaryClass = {astro-ph.GA},
       adsurl = {https://ui.adsabs.harvard.edu/abs/2024A&A...684A.133A}
}

@ARTICLE{2023ApJ...951...93C,
       author = {{Chen}, Bo-Yan and {Bower}, Geoffrey C. and {Dexter}, Jason and {Markoff}, Sera and {Ridenour}, Anthony and {Gurwell}, Mark A. and {Rao}, Ramprasad and {Wallstr{\"o}m}, Sofia H.~J.},
        title = "{Testing the Linear Relationship between Black Hole Mass and Variability Timescale in Low-luminosity AGNs at Submillimeter Wavelengths}",
      journal = {\apj},
         year = 2023,
        month = jul,
       volume = {951},
       number = {2},
          eid = {93},
        pages = {93},
          doi = {10.3847/1538-4357/acd250},
archivePrefix = {arXiv},
       eprint = {2305.06529},
 primaryClass = {astro-ph.HE},
       adsurl = {https://ui.adsabs.harvard.edu/abs/2023ApJ...951...93C}
}

@ARTICLE{2015ApJ...802...69B,
       author = {{Bower}, Geoffrey C. and {Markoff}, Sera and {Dexter}, Jason and {Gurwell}, Mark A. and {Moran}, James M. and {Brunthaler}, Andreas and {Falcke}, Heino and {Fragile}, P. Chris and {Maitra}, Dipankar and {Marrone}, Dan and {Peck}, Alison and {Rushton}, Anthony and {Wright}, Melvyn C.~H.},
        title = "{Radio and Millimeter Monitoring of Sgr A*: Spectrum, Variability, and Constraints on the G2 Encounter}",
      journal = {\apj},
         year = 2015,
        month = mar,
       volume = {802},
       number = {1},
          eid = {69},
        pages = {69},
          doi = {10.1088/0004-637X/802/1/69},
archivePrefix = {arXiv},
       eprint = {1502.06534},
 primaryClass = {astro-ph.HE},
       adsurl = {https://ui.adsabs.harvard.edu/abs/2015ApJ...802...69B}
}

@ARTICLE{2022A&A...658A.172F,
       author = {{Fahrion}, Katja and {Leaman}, Ryan and {Lyubenova}, Mariya and {van de Ven}, Glenn},
        title = "{Disentangling the formation mechanisms of nuclear star clusters}",
      journal = {\aap},
         year = 2022,
        month = feb,
       volume = {658},
          eid = {A172},
        pages = {A172},
          doi = {10.1051/0004-6361/202039778},
archivePrefix = {arXiv},
       eprint = {2112.05610},
 primaryClass = {astro-ph.GA},
       adsurl = {https://ui.adsabs.harvard.edu/abs/2022A&A...658A.172F}
}

@ARTICLE{2020AA...634A..71G,
       author = {{Gallego-Cano}, E. and {Sch{\"o}del}, R. and {Nogueras-Lara}, F. and {Dong}, H. and {Shahzamanian}, B. and {Fritz}, T.~K. and {Gallego-Calvente}, A.~T. and {Neumayer}, N.},
        title = "{New constraints on the structure of the nuclear stellar cluster of the Milky Way from star counts and MIR imaging}",
      journal = {\aap},
         year = 2020,
        month = feb,
       volume = {634},
          eid = {A71},
        pages = {A71},
          doi = {10.1051/0004-6361/201935303},
archivePrefix = {arXiv},
       eprint = {2001.08182},
 primaryClass = {astro-ph.GA},
       adsurl = {https://ui.adsabs.harvard.edu/abs/2020A&A...634A..71G}
}

@ARTICLE{2010PASA...27..449N,
       author = {{Neumayer}, Nadine},
        title = "{The Supermassive Black Hole at the Heart of Centaurus A: Revealed by the Kinematics of Gas and Stars}",
      journal = {\pasa},
         year = 2010,
        month = oct,
       volume = {27},
       number = {4},
        pages = {449-456},
          doi = {10.1071/AS09080},
archivePrefix = {arXiv},
       eprint = {1002.0965},
 primaryClass = {astro-ph.CO},
       adsurl = {https://ui.adsabs.harvard.edu/abs/2010PASA...27..449N}
}

@ARTICLE{2020A&A...641A.102S,
       author = {{Sch{\"o}del}, R. and {Nogueras-Lara}, F. and {Gallego-Cano}, E. and {Shahzamanian}, B. and {Gallego-Calvente}, A.~T. and {Gardini}, A.},
        title = "{The Milky Way's nuclear star cluster: Old, metal-rich, and cuspy. Structure and star formation history from deep imaging}",
      journal = {\aap},
         year = 2020,
        month = sep,
       volume = {641},
          eid = {A102},
        pages = {A102},
          doi = {10.1051/0004-6361/201936688},
archivePrefix = {arXiv},
       eprint = {2007.15950},
 primaryClass = {astro-ph.GA},
       adsurl = {https://ui.adsabs.harvard.edu/abs/2020A&A...641A.102S}
}

@ARTICLE{2017ApJ...835...77M,
       author = {{Marigo}, Paola and {Girardi}, L{\'e}o and {Bressan}, Alessandro and {Rosenfield}, Philip and {Aringer}, Bernhard and {Chen}, Yang and {Dussin}, Marco and {Nanni}, Ambra and {Pastorelli}, Giada and {Rodrigues}, Tha{\'\i}se S. and {Trabucchi}, Michele and {Bladh}, Sara and {Dalcanton}, Julianne and {Groenewegen}, Martin A.~T. and {Montalb{\'a}n}, Josefina and {Wood}, Peter R.},
        title = "{A New Generation of PARSEC-COLIBRI Stellar Isochrones Including the TP-AGB Phase}",
      journal = {\apj},
         year = 2017,
        month = jan,
       volume = {835},
       number = {1},
          eid = {77},
        pages = {77},
          doi = {10.3847/1538-4357/835/1/77},
archivePrefix = {arXiv},
       eprint = {1701.08510},
 primaryClass = {astro-ph.SR},
       adsurl = {https://ui.adsabs.harvard.edu/abs/2017ApJ...835...77M}
}

@ARTICLE{2012ApJ...755..155S,
       author = {{Schartmann}, M. and {Burkert}, A. and {Alig}, C. and {Gillessen}, S. and {Genzel}, R. and {Eisenhauer}, F. and {Fritz}, T.~K.},
        title = "{Simulations of the Origin and Fate of the Galactic Center Cloud G2}",
      journal = {\apj},
         year = 2012,
        month = aug,
       volume = {755},
       number = {2},
          eid = {155},
        pages = {155},
          doi = {10.1088/0004-637X/755/2/155},
archivePrefix = {arXiv},
       eprint = {1203.6356},
 primaryClass = {astro-ph.GA},
       adsurl = {https://ui.adsabs.harvard.edu/abs/2012ApJ...755..155S}
}

@ARTICLE{2022ApJ...931...39M,
       author = {{M{\"u}ller}, Ana Laura and {Naddaf}, Mohammad-Hassan and {Zaja{\v{c}}ek}, Michal and {Czerny}, Bo{\.z}ena and {Araudo}, Anabella and {Karas}, Vladim{\'\i}r},
        title = "{Nonthermal Emission from Fall-back Clouds in the Broad-line Region of Active Galactic Nuclei}",
      journal = {\apj},
         year = 2022,
        month = may,
       volume = {931},
       number = {1},
          eid = {39},
        pages = {39},
          doi = {10.3847/1538-4357/ac660a},
archivePrefix = {arXiv},
       eprint = {2204.05361},
 primaryClass = {astro-ph.HE},
       adsurl = {https://ui.adsabs.harvard.edu/abs/2022ApJ...931...39M}
}

@ARTICLE{2002ApJ...580L.171M,
       author = {{Mandel}, Kaisey and {Agol}, Eric},
        title = "{Analytic Light Curves for Planetary Transit Searches}",
      journal = {\apjl},
         year = 2002,
        month = dec,
       volume = {580},
       number = {2},
        pages = {L171-L175},
          doi = {10.1086/345520},
archivePrefix = {arXiv},
       eprint = {astro-ph/0210099},
 primaryClass = {astro-ph},
       adsurl = {https://ui.adsabs.harvard.edu/abs/2002ApJ...580L.171M}
}

@ARTICLE{2019A&A...625L..10G,
       author = {{GRAVITY Collaboration} and {Abuter}, R. and {Amorim}, A. and {Baub{\"o}ck}, M. and {Berger}, J.~P. and {Bonnet}, H. and {Brandner}, W. and {Cl{\'e}net}, Y. and {Coud{\'e} Du Foresto}, V. and {de Zeeuw}, P.~T. and {Dexter}, J. and {Duvert}, G. and {Eckart}, A. and {Eisenhauer}, F. and {F{\"o}rster Schreiber}, N.~M. and {Garcia}, P. and {Gao}, F. and {Gendron}, E. and {Genzel}, R. and {Gerhard}, O. and {Gillessen}, S. and {Habibi}, M. and {Haubois}, X. and {Henning}, T. and {Hippler}, S. and {Horrobin}, M. and {Jim{\'e}nez-Rosales}, A. and {Jocou}, L. and {Kervella}, P. and {Lacour}, S. and {Lapeyr{\`e}re}, V. and {Le Bouquin}, J.-B. and {L{\'e}na}, P. and {Ott}, T. and {Paumard}, T. and {Perraut}, K. and {Perrin}, G. and {Pfuhl}, O. and {Rabien}, S. and {Rodriguez Coira}, G. and {Rousset}, G. and {Scheithauer}, S. and {Sternberg}, A. and {Straub}, O. and {Straubmeier}, C. and {Sturm}, E. and {Tacconi}, L.~J. and {Vincent}, F. and {von Fellenberg}, S. and {Waisberg}, I. and {Widmann}, F. and {Wieprecht}, E. and {Wiezorrek}, E. and {Woillez}, J. and {Yazici}, S.},
        title = "{A geometric distance measurement to the Galactic center black hole with 0.3\% uncertainty}",
      journal = {\aap},
         year = 2019,
        month = may,
       volume = {625},
          eid = {L10},
        pages = {L10},
          doi = {10.1051/0004-6361/201935656},
archivePrefix = {arXiv},
       eprint = {1904.05721},
 primaryClass = {astro-ph.GA},
       adsurl = {https://ui.adsabs.harvard.edu/abs/2019A&A...625L..10G}
}

@ARTICLE{2023MNRAS.519..948L,
       author = {{Leung}, Henry W. and {Bovy}, Jo and {Mackereth}, J. Ted and {Hunt}, Jason A.~S. and {Lane}, Richard R. and {Wilson}, John C.},
        title = "{A measurement of the distance to the Galactic centre using the kinematics of bar stars}",
      journal = {\mnras},
         year = 2023,
        month = feb,
       volume = {519},
       number = {1},
        pages = {948-960},
          doi = {10.1093/mnras/stac3529},
archivePrefix = {arXiv},
       eprint = {2204.12551},
 primaryClass = {astro-ph.GA},
       adsurl = {https://ui.adsabs.harvard.edu/abs/2023MNRAS.519..948L}
}

@ARTICLE{2019ApJ...875L...1E,
       author = {{Event Horizon Telescope Collaboration} and {Akiyama}, Kazunori and {Alberdi}, Antxon and {Alef}, Walter and {Asada}, Keiichi and {Azulay}, Rebecca and {Baczko}, Anne-Kathrin and {Ball}, David and {Balokovi{\'c}}, Mislav and {Barrett}, John and {Bintley}, Dan and {Blackburn}, Lindy and {Boland}, Wilfred and {Bouman}, Katherine L. and {Bower}, Geoffrey C. and {Bremer}, Michael and {Brinkerink}, Christiaan D. and {Brissenden}, Roger and {Britzen}, Silke and {Broderick}, Avery E. and {Broguiere}, Dominique and {Bronzwaer}, Thomas and {Byun}, Do-Young and {Carlstrom}, John E. and {Chael}, Andrew and {Chan}, Chi-kwan and {Chatterjee}, Shami and {Chatterjee}, Koushik and {Chen}, Ming-Tang and {Chen}, Yongjun and {Cho}, Ilje and {Christian}, Pierre and {Conway}, John E. and {Cordes}, James M. and {Crew}, Geoffrey B. and {Cui}, Yuzhu and {Davelaar}, Jordy and {De Laurentis}, Mariafelicia and {Deane}, Roger and {Dempsey}, Jessica and {Desvignes}, Gregory and {Dexter}, Jason and {Doeleman}, Sheperd S. and {Eatough}, Ralph P. and {Falcke}, Heino and {Fish}, Vincent L. and {Fomalont}, Ed and {Fraga-Encinas}, Raquel and {Freeman}, William T. and {Friberg}, Per and {Fromm}, Christian M. and {G{\'o}mez}, Jos{\'e} L. and {Galison}, Peter and {Gammie}, Charles F. and {Garc{\'\i}a}, Roberto and {Gentaz}, Olivier and {Georgiev}, Boris and {Goddi}, Ciriaco and {Gold}, Roman and {Gu}, Minfeng and {Gurwell}, Mark and {Hada}, Kazuhiro and {Hecht}, Michael H. and {Hesper}, Ronald and {Ho}, Luis C. and {Ho}, Paul and {Honma}, Mareki and {Huang}, Chih-Wei L. and {Huang}, Lei and {Hughes}, David H. and {Ikeda}, Shiro and {Inoue}, Makoto and {Issaoun}, Sara and {James}, David J. and {Jannuzi}, Buell T. and {Janssen}, Michael and {Jeter}, Britton and {Jiang}, Wu and {Johnson}, Michael D. and {Jorstad}, Svetlana and {Jung}, Taehyun and {Karami}, Mansour and {Karuppusamy}, Ramesh and {Kawashima}, Tomohisa and {Keating}, Garrett K. and {Kettenis}, Mark and {Kim}, Jae-Young and {Kim}, Junhan and {Kim}, Jongsoo and {Kino}, Motoki and {Koay}, Jun Yi and {Koch}, Patrick M. and {Koyama}, Shoko and {Kramer}, Michael and {Kramer}, Carsten and {Krichbaum}, Thomas P. and {Kuo}, Cheng-Yu and {Lauer}, Tod R. and {Lee}, Sang-Sung and {Li}, Yan-Rong and {Li}, Zhiyuan and {Lindqvist}, Michael and {Liu}, Kuo and {Liuzzo}, Elisabetta and {Lo}, Wen-Ping and {Lobanov}, Andrei P. and {Loinard}, Laurent and {Lonsdale}, Colin and {Lu}, Ru-Sen and {MacDonald}, Nicholas R. and {Mao}, Jirong and {Markoff}, Sera and {Marrone}, Daniel P. and {Marscher}, Alan P. and {Mart{\'\i}-Vidal}, Iv{\'a}n and {Matsushita}, Satoki and {Matthews}, Lynn D. and {Medeiros}, Lia and {Menten}, Karl M. and {Mizuno}, Yosuke and {Mizuno}, Izumi and {Moran}, James M. and {Moriyama}, Kotaro and {Moscibrodzka}, Monika and {M{\"u}ller}, Cornelia and {Nagai}, Hiroshi and {Nagar}, Neil M. and {Nakamura}, Masanori and {Narayan}, Ramesh and {Narayanan}, Gopal and {Natarajan}, Iniyan and {Neri}, Roberto and {Ni}, Chunchong and {Noutsos}, Aristeidis and {Okino}, Hiroki and {Olivares}, H{\'e}ctor and {Ortiz-Le{\'o}n}, Gisela N. and {Oyama}, Tomoaki and {{\"O}zel}, Feryal and {Palumbo}, Daniel C.~M. and {Patel}, Nimesh and {Pen}, Ue-Li and {Pesce}, Dominic W. and {Pi{\'e}tu}, Vincent and {Plambeck}, Richard and {PopStefanija}, Aleksandar and {Porth}, Oliver and {Prather}, Ben and {Preciado-L{\'o}pez}, Jorge A. and {Psaltis}, Dimitrios and {Pu}, Hung-Yi and {Ramakrishnan}, Venkatessh and {Rao}, Ramprasad and {Rawlings}, Mark G. and {Raymond}, Alexander W. and {Rezzolla}, Luciano and {Ripperda}, Bart and {Roelofs}, Freek and {Rogers}, Alan and {Ros}, Eduardo and {Rose}, Mel and {Roshanineshat}, Arash and {Rottmann}, Helge and {Roy}, Alan L. and {Ruszczyk}, Chet and {Ryan}, Benjamin R. and {Rygl}, Kazi L.~J. and {S{\'a}nchez}, Salvador and {S{\'a}nchez-Arguelles}, David and {Sasada}, Mahito and {Savolainen}, Tuomas and {Schloerb}, F. Peter and {Schuster}, Karl-Friedrich and {Shao}, Lijing and {Shen}, Zhiqiang and {Small}, Des and {Sohn}, Bong Won and {SooHoo}, Jason and {Tazaki}, Fumie and {Tiede}, Paul and {Tilanus}, Remo P.~J. and {Titus}, Michael and {Toma}, Kenji and {Torne}, Pablo and {Trent}, Tyler and {Trippe}, Sascha and {Tsuda}, Shuichiro and {van Bemmel}, Ilse and {van Langevelde}, Huib Jan and {van Rossum}, Daniel R. and {Wagner}, Jan and {Wardle}, John and {Weintroub}, Jonathan and {Wex}, Norbert and {Wharton}, Robert and {Wielgus}, Maciek and {Wong}, George N. and {Wu}, Qingwen and {Young}, Ken and {Young}, Andr{\'e}},
        title = "{First M87 Event Horizon Telescope Results. I. The Shadow of the Supermassive Black Hole}",
      journal = {\apjl},
         year = 2019,
        month = apr,
       volume = {875},
       number = {1},
          eid = {L1},
        pages = {L1},
          doi = {10.3847/2041-8213/ab0ec7},
archivePrefix = {arXiv},
       eprint = {1906.11238},
 primaryClass = {astro-ph.GA},
       adsurl = {https://ui.adsabs.harvard.edu/abs/2019ApJ...875L...1E}
}

@ARTICLE{2022ApJ...930L..12E,
       author = {{Event Horizon Telescope Collaboration} and {Akiyama}, Kazunori and {Alberdi}, Antxon and {Alef}, Walter and {Algaba}, Juan Carlos and {Anantua}, Richard and {Asada}, Keiichi and {Azulay}, Rebecca and {Bach}, Uwe and {Baczko}, Anne-Kathrin and {Ball}, David and {Balokovi{\'c}}, Mislav and {Barrett}, John and {Baub{\"o}ck}, Michi and {Benson}, Bradford A. and {Bintley}, Dan and {Blackburn}, Lindy and {Blundell}, Raymond and {Bouman}, Katherine L. and {Bower}, Geoffrey C. and {Boyce}, Hope and {Bremer}, Michael and {Brinkerink}, Christiaan D. and {Brissenden}, Roger and {Britzen}, Silke and {Broderick}, Avery E. and {Broguiere}, Dominique and {Bronzwaer}, Thomas and {Bustamante}, Sandra and {Byun}, Do-Young and {Carlstrom}, John E. and {Ceccobello}, Chiara and {Chael}, Andrew and {Chan}, Chi-kwan and {Chatterjee}, Koushik and {Chatterjee}, Shami and {Chen}, Ming-Tang and {Chen}, Yongjun and {Cheng}, Xiaopeng and {Cho}, Ilje and {Christian}, Pierre and {Conroy}, Nicholas S. and {Conway}, John E. and {Cordes}, James M. and {Crawford}, Thomas M. and {Crew}, Geoffrey B. and {Cruz-Osorio}, Alejandro and {Cui}, Yuzhu and {Davelaar}, Jordy and {De Laurentis}, Mariafelicia and {Deane}, Roger and {Dempsey}, Jessica and {Desvignes}, Gregory and {Dexter}, Jason and {Dhruv}, Vedant and {Doeleman}, Sheperd S. and {Dougal}, Sean and {Dzib}, Sergio A. and {Eatough}, Ralph P. and {Emami}, Razieh and {Falcke}, Heino and {Farah}, Joseph and {Fish}, Vincent L. and {Fomalont}, Ed and {Ford}, H. Alyson and {Fraga-Encinas}, Raquel and {Freeman}, William T. and {Friberg}, Per and {Fromm}, Christian M. and {Fuentes}, Antonio and {Galison}, Peter and {Gammie}, Charles F. and {Garc{\'\i}a}, Roberto and {Gentaz}, Olivier and {Georgiev}, Boris and {Goddi}, Ciriaco and {Gold}, Roman and {G{\'o}mez-Ruiz}, Arturo I. and {G{\'o}mez}, Jos{\'e} L. and {Gu}, Minfeng and {Gurwell}, Mark and {Hada}, Kazuhiro and {Haggard}, Daryl and {Haworth}, Kari and {Hecht}, Michael H. and {Hesper}, Ronald and {Heumann}, Dirk and {Ho}, Luis C. and {Ho}, Paul and {Honma}, Mareki and {Huang}, Chih-Wei L. and {Huang}, Lei and {Hughes}, David H. and {Ikeda}, Shiro and {Impellizzeri}, C.~M. Violette and {Inoue}, Makoto and {Issaoun}, Sara and {James}, David J. and {Jannuzi}, Buell T. and {Janssen}, Michael and {Jeter}, Britton and {Jiang}, Wu and {Jim{\'e}nez-Rosales}, Alejandra and {Johnson}, Michael D. and {Jorstad}, Svetlana and {Joshi}, Abhishek V. and {Jung}, Taehyun and {Karami}, Mansour and {Karuppusamy}, Ramesh and {Kawashima}, Tomohisa and {Keating}, Garrett K. and {Kettenis}, Mark and {Kim}, Dong-Jin and {Kim}, Jae-Young and {Kim}, Jongsoo and {Kim}, Junhan and {Kino}, Motoki and {Koay}, Jun Yi and {Kocherlakota}, Prashant and {Kofuji}, Yutaro and {Koch}, Patrick M. and {Koyama}, Shoko and {Kramer}, Carsten and {Kramer}, Michael and {Krichbaum}, Thomas P. and {Kuo}, Cheng-Yu and {La Bella}, Noemi and {Lauer}, Tod R. and {Lee}, Daeyoung and {Lee}, Sang-Sung and {Leung}, Po Kin and {Levis}, Aviad and {Li}, Zhiyuan and {Lico}, Rocco and {Lindahl}, Greg and {Lindqvist}, Michael and {Lisakov}, Mikhail and {Liu}, Jun and {Liu}, Kuo and {Liuzzo}, Elisabetta and {Lo}, Wen-Ping and {Lobanov}, Andrei P. and {Loinard}, Laurent and {Lonsdale}, Colin J. and {Lu}, Ru-Sen and {Mao}, Jirong and {Marchili}, Nicola and {Markoff}, Sera and {Marrone}, Daniel P. and {Marscher}, Alan P. and {Mart{\'\i}-Vidal}, Iv{\'a}n and {Matsushita}, Satoki and {Matthews}, Lynn D. and {Medeiros}, Lia and {Menten}, Karl M. and {Michalik}, Daniel and {Mizuno}, Izumi and {Mizuno}, Yosuke and {Moran}, James M. and {Moriyama}, Kotaro and {Moscibrodzka}, Monika and {M{\"u}ller}, Cornelia and {Mus}, Alejandro and {Musoke}, Gibwa and {Myserlis}, Ioannis and {Nadolski}, Andrew and {Nagai}, Hiroshi and {Nagar}, Neil M. and {Nakamura}, Masanori and {Narayan}, Ramesh and {Narayanan}, Gopal and {Natarajan}, Iniyan and {Nathanail}, Antonios and {Fuentes}, Santiago Navarro and {Neilsen}, Joey and {Neri}, Roberto and {Ni}, Chunchong and {Noutsos}, Aristeidis and {Nowak}, Michael A. and {Oh}, Junghwan and {Okino}, Hiroki and {Olivares}, H{\'e}ctor and {Ortiz-Le{\'o}n}, Gisela N. and {Oyama}, Tomoaki and {{\"O}zel}, Feryal and {Palumbo}, Daniel C.~M. and {Paraschos}, Georgios Filippos and {Park}, Jongho and {Parsons}, Harriet and {Patel}, Nimesh and {Pen}, Ue-Li and {Pesce}, Dominic W. and {Pi{\'e}tu}, Vincent and {Plambeck}, Richard and {PopStefanija}, Aleksandar and {Porth}, Oliver and {P{\"o}tzl}, Felix M. and {Prather}, Ben and {Preciado-L{\'o}pez}, Jorge A. and {Psaltis}, Dimitrios},
        title = "{First Sagittarius A* Event Horizon Telescope Results. I. The Shadow of the Supermassive Black Hole in the Center of the Milky Way}",
      journal = {\apjl},
         year = 2022,
        month = may,
       volume = {930},
       number = {2},
          eid = {L12},
        pages = {L12},
          doi = {10.3847/2041-8213/ac6674},
       adsurl = {https://ui.adsabs.harvard.edu/abs/2022ApJ...930L..12E}
}

@ARTICLE{2008ApJ...678..780G,
       author = {{Gualandris}, Alessia and {Merritt}, David},
        title = "{Ejection of Supermassive Black Holes from Galaxy Cores}",
      journal = {\apj},
         year = 2008,
        month = may,
       volume = {678},
       number = {2},
        pages = {780-797},
          doi = {10.1086/586877},
archivePrefix = {arXiv},
       eprint = {0708.0771},
 primaryClass = {astro-ph},
       adsurl = {https://ui.adsabs.harvard.edu/abs/2008ApJ...678..780G}
}

@ARTICLE{1998A&A...330...79L,
       author = {{Lobanov}, A.~P.},
        title = "{Ultracompact jets in active galactic nuclei}",
      journal = {\aap},
         year = 1998,
        month = feb,
       volume = {330},
        pages = {79-89},
          doi = {10.48550/arXiv.astro-ph/9712132},
archivePrefix = {arXiv},
       eprint = {astro-ph/9712132},
 primaryClass = {astro-ph},
       adsurl = {https://ui.adsabs.harvard.edu/abs/1998A&A...330...79L}
}

@ARTICLE{2010ApJ...717L...6B,
       author = {{Batcheldor}, D. and {Robinson}, A. and {Axon}, D.~J. and {Perlman}, E.~S. and {Merritt}, D.},
        title = "{A Displaced Supermassive Black Hole in M87}",
      journal = {\apjl},
         year = 2010,
        month = jul,
       volume = {717},
       number = {1},
        pages = {L6-L10},
          doi = {10.1088/2041-8205/717/1/L6},
archivePrefix = {arXiv},
       eprint = {1005.2173},
 primaryClass = {astro-ph.CO},
       adsurl = {https://ui.adsabs.harvard.edu/abs/2010ApJ...717L...6B}
}

@ARTICLE{2014ApJ...795..146L,
       author = {{Lena}, D. and {Robinson}, A. and {Marconi}, A. and {Axon}, D.~J. and {Capetti}, A. and {Merritt}, D. and {Batcheldor}, D.},
        title = "{Recoiling Supermassive Black Holes: A Search in the Nearby Universe}",
      journal = {\apj},
         year = 2014,
        month = nov,
       volume = {795},
       number = {2},
          eid = {146},
        pages = {146},
          doi = {10.1088/0004-637X/795/2/146},
archivePrefix = {arXiv},
       eprint = {1409.3976},
 primaryClass = {astro-ph.GA},
       adsurl = {https://ui.adsabs.harvard.edu/abs/2014ApJ...795..146L}
}

@ARTICLE{2009ApJ...699.1690M,
       author = {{Merritt}, David and {Schnittman}, Jeremy D. and {Komossa}, S.},
        title = "{Hypercompact Stellar Systems Around Recoiling Supermassive Black Holes}",
      journal = {\apj},
         year = 2009,
        month = jul,
       volume = {699},
       number = {2},
        pages = {1690-1710},
          doi = {10.1088/0004-637X/699/2/1690},
archivePrefix = {arXiv},
       eprint = {0809.5046},
 primaryClass = {astro-ph},
       adsurl = {https://ui.adsabs.harvard.edu/abs/2009ApJ...699.1690M}
}

@ARTICLE{2021NatAs...5.1017J,
       author = {{Janssen}, Michael and {Falcke}, Heino and {Kadler}, Matthias and {Ros}, Eduardo and {Wielgus}, Maciek and {Akiyama}, Kazunori and {Balokovi{\'c}}, Mislav and {Blackburn}, Lindy and {Bouman}, Katherine L. and {Chael}, Andrew and {Chan}, Chi-kwan and {Chatterjee}, Koushik and {Davelaar}, Jordy and {Edwards}, Philip G. and {Fromm}, Christian M. and {G{\'o}mez}, Jos{\'e} L. and {Goddi}, Ciriaco and {Issaoun}, Sara and {Johnson}, Michael D. and {Kim}, Junhan and {Koay}, Jun Yi and {Krichbaum}, Thomas P. and {Liu}, Jun and {Liuzzo}, Elisabetta and {Markoff}, Sera and {Markowitz}, Alex and {Marrone}, Daniel P. and {Mizuno}, Yosuke and {M{\"u}ller}, Cornelia and {Ni}, Chunchong and {Pesce}, Dominic W. and {Ramakrishnan}, Venkatessh and {Roelofs}, Freek and {Rygl}, Kazi L.~J. and {van Bemmel}, Ilse and {Event Horizon Telescope Collaboration} and {Alberdi}, Antxon and {Alef}, Walter and {Algaba}, Juan Carlos and {Anantua}, Richard and {Asada}, Keiichi and {Azulay}, Rebecca and {Baczko}, Anne-Kathrin and {Ball}, David and {Ball}, David and {Barrett}, John and {Benson}, Bradford A. and {Bintley}, Dan and {Bintley}, Dan and {Blundell}, Raymond and {Boland}, Wilfred and {Boland}, Wilfred and {Bower}, Geoffrey C. and {Boyce}, Hope and {Bremer}, Michael and {Brinkerink}, Christiaan D. and {Brissenden}, Roger and {Britzen}, Silke and {Broderick}, Avery E. and {Broguiere}, Dominique and {Bronzwaer}, Thomas and {Byun}, Do-Young and {Carlstrom}, John E. and {Chatterjee}, Shami and {Chen}, Ming-Tang and {Chen}, Yongjun and {Chesler}, Paul M. and {Cho}, Ilje and {Christian}, Pierre and {Conway}, John E. and {Cordes}, James M. and {Crawford}, Thomas M. and {Crew}, Geoffrey B. and {Cruz-Osorio}, Alejandro and {Cui}, Yuzhu and {Cui}, Yuzhu and {De Laurentis}, Mariafelicia and {Deane}, Roger and {Dempsey}, Jessica and {Desvignes}, Gregory and {Dexter}, Jason and {Doeleman}, Sheperd S. and {Eatough}, Ralph P. and {Farah}, Joseph and {Farah}, Joseph and {Fish}, Vincent L. and {Fomalont}, Ed and {Ford}, H. Alyson and {Fraga-Encinas}, Raquel and {Friberg}, Per and {Friberg}, Per and {Fuentes}, Antonio and {Galison}, Peter and {Gammie}, Charles F. and {Garc{\'\i}a}, Roberto and {Gelles}, Zachary and {Gentaz}, Olivier and {Georgiev}, Boris and {Georgiev}, Boris and {Gold}, Roman and {Gold}, Roman and {G{\'o}mez-Ruiz}, Arturo I. and {Gu}, Minfeng and {Gurwell}, Mark and {Hada}, Kazuhiro and {Haggard}, Daryl and {Hecht}, Michael H. and {Hesper}, Ronald and {Himwich}, Elizabeth and {Ho}, Luis C. and {Ho}, Paul and {Honma}, Mareki and {Huang}, Chih-Wei L. and {Huang}, Lei and {Hughes}, David H. and {Ikeda}, Shiro and {Inoue}, Makoto and {Inoue}, Makoto and {James}, David J. and {Jannuzi}, Buell T. and {Jeter}, Britton and {Jiang}, Wu and {Jimenez-Rosales}, Alejandra and {Jorstad}, Svetlana and {Jung}, Taehyun and {Karami}, Mansour and {Karuppusamy}, Ramesh and {Kawashima}, Tomohisa and {Keating}, Garrett K. and {Kettenis}, Mark and {Kim}, Dong-Jin and {Kim}, Jae-Young and {Kim}, Jongsoo and {Kino}, Motoki and {Kofuji}, Yutaro and {Koyama}, Shoko and {Kramer}, Michael and {Kramer}, Carsten and {Kuo}, Cheng-Yu and {Lauer}, Tod R. and {Lee}, Sang-Sung and {Levis}, Aviad and {Li}, Yan-Rong and {Li}, Zhiyuan and {Lindqvist}, Michael and {Lico}, Rocco and {Lindahl}, Greg and {Liu}, Kuo and {Lo}, Wen-Ping and {Lobanov}, Andrei P. and {Loinard}, Laurent and {Lonsdale}, Colin and {Lu}, Ru-Sen and {MacDonald}, Nicholas R. and {Mao}, Jirong and {Marchili}, Nicola and {Marscher}, Alan P. and {Mart{\'\i}-Vidal}, Iv{\'a}n and {Matsushita}, Satoki and {Matthews}, Lynn D. and {Medeiros}, Lia and {Menten}, Karl M. and {Mizuno}, Izumi and {Moran}, James M. and {Moriyama}, Kotaro and {Moscibrodzka}, Monika and {Moscibrodzka}, Monika and {Musoke}, Gibwa and {Mej{\'\i}as}, Alejandro Mus and {Nagai}, Hiroshi and {Nagar}, Neil M. and {Nakamura}, Masanori and {Narayan}, Ramesh and {Narayanan}, Gopal and {Natarajan}, Iniyan and {Nathanail}, Antonios and {Neilsen}, Joey and {Neri}, Roberto and {Noutsos}, Aristeidis and {Nowak}, Michael A. and {Okino}, Hiroki and {Olivares}, H{\'e}ctor and {Ortiz-Le{\'o}n}, Gisela N. and {Oyama}, Tomoaki and {{\"O}zel}, Feryal and {Palumbo}, Daniel C.~M. and {Park}, Jongho and {Patel}, Nimesh and {Pen}, Ue-Li and {Pi{\'e}tu}, Vincent and {Plambeck}, Richard and {PopStefanija}, Aleksandar and {Porth}, Oliver and {P{\"o}tzl}, Felix M. and {Prather}, Ben and {Preciado-L{\'o}pez}, Jorge A. and {Psaltis}, Dimitrios and {Pu}, Hung-Yi and {Pu}, Hung-Yi and {Rao}, Ramprasad},
        title = "{Event Horizon Telescope observations of the jet launching and collimation in Centaurus A}",
      journal = {Nature Astronomy},
         year = 2021,
        month = jul,
       volume = {5},
        pages = {1017-1028},
          doi = {10.1038/s41550-021-01417-w},
archivePrefix = {arXiv},
       eprint = {2111.03356},
 primaryClass = {astro-ph.GA},
       adsurl = {https://ui.adsabs.harvard.edu/abs/2021NatAs...5.1017J}
}

@ARTICLE{2021ApJ...910L..14G,
       author = {{Goddi}, Ciriaco and {Mart{\'\i}-Vidal}, Iv{\'a}n and {Messias}, Hugo and {Bower}, Geoffrey C. and {Broderick}, Avery E. and {Dexter}, Jason and {Marrone}, Daniel P. and {Moscibrodzka}, Monika and {Nagai}, Hiroshi and {Algaba}, Juan Carlos and {Asada}, Keiichi and {Crew}, Geoffrey B. and {G{\'o}mez}, Jos{\'e} L. and {Impellizzeri}, C.~M. Violette and {Janssen}, Michael and {Kadler}, Matthias and {Krichbaum}, Thomas P. and {Lico}, Rocco and {Matthews}, Lynn D. and {Nathanail}, Antonios and {Ricarte}, Angelo and {Ros}, Eduardo and {Younsi}, Ziri and {Akiyama}, Kazunori and {Alberdi}, Antxon and {Alef}, Walter and {Anantua}, Richard and {Azulay}, Rebecca and {Baczko}, Anne-Kathrin and {Ball}, David and {Balokovi{\'c}}, Mislav and {Barrett}, John and {Benson}, Bradford A. and {Bintley}, Dan and {Blackburn}, Lindy and {Blundell}, Raymond and {Boland}, Wilfred and {Bouman}, Katherine L. and {Boyce}, Hope and {Bremer}, Michael and {Brinkerink}, Christiaan D. and {Brissenden}, Roger and {Britzen}, Silke and {Broguiere}, Dominique and {Bronzwaer}, Thomas and {Byun}, Do-Young and {Carlstrom}, John E. and {Chael}, Andrew and {Chan}, Chi-kwan and {Chatterjee}, Shami and {Chatterjee}, Koushik and {Chen}, Ming-Tang and {Chen}, Yongjun and {Chesler}, Paul M. and {Cho}, Ilje and {Christian}, Pierre and {Conway}, John E. and {Cordes}, James M. and {Crawford}, Thomas M. and {Cruz-Osorio}, Alejandro and {Cui}, Yuzhu and {Davelaar}, Jordy and {De Laurentis}, Mariafelicia and {Deane}, Roger and {Dempsey}, Jessica and {Desvignes}, Gregory and {Doeleman}, Sheperd S. and {Eatough}, Ralph P. and {Falcke}, Heino and {Farah}, Joseph and {Fish}, Vincent L. and {Fomalont}, Ed and {Ford}, H. Alyson and {Fraga-Encinas}, Raquel and {Freeman}, William T. and {Friberg}, Per and {Fromm}, Christian M. and {Fuentes}, Antonio and {Galison}, Peter and {Gammie}, Charles F. and {Garc{\'\i}a}, Roberto and {Gentaz}, Olivier and {Georgiev}, Boris and {Gold}, Roman and {G{\'o}mez-Ruiz}, Arturo I. and {Gu}, Minfeng and {Gurwell}, Mark and {Hada}, Kazuhiro and {Haggard}, Daryl and {Hecht}, Michael H. and {Hesper}, Ronald and {Ho}, Luis C. and {Ho}, Paul and {Honma}, Mareki and {Huang}, Chih-Wei L. and {Huang}, Lei and {Hughes}, David H. and {Inoue}, Makoto and {Issaoun}, Sara and {James}, David J. and {Jannuzi}, Buell T. and {Jeter}, Britton and {Jiang}, Wu and {Jimenez-Rosales}, Alejandra and {Johnson}, Michael D. and {Jorstad}, Svetlana and {Jung}, Taehyun and {Karami}, Mansour and {Karuppusamy}, Ramesh and {Kawashima}, Tomohisa and {Keating}, Garrett K. and {Kettenis}, Mark and {Kim}, Dong-Jin and {Kim}, Jae-Young and {Kim}, Jongsoo and {Kim}, Junhan and {Kino}, Motoki and {Koay}, Jun Yi and {Kofuji}, Yutaro and {Koch}, Patrick M. and {Koyama}, Shoko and {Kramer}, Michael and {Kramer}, Carsten and {Kuo}, Cheng-Yu and {Lauer}, Tod R. and {Lee}, Sang-Sung and {Levis}, Aviad and {Li}, Yan-Rong and {Li}, Zhiyuan and {Lindqvist}, Michael and {Lindahl}, Greg and {Liu}, Jun and {Liu}, Kuo and {Liuzzo}, Elisabetta and {Lo}, Wen-Ping and {Lobanov}, Andrei P. and {Loinard}, Laurent and {Lonsdale}, Colin and {Lu}, Ru-Sen and {MacDonald}, Nicholas R. and {Mao}, Jirong and {Marchili}, Nicola and {Markoff}, Sera and {Marscher}, Alan P. and {Matsushita}, Satoki and {Medeiros}, Lia and {Menten}, Karl M. and {Mizuno}, Izumi and {Mizuno}, Yosuke and {Moran}, James M. and {Moriyama}, Kotaro and {M{\"u}ller}, Cornelia and {Musoke}, Gibwa and {Mej{\'\i}as}, Alejandro Mus and {Nagar}, Neil M. and {Nakamura}, Masanori and {Narayan}, Ramesh and {Narayanan}, Gopal and {Natarajan}, Iniyan and {Neilsen}, Joey and {Neri}, Roberto and {Ni}, Chunchong and {Noutsos}, Aristeidis and {Nowak}, Michael A. and {Okino}, Hiroki and {Olivares}, H{\'e}ctor and {Ortiz-Le{\'o}n}, Gisela N. and {Oyama}, Tomoaki and {{\"O}zel}, Feryal and {Palumbo}, Daniel C.~M. and {Park}, Jongho and {Patel}, Nimesh and {Pen}, Ue-Li and {Pesce}, Dominic W. and {Pi{\'e}tu}, Vincent and {Plambeck}, Richard and {PopStefanija}, Aleksandar and {Porth}, Oliver and {P{\"o}tzl}, Felix M. and {Prather}, Ben and {Preciado-L{\'o}pez}, Jorge A. and {Psaltis}, Dimitrios and {Pu}, Hung-Yi and {Ramakrishnan}, Venkatessh and {Rao}, Ramprasad and {Rawlings}, Mark G. and {Raymond}, Alexander W. and {Rezzolla}, Luciano and {Ripperda}, Bart and {Roelofs}, Freek and {Rogers}, Alan and {Rose}, Mel and {Roshanineshat}, Arash and {Rottmann}, Helge and {Roy}, Alan L. and {Ruszczyk}, Chet and {Rygl}, Kazi L.~J. and {S{\'a}nchez}, Salvador and {S{\'a}nchez-Arguelles}, David and {Sasada}, Mahito},
        title = "{Polarimetric Properties of Event Horizon Telescope Targets from ALMA}",
      journal = {\apjl},
         year = 2021,
        month = mar,
       volume = {910},
       number = {1},
          eid = {L14},
        pages = {L14},
          doi = {10.3847/2041-8213/abee6a},
archivePrefix = {arXiv},
       eprint = {2105.02272},
 primaryClass = {astro-ph.GA},
       adsurl = {https://ui.adsabs.harvard.edu/abs/2021ApJ...910L..14G}
}

@ARTICLE{2025ApJ...981..126F,
       author = {{Ford}, Nicole M. and {Nowak}, Michael and {Ramakrishnan}, Venkatessh and {Haggard}, Daryl and {Dage}, Kristen and {Nair}, Dhanya G. and {Chan}, Chi-kwan},
        title = "{Tracking X-Ray Variability in Next-generation EHT Low-luminosity Active Galactic Nucleus Targets}",
      journal = {\apj},
         year = 2025,
        month = mar,
       volume = {981},
       number = {2},
          eid = {126},
        pages = {126},
          doi = {10.3847/1538-4357/adae0f},
archivePrefix = {arXiv},
       eprint = {2501.14871},
 primaryClass = {astro-ph.HE},
       adsurl = {https://ui.adsabs.harvard.edu/abs/2025ApJ...981..126F}
}

@ARTICLE{2013ApJ...765...63D,
       author = {{Doi}, Akihiro and {Kohno}, Kotaro and {Nakanishi}, Kouichiro and {Kameno}, Seiji and {Inoue}, Makoto and {Hada}, Kazuhiro and {Sorai}, Kazuo},
        title = "{Nuclear Radio Jet from a Low-luminosity Active Galactic Nucleus in NGC 4258}",
      journal = {\apj},
         year = 2013,
        month = mar,
       volume = {765},
       number = {1},
          eid = {63},
        pages = {63},
          doi = {10.1088/0004-637X/765/1/63},
archivePrefix = {arXiv},
       eprint = {1301.4757},
 primaryClass = {astro-ph.GA},
       adsurl = {https://ui.adsabs.harvard.edu/abs/2013ApJ...765...63D}
}

@ARTICLE{Vedanthamb,
       author = {{Vedantham}, H.~K. and {Readhead}, A.~C.~S. and {Hovatta}, T. and {Koopmans}, L.~V.~E. and {Pearson}, T.~J. and {Blandford}, R.~D. and {Gurwell}, M.~A. and {L{\"a}hteenm{\"a}ki}, A. and {Max-Moerbeck}, W. and {Pavlidou}, V. and {Ravi}, V. and {Reeves}, R.~A. and {Richards}, J.~L. and {Tornikoski}, M. and {Zensus}, J.~A.},
        title = "{The Peculiar Light Curve of J1415+1320: A Case Study in Extreme Scattering Events}",
      journal = {\apj},
         year = 2017,
        month = aug,
       volume = {845},
       number = {2},
          eid = {90},
        pages = {90},
          doi = {10.3847/1538-4357/aa7741},
archivePrefix = {arXiv},
       eprint = {1702.05519},
 primaryClass = {astro-ph.GA},
       adsurl = {https://ui.adsabs.harvard.edu/abs/2017ApJ...845...90V}
}

@ARTICLE{1930PhRv...36..823U,
       author = {{Uhlenbeck}, G.~E. and {Ornstein}, L.~S.},
        title = "{On the Theory of the Brownian Motion}",
      journal = {Physical Review},
         year = 1930,
        month = sep,
       volume = {36},
       number = {5},
        pages = {823-841},
          doi = {10.1103/PhysRev.36.823},
       adsurl = {https://ui.adsabs.harvard.edu/abs/1930PhRv...36..823U}
}

@ARTICLE{2009ApJ...698..895K,
       author = {{Kelly}, Brandon C. and {Bechtold}, Jill and {Siemiginowska}, Aneta},
        title = "{Are the Variations in Quasar Optical Flux Driven by Thermal Fluctuations?}",
      journal = {\apj},
         year = 2009,
        month = jun,
       volume = {698},
       number = {1},
        pages = {895-910},
          doi = {10.1088/0004-637X/698/1/895},
archivePrefix = {arXiv},
       eprint = {0903.5315},
 primaryClass = {astro-ph.CO},
       adsurl = {https://ui.adsabs.harvard.edu/abs/2009ApJ...698..895K}
}

@ARTICLE{2017A&A...597A.128K,
       author = {{Koz{\l}owski}, Szymon},
        title = "{Limitations on the recovery of the true AGN variability parameters using damped random walk modeling}",
      journal = {\aap},
         year = 2017,
        month = jan,
       volume = {597},
          eid = {A128},
        pages = {A128},
          doi = {10.1051/0004-6361/201629890},
archivePrefix = {arXiv},
       eprint = {1611.08248},
 primaryClass = {astro-ph.GA},
       adsurl = {https://ui.adsabs.harvard.edu/abs/2017A&A...597A.128K}
}

@ARTICLE{2018ApJ...863...15W,
       author = {{Witzel}, G. and {Martinez}, G. and {Hora}, J. and {Willner}, S.~P. and {Morris}, M.~R. and {Gammie}, C. and {Becklin}, E.~E. and {Ashby}, M.~L.~N. and {Baganoff}, F. and {Carey}, S. and {Do}, T. and {Fazio}, G.~G. and {Ghez}, A. and {Glaccum}, W.~J. and {Haggard}, D. and {Herrero-Illana}, R. and {Ingalls}, J. and {Narayan}, R. and {Smith}, H.~A.},
        title = "{Variability Timescale and Spectral Index of Sgr A* in the Near Infrared: Approximate Bayesian Computation Analysis of the Variability of the Closest Supermassive Black Hole}",
      journal = {\apj},
         year = 2018,
        month = aug,
       volume = {863},
       number = {1},
          eid = {15},
        pages = {15},
          doi = {10.3847/1538-4357/aace62},
archivePrefix = {arXiv},
       eprint = {1806.00479},
 primaryClass = {astro-ph.HE},
       adsurl = {https://ui.adsabs.harvard.edu/abs/2018ApJ...863...15W}
}

@ARTICLE{2021MNRAS.503.1703I,
       author = {{Ingram}, Adam and {Motta}, Sara E. and {Aigrain}, Suzanne and {Karastergiou}, Aris},
        title = "{A self-lensing binary massive black hole interpretation of quasi-periodic eruptions}",
      journal = {\mnras},
         year = 2021,
        month = may,
       volume = {503},
       number = {2},
        pages = {1703-1716},
          doi = {10.1093/mnras/stab609},
archivePrefix = {arXiv},
       eprint = {2103.00017},
 primaryClass = {astro-ph.HE},
       adsurl = {https://ui.adsabs.harvard.edu/abs/2021MNRAS.503.1703I}
}

@ARTICLE{2023Galax..11...61J,
       author = {{Johnson}, Michael D. and {Akiyama}, Kazunori and {Blackburn}, Lindy and {Bouman}, Katherine L. and {Broderick}, Avery E. and {Cardoso}, Vitor and {Fender}, Rob P. and {Fromm}, Christian M. and {Galison}, Peter and {G{\'o}mez}, Jos{\'e} L. and {Haggard}, Daryl and {Lister}, Matthew L. and {Lobanov}, Andrei P. and {Markoff}, Sera and {Narayan}, Ramesh and {Natarajan}, Priyamvada and {Nichols}, Tiffany and {Pesce}, Dominic W. and {Younsi}, Ziri and {Chael}, Andrew and {Chatterjee}, Koushik and {Chaves}, Ryan and {Doboszewski}, Juliusz and {Dodson}, Richard and {Doeleman}, Sheperd S. and {Elder}, Jamee and {Fitzpatrick}, Garret and {Haworth}, Kari and {Houston}, Janice and {Issaoun}, Sara and {Kovalev}, Yuri Y. and {Levis}, Aviad and {Lico}, Rocco and {Marcoci}, Alexandru and {Martens}, Niels C.~M. and {Nagar}, Neil M. and {Oppenheimer}, Aaron and {Palumbo}, Daniel C.~M. and {Ricarte}, Angelo and {Rioja}, Mar{\'\i}a J. and {Roelofs}, Freek and {Thresher}, Ann C. and {Tiede}, Paul and {Weintroub}, Jonathan and {Wielgus}, Maciek},
        title = "{Key Science Goals for the Next-Generation Event Horizon Telescope}",
      journal = {Galaxies},
         year = 2023,
        month = apr,
       volume = {11},
       number = {3},
          eid = {61},
        pages = {61},
          doi = {10.3390/galaxies11030061},
archivePrefix = {arXiv},
       eprint = {2304.11188},
 primaryClass = {astro-ph.HE},
       adsurl = {https://ui.adsabs.harvard.edu/abs/2023Galax..11...61J}
}

@ARTICLE{1977ApJ...216..883B,
       author = {{Bahcall}, J.~N. and {Wolf}, R.~A.},
        title = "{The star distribution around a massive black hole in a globular cluster. II. Unequal star masses.}",
      journal = {\apj},
         year = 1977,
        month = sep,
       volume = {216},
        pages = {883-907},
          doi = {10.1086/155534},
       adsurl = {https://ui.adsabs.harvard.edu/abs/1977ApJ...216..883B}
}

@ARTICLE{2022RvMP...94b0501G,
       author = {{Genzel}, Reinhard},
        title = "{Nobel Lecture: A forty-year journey$^{*}$}",
      journal = {Reviews of Modern Physics},
         year = 2022,
        month = apr,
       volume = {94},
       number = {2},
          eid = {020501},
        pages = {020501},
          doi = {10.1103/RevModPhys.94.020501},
       adsurl = {https://ui.adsabs.harvard.edu/abs/2022RvMP...94b0501G}
}

@ARTICLE{2013ApJ...762...35B,
       author = {{B{\'e}ky}, Bence and {Kocsis}, Bence},
        title = "{Stellar Transits in Active Galactic Nuclei}",
      journal = {\apj},
         year = 2013,
        month = jan,
       volume = {762},
       number = {1},
          eid = {35},
        pages = {35},
          doi = {10.1088/0004-637X/762/1/35},
archivePrefix = {arXiv},
       eprint = {1210.4159},
 primaryClass = {astro-ph.GA},
       adsurl = {https://ui.adsabs.harvard.edu/abs/2013ApJ...762...35B}
}

@ARTICLE{2005AJ....130.2473K,
       author = {{Kovalev}, Y.~Y. and {Kellermann}, K.~I. and {Lister}, M.~L. and {Homan}, D.~C. and {Vermeulen}, R.~C. and {Cohen}, M.~H. and {Ros}, E. and {Kadler}, M. and {Lobanov}, A.~P. and {Zensus}, J.~A. and {Kardashev}, N.~S. and {Gurvits}, L.~I. and {Aller}, M.~F. and {Aller}, H.~D.},
        title = "{Sub-Milliarcsecond Imaging of Quasars and Active Galactic Nuclei. IV. Fine-Scale Structure}",
      journal = {\aj},
         year = 2005,
        month = dec,
       volume = {130},
       number = {6},
        pages = {2473-2505},
          doi = {10.1086/497430},
archivePrefix = {arXiv},
       eprint = {astro-ph/0505536},
 primaryClass = {astro-ph},
       adsurl = {https://ui.adsabs.harvard.edu/abs/2005AJ....130.2473K}
}

@INPROCEEDINGS{2008ASPC..386..186K,
       author = {{Krichbaum}, T.~P. and {Lee}, S.~S. and {Lobanov}, A.~P. and {Marscher}, A.~P. and {Gurwell}, M.~A.},
        title = "{How Compact are the Cores of AGN? Sub-Parsec Scale Imaging with VLBI at Millimeter Wavelength}",
    booktitle = {Extragalactic Jets: Theory and Observation from Radio to Gamma Ray},
         year = 2008,
       editor = {{Rector}, T.~A. and {De Young}, D.~S.},
       series = {Astronomical Society of the Pacific Conference Series},
       volume = {386},
        month = jun,
        pages = {186},
          doi = {10.48550/arXiv.0708.3915},
archivePrefix = {arXiv},
       eprint = {0708.3915},
 primaryClass = {astro-ph},
       adsurl = {https://ui.adsabs.harvard.edu/abs/2008ASPC..386..186K}
}

@BOOK{2013degn.book.....M,
       author = {{Merritt}, David},
        title = "{Dynamics and Evolution of Galactic Nuclei (Princeton: Princeton University Press)}",
         year = 2013,
       adsurl = {https://ui.adsabs.harvard.edu/abs/2013degn.book.....M}
}

@ARTICLE{2009ApJ...698..198G,
       author = {{G{\"u}ltekin}, Kayhan and {Richstone}, Douglas O. and {Gebhardt}, Karl and {Lauer}, Tod R. and {Tremaine}, Scott and {Aller}, M.~C. and {Bender}, Ralf and {Dressler}, Alan and {Faber}, S.~M. and {Filippenko}, Alexei V. and {Green}, Richard and {Ho}, Luis C. and {Kormendy}, John and {Magorrian}, John and {Pinkney}, Jason and {Siopis}, Christos},
        title = "{The M-{\ensuremath{\sigma}} and M-L Relations in Galactic Bulges, and Determinations of Their Intrinsic Scatter}",
      journal = {\apj},
         year = 2009,
        month = jun,
       volume = {698},
       number = {1},
        pages = {198-221},
          doi = {10.1088/0004-637X/698/1/198},
archivePrefix = {arXiv},
       eprint = {0903.4897},
 primaryClass = {astro-ph.GA},
       adsurl = {https://ui.adsabs.harvard.edu/abs/2009ApJ...698..198G}
}

@ARTICLE{2009MNRAS.394..660C,
       author = {{Cappellari}, Michele and {Neumayer}, N. and {Reunanen}, J. and {van der Werf}, P.~P. and {de Zeeuw}, P.~T. and {Rix}, H. -W.},
        title = "{The mass of the black hole in Centaurus A from SINFONI AO-assisted integral-field observations of stellar kinematics}",
      journal = {\mnras},
         year = 2009,
        month = apr,
       volume = {394},
       number = {2},
        pages = {660-674},
          doi = {10.1111/j.1365-2966.2008.14377.x},
archivePrefix = {arXiv},
       eprint = {0812.1000},
 primaryClass = {astro-ph},
       adsurl = {https://ui.adsabs.harvard.edu/abs/2009MNRAS.394..660C}
}

@ARTICLE{2020ApJ...903..140Z,
       author = {{Zaja{\v{c}}ek}, Michal and {Araudo}, Anabella and {Karas}, Vladim{\'\i}r and {Czerny}, Bo{\.z}ena and {Eckart}, Andreas},
        title = "{Depletion of Bright Red Giants in the Galactic Center during Its Active Phases}",
      journal = {\apj},
         year = 2020,
        month = nov,
       volume = {903},
       number = {2},
          eid = {140},
        pages = {140},
          doi = {10.3847/1538-4357/abbd94},
archivePrefix = {arXiv},
       eprint = {2009.14364},
 primaryClass = {astro-ph.HE},
       adsurl = {https://ui.adsabs.harvard.edu/abs/2020ApJ...903..140Z}
}
\bibliographystyle{aasjournalv7}



\end{document}